\documentclass[
	envcountsect, % For compability with cleverref
	twocolumn
	]{svjour3}

\usepackage{newtxtext}

\usepackage[utf8]{inputenc}

\usepackage{amsmath}

\usepackage{graphicx}

\usepackage{xspace}

\usepackage{units}

\usepackage{color}
\newenvironment{redEnv}{\color{red}}{}

\newenvironment{magentaEnv}{\color{magenta}}{}

\newcommand{\todoDiss}[1]{}%{\begin{greenEnv}#1\end{greenEnv}}

\usepackage{booktabs}

\usepackage{subfigure}

\usepackage{hyperref}

\usepackage{cleveref}
\crefrangeformat{equation}{(#3#1#4) to~(#5#2#6)}
\crefmultiformat{equation}{(#2#1#3)}{ and~(#2#1#3)}{, (#2#1#3)}{ and~(#2#1#3)}

\usepackage[numbers,sort&compress]{natbib}

\usepackage{rotating}

\usepackage{xurl}

\makeatletter
\define@key{Gin}{resolution}{\pdfimageresolution=#1\relax}
\makeatother

\newcommand{\alignVcenter}[1]{\raisebox{-0.5\height}{#1}}
\usepackage{etoolbox}
\usepackage{xifthen}
\usepackage{bm}
\usepackage{amssymb}
\usepackage{mathtools}

\newcommand{\ie}{i.e.~}
\newcommand{\eg}{e.g.~}
\newcommand{\cf}{cf.~}

\newcommand{\sutsu}{surface-to-surface\xspace}

\newcommand{\btsFull}{beam-to-solid\xspace}
\newcommand{\btsvc}{BTS-TRANS\xspace}
\newcommand{\btsvcFull}{beam-to-solid volume coupling\xspace}
\newcommand{\BtsvcFull}{Beam-to-solid volume coupling\xspace}
\newcommand{\btsvrc}{BTS-FULL\xspace}
\newcommand{\btsvrcFull}{full beam-to-solid volume coupling\xspace}
\newcommand{\BtsvrcFull}{Full beam-to-solid volume coupling\xspace}

\newcommand{\gptslong}{Gauss point-to-segment\xspace}
\newcommand{\gp}{Gauss point\xspace}
\newcommand{\gpts}{GPTS\xspace}

\newcommand{\sr}{Simo--Reissner\xspace}
\newcommand{\kl}{Kirchhoff--Love\xspace}

\newcommand{\cs}[1][]{cross-section#1\xspace}
\newcommand{\pg}{Petrov--Galerkin\xspace}

\newcommand{\gale}{Gauss--Legendre\xspace}
\newcommand{\pk}{Piola--Kirchhoff\xspace}
\newcommand{\svk}{Saint Venant--Kirchhoff\xspace}
\newcommand{\boltz}{Boltzmann\xspace}
\newcommand{\nr}{Newton--Raphson\xspace}
\newcommand{\cosserat}{Cosserat\xspace}

\newcommand{\gexact}{geometrically exact\xspace}
\newcommand{\Gexact}{Geometrically exact\xspace}

\newcommand{\hex}[1]{\emph{hex#1}\xspace}
\newcommand{\tet}[1]{\emph{tet#1}\xspace}
\newcommand{\xfem}{XFEM\xspace}

\newcommand{\fad}{FAD\xspace}

\newcommand{\R}[1]{\mathbb{R}^{#1}}
\newcommand{\C}[1]{\ensuremath{C^{#1}}}
\newcommand{\SO}{SO^3}
\newcommand{\so}{so^3}
\newcommand{\order}[1]{\mathcal O(#1)}
\newcommand{\ei}{\tns{e}_i}
\newcommand{\ex}{\tns{e}_1}
\newcommand{\ey}{\tns{e}_2}
\newcommand{\ez}{\tns{e}_3}

\newcommand{\pfrac}[2]{\frac{\partial #1}{\partial #2}}
\newcommand{\ptfrac}[2]{\frac{\mathrm d #1}{\mathrm d #2}}
\newcommand{\tn}[2]{%
\ifnumcomp{#1}{=}{1}{\underline{\boldsymbol{#2}}}% First order tensor
{\ifnumcomp{#1}{=}{2}{\underline{\boldsymbol{#2}}}% Second order tensor
{\ifnumcomp{#1}{>}{2}{Higher order tensor not yet implemented!}% Higher order tensors
{Wrong tensor order given}%
}}}
\newcommand{\tns}[1]{\tn{1}{#1}}
\newcommand{\tnss}[1]{\tn{2}{#1}}
\newcommand{\tnsO}{\tns{0}} % Zero first order tensor
\newcommand{\tnssI}{\tnss{I}} % Identity second order tensor
\newcommand{\vv}[1]{\boldsymbol{\mathsf{#1}}}
\newcommand{\mat}[1]{\boldsymbol{\mathsf{#1}}}

\newcommand{\vectO}{\vv{0}}
\newcommand{\matO}{\mat{0}}
\newcommand{\matI}{\mat{I}}
\newcommand{\norm}[1]{\left\|#1\right\|}
\newcommand{\normalize}[1]{\frac{#1}{\norm{#1}}}

\newcommand{\tr}{^{\mathrm T}}
\newcommand{\trinv}{^{-\mathrm T}}
\newcommand{\inv}{^{-1}}
\newcommand{\dyad}{\otimes}

\renewcommand{\vector}[1]{\begin{bmatrix}#1\end{bmatrix}}
\renewcommand{\matrix}[1]{\begin{bmatrix}#1\end{bmatrix}}

\newcommand{\brackets}[4][]{%
\ifthenelse{\isempty{#1}}{%
\left#2#4\right#3%
}{%
\ifnumcomp{#1}{=}{0}{#2#4#3}% Default size
{\ifnumcomp{#1}{=}{1}{\bigl#2#4\bigr#3}% bigl
{\ifnumcomp{#1}{=}{2}{\Bigl#2#4\Bigr#3}% Bigl
{\ifnumcomp{#1}{=}{3}{\biggl#2#4\biggr#3}% biggl
{\ifnumcomp{#1}{=}{4}{\Biggl#2#4\Biggr#3}% Biggl
{size not supported}
}}}}}}

\newcommand{\br}[2][]{\brackets[#1]{(}{)}{#2}}
\newcommand{\at}[2][]{\brackets[#1]{.}{|}{#2}}

\newcommand{\encapsulate}[1]{{#1}}

\newcommand{\placeholder}{(\cdot)}

\DeclareMathOperator{\nl}{nl}

\newcommand{\letterbeam}{B}
\newcommand{\lettersolid}{S}

\newcommand{\letterbeamcenterline}{r}
\newcommand{\letterbeamrot}{\theta}

\newcommand{\esolidName}{e}
\newcommand{\esolid}{{(\esolidName)}}
\newcommand{\ebeamName}{f}
\newcommand{\ebeam}{{(\ebeamName)}}
\newcommand{\ebeamsolid}{{(\esolidName,\ebeamName)}}

\newcommand{\nedofsolid}{n_{\text{dof}}^\esolid}
\newcommand{\nedoflagrange}{n_{\text{dof},\lambda}^\ebeam}
\newcommand{\nenlagrange}{n_{\lambda}^\ebeam}

\newcommand{\nesolid}{n_{\text{el},\lettersolid}}
\newcommand{\nebeam}{n_{\text{el},\letterbeam}}

\newcommand{\sbeam}{s}

\newcommand{\error}{\norm{e}_{L2}}
\newcommand{\errorRel}{\norm{e}_{L2,\text{rel}}}

\newcommand{\qGeneral}{\vv{q}}
\newcommand{\dqGeneral}{\delta\qGeneral}

\newcommand{\domainSolidref}{\Omega_{\lettersolid,0}}
\newcommand{\domainSolid}{\Omega_\lettersolid}
\newcommand{\domainSolidh}{\Omega_{\lettersolid,h}}
\newcommand{\domainSolidhe}{\domainSolidh^\esolid}
\newcommand{\domainSolidSurfaceref}{\partial\domainSolidref}
\newcommand{\domainSolidSurface}{\partial\domainSolid}
\newcommand{\domainSolidNeumann}{\Gamma_\sigma}

\newcommand{\domainBeamref}[1][]{\Omega_{\letterbeam#1,0}}
\newcommand{\domainBeam}[1][]{\Omega_{\letterbeam#1}}
\newcommand{\domainBeamh}{\Omega_{\letterbeam,h}}
\newcommand{\domainBeamhe}{\domainBeamh^\ebeam}

\newcommand{\domainCoupling}{\Gamma_{c}}
\newcommand{\domainCouplingh}{\Gamma_{c,h}}
\newcommand{\domainCouplinghe}{\Gamma_{c,h}^{\ebeamsolid}}

\newcommand{\twotothree}{\text{2D-3D}}
\newcommand{\domainCouplingFull}{\Gamma_{\twotothree}}
\newcommand{\domainCouplingFullh}{\Gamma_{\twotothree,h}}
\newcommand{\intSolid}[1]{\int_{\domainSolidref}{ #1 \,\mathrm d V_0}\,}
\newcommand{\intSolidNeumann}[1]{\int_{\domainSolidNeumann}{ #1 \,\mathrm d A_0}\,}
\newcommand{\intBeamCenterline}[1]{\int_{\domainBeamref}{ #1 \,\mathrm d \sbeam}\,}
\newcommand{\intCoupling}[1]{\int_{\domainCoupling}{ #1 \,\mathrm d \sbeam}\,}
\newcommand{\intCouplingh}[1]{\int_{\domainCouplingh}{ #1 \,\mathrm d \sbeam}\,}
\newcommand{\intCouplinghe}[1]{\int_{\domainCouplinghe}{ #1 \,\mathrm d \sbeam}\,}
\newcommand{\intCouplingheOpen}[1]{\int_{\domainCouplinghe}{ #1}}
\newcommand{\intCouplingheClose}[1]{{#1 \,\mathrm d \sbeam}\,}
\newcommand{\intCouplingFull}[1]{\int_{\domainCouplingFull}{ #1 \,\mathrm d A_0}\,}
\newcommand{\intCouplingFullOpen}[1]{\int_{\domainCouplingFull}{ #1}}
\newcommand{\intCouplingFullClose}[1]{{ #1 \,\mathrm d A_0}\,}
\newcommand{\intCouplingFullh}[1]{\int_{\domainCouplingFullh}{ #1 \,\mathrm d A_0}\,}
\newcommand{\intCouplingFullhOpen}[1]{\int_{\domainCouplingFullh}{ #1}}
\newcommand{\intCouplingFullhClose}[1]{{#1 \,\mathrm d A_0}\,}
\newcommand{\indexGauss}{i}
\newcommand{\numberGauss}{n_{\text{GP}}}
\newcommand{\weightGauss}{w_{\indexGauss}}
\newcommand{\intGaussLegendre}[1]{\sum_{\indexGauss=1}^{\numberGauss}{ #1 }}

\newcommand{\nameinternal}{\text{int}}
\newcommand{\nameexternal}{\text{ext}}
\newcommand{\rotMat}{\tnss{R}}
\newcommand{\rotMatArbitrary}{\tnss{R}^{*}}
\newcommand{\rotMatNormal}{\tnss{R}^{\normal}}
\newcommand{\rotMatBeam}{\tnss{R}_{\letterbeam}}
\newcommand{\rotMatBeamtwoD}{\tnss{R}_{\letterbeam}^{\text{2D}}}
\newcommand{\rotMatSolid}{\tnss{R}_{\lettersolid}}
\newcommand{\rotMatSolidtwoD}{\tnss{R}_{\lettersolid}^{\text{2D}}}
\newcommand{\strechtwoD}{\tnss{v}^{\text{2D}}}

\DeclareMathOperator{\rv}{rv}

\newcommand{\nameRef}{\text{ref}}

\newcommand{\triad}{\tnss{\Lambda}}
\newcommand{\triadbeam}{\triad_{\letterbeam}}
\newcommand{\triadbeamO}{\triad_{\letterbeam,0}}
\newcommand{\triadsolid}{\triad_{\lettersolid}}

\newcommand{\triadsolidO}{\triad_{\lettersolid,0}}
\newcommand{\gtriad}[1]{\tns{g}_{#1}}

\newcommand{\dgtriad}[1]{\delta\tns{g}_{#1}}
\newcommand{\gtriadbeam}[1]{\tns{g}_{\letterbeam#1}}
\newcommand{\gtriadbeamO}[1]{\tns{g}_{\letterbeam#1,0}}

\newcommand{\rotvec}{\tns{\psi}}
\newcommand{\rotvecaxis}{\tns{e}_{\psi}}
\newcommand{\rotvecnorm}{\psi}
\newcommand{\drotvec}{\delta\rotvec}
\newcommand{\rotvecbeam}{\rotvec_{\letterbeam}}
\newcommand{\rotvecbeamScalar}{\psi_{\letterbeam}}
\newcommand{\rotvecsolid}{\rotvec_{\lettersolid}}

\newcommand{\drotvecsolid}{\drotvec_{\lettersolid}}

\newcommand{\rotvecbeamsolid}{\rotvec_{\lettersolid\letterbeam}}

\newcommand{\rotmult}{\tns{\theta}}
\newcommand{\drotmult}{\delta\rotmult}
\newcommand{\rotmultbeam}{\tns{\theta}_{\letterbeam}}
\newcommand{\drotmultbeam}{\delta\rotmultbeam}
\newcommand{\rotmultsolid}{\tns{\theta}_{\lettersolid}}
\newcommand{\drotmultsolid}{\delta\rotmultsolid}

\newcommand{\drotmultbeamsolid}{\delta\rotmultsolid}

\newcommand{\objectiveVariation}{\delta_{o}}

\newcommand{\Sskew}[1][]{\tnss{S}\ifthenelse{\isempty{#1}}{}{\br{#1}}}
\newcommand{\Ttrans}{\tnss{T}}
\newcommand{\rotPenaltyTns}{\tnss{c}}
\newcommand{\penRot}{\epsilon_{\letterbeamrot}}
\newcommand{\penTwoThree}{\epsilon_{\twotothree}}
\newcommand{\rCS}{\tns{r}_{\text{CS}}}

\newcommand{\gtriadsolid}[1]{\tns{g}_{\lettersolid #1}}
\newcommand{\gtriadsolidO}[1]{\tns{g}_{\lettersolid #1,0}}
\newcommand{\dgtriadsolid}[1]{\delta\gtriadsolid{#1}}
\newcommand{\namepolar}{\text{POL}}
\newcommand{\triadsolidpolar}{\triad_{\lettersolid,\namepolar}}

\newcommand{\namefixfiber}[1]{{\ensuremath{\text{DIR}_\text{#1}}}}

\newcommand{\triadsolidfixfiber}[1]{\triad_{\lettersolid,\namefixfiber{#1}}}

\newcommand{\namefixtriad}{\text{ORT}}

\newcommand{\nameaverage}{{\ensuremath{\text{AVG}}}}
\newcommand{\gtriadsolidaverage}{\tns{g}_{\lettersolid,\nameaverage}}

\newcommand{\triadsolidaverage}{\triad_{\lettersolid,\nameaverage}}
\newcommand{\triadsolidaverageRef}{\triad_{\lettersolid,\nameaverage,\nameRef}}

\newcommand{\nameSolidTriadShort}{STR}
\newcommand{\solidTriadPolar}{\ensuremath{\text{\nameSolidTriadShort-\namepolar}}\xspace}
\newcommand{\solidTriadFixFiber}[1]{\ensuremath{\text{\nameSolidTriadShort-\namefixfiber{#1}}}\xspace}
\newcommand{\solidTriadFixTriad}{\ensuremath{\text{\nameSolidTriadShort-\namefixtriad}}\xspace}
\newcommand{\solidTriadAverage}{\ensuremath{\text{\nameSolidTriadShort-\nameaverage}}\xspace}

\newcommand{\normal}{\tns{n}}

\newcommand{\rotmultbeamh}{\rotmult_{\letterbeam,h}}
\newcommand{\drotmultbeamh}{\delta\rotmultbeamh}
\newcommand{\drotmultbeamhe}{\encapsulate{\drotmultbeamh^\ebeam}}
\newcommand{\Drotmultbeamh}{\Delta\rotmultbeamh}
\newcommand{\Drotmultbeamhe}{\Drotmultbeamh^\ebeam}

\newcommand{\couplingpotentialRotGPname}{{\epsilon_\letterbeamrot}}
\newcommand{\couplingpotentialRotGP}{\pi_{\couplingpotentialRotGPname}}
\newcommand{\dcouplingPotentialRotGPh}{\delta\Pi_{\couplingpotentialRotGPname,h}}
\newcommand{\dcouplingPotentialRotGPhe}{\dcouplingPotentialRotGPh^\ebeamsolid}
\newcommand{\couplingPotentialRotGP}{\Pi_{\couplingpotentialRotGPname}}
\newcommand{\dcouplingPotentialRotGP}{\delta\couplingPotentialRotGP}

\newcommand{\couplingPotentialFull}{\Pi_{\epsilon,\twotothree}}
\newcommand{\dcouplingPotentialFull}{\delta\couplingPotentialFull}
\newcommand{\dcouplingPotentialFullh}{\delta{\Pi_{\epsilon,\twotothree,h}}}

\newcommand{\couplingPotentialFullLagrange}{\Pi_{\lambda,\twotothree}}
\newcommand{\dcouplingPotentialFullLagrange}{\delta\couplingPotentialFullLagrange}

\newcommand{\rotMortarName}{\lambda_\letterbeamrot}
\newcommand{\couplingPotentialRotMortar}{\Pi_{\rotMortarName}}
\newcommand{\dcouplingPotentialRotMortar}{\delta\couplingPotentialRotMortar}
\newcommand{\dWrotLambda}{\delta W_{\rotMortarName}}

\newcommand{\dWrotLambdahe}{\delta W_{\rotMortarName,h}^\ebeamsolid}
\newcommand{\dWrotC}{\delta W_{C_\letterbeamrot}}

\newcommand{\dWrotChe}{\delta W_{C_\letterbeamrot,h}^\ebeamsolid}

\newcommand{\DQbeamrot}{\Delta\vv{D}^\letterbeam_{\letterbeamrot}}%{\Delta\Qbeamrot}
\newcommand{\qbeamrot}{\hat{\vv{\theta}}_\letterbeam}
\newcommand{\dqbeamrot}{\delta\qbeamrot}
\newcommand{\dqbeamrote}{\encapsulate{\delta\qbeamrot^\ebeam}}
\newcommand{\Dqbeamrot}{\Delta\qbeamrot}
\newcommand{\Dqbeamrote}{\Dqbeamrot^\ebeam}

\newcommand{\qbeampos}{\vv{d}^\letterbeam}
\newcommand{\Dqbeampos}{\Delta\qbeampos}

\newcommand{\rbeamne}{\hat{\rbeam}^\ebeam}
\newcommand{\rbeampne}{\hat{\tns{t}}^\ebeam}
\newcommand{\rotvecbeamne}{\hat{\rotvec}^\ebeam}
\newcommand{\drotvecbeamne}{\delta\hat{\tns{\theta}}^\ebeam}
\newcommand{\Drotvecbeamne}{\Delta\hat{\tns{\theta}}^\ebeam}

\newcommand{\triadbeamh}{\triad_{\letterbeam,h}}
\newcommand{\triadbeamhe}{\triadbeamh^\ebeam}
\newcommand{\rotvecbeamh}{\rotvec_{\letterbeam,h}}
\newcommand{\rotvecsolidh}{\rotvec_{\lettersolid,h}}

\newcommand{\rotvecbeamsolidh}{\rotvec_{\lettersolid\letterbeam,h}}

\newcommand{\Lrot}{\mat{L}}
\newcommand{\Lrote}{\encapsulate{\Lrot^\ebeam}}
\newcommand{\Lrotne}{L^\ebeam}
\newcommand{\Irot}{\tilde{\mat{I}}}
\newcommand{\Irote}{\Irot^\ebeam}
\newcommand{\Irotne}{\tilde{\tnss{I}}^\ebeam}

\newcommand{\fcbeamRotGP}{\vv{f}_{c,\text{GP}}^{\letterbeam}}
\newcommand{\fcsolidRotGP}{\vv{f}_{c,\text{GP}}^{\lettersolid}}
\newcommand{\rcsolidRotGP}{\vv{r}_{c,\text{GP}}^\lettersolid}
\newcommand{\rcbeamRotGP}{\vv{r}_{c,\text{GP}}^\letterbeam}

\newcommand{\fcbeamRotTtT}{\vv{f}_{c,\letterbeamrot,\twotothree}^{\letterbeam}}
\newcommand{\fcbeamPosTtT}{\vv{f}_{c,\letterbeamcenterline,\twotothree}^{\letterbeam}}
\newcommand{\fcsolidRotTtT}{\vv{f}_{c,\twotothree}^{\lettersolid}}
\newcommand{\rcsolidRotTtT}{\vv{r}_{c,\twotothree}^\lettersolid}
\newcommand{\rcbeamRotTtT}{\vv{r}_{c,\letterbeamrot,\twotothree}^\letterbeam}
\newcommand{\rcbeamPosTtT}{\vv{r}_{c,\letterbeamcenterline,\twotothree}^\letterbeam}

\newcommand{\fcbeamRot}{\vv{f}_{c,\rotMortarName}^{\letterbeam}}

\newcommand{\fcsolidRot}{\vv{f}_{c,\rotMortarName}^{\lettersolid}}

\newcommand{\gcRot}{\vv{g}_{c,\rotMortarName}}

\newcommand{\RcsolidRot}{\vv{R}_{c,\rotMortarName}^\lettersolid}
\newcommand{\rcsolidRot}{\vv{r}_{c,\rotMortarName}^\lettersolid}
\newcommand{\RcbeamRot}{\vv{R}_{c,\rotMortarName}^\letterbeam}
\newcommand{\rcbeamRot}{\vv{r}_{c,\rotMortarName}^\letterbeam}
\newcommand{\RcRot}{\vv{R}_{c,\rotMortarName}}
\newcommand{\rcRot}{\vv{r}_{c,\rotMortarName}}

\newcommand{\rcx}{\vv{r}_{c,\rotMortarName}^{\placeholder}}
\newcommand{\Rcx}{\vv{R}_{c,\rotMortarName}^{\placeholder}}
\newcommand{\qcxx}{\mat{q}_{\placeholder}}
\newcommand{\Qcxx}{\mat{Q}_{\placeholder}}

\newcommand{\qcss}{\mat{q}_{\lettersolid\lettersolid}}
\newcommand{\qcsb}{\mat{q}_{\lettersolid\letterbeam}}
\newcommand{\qcsl}{\mat{q}_{\lettersolid\rotMortarName}}
\newcommand{\Qcss}{\mat{Q}_{\lettersolid\lettersolid}}
\newcommand{\Qcsb}{\mat{Q}_{\lettersolid\letterbeam}}
\newcommand{\Qcsl}{\mat{Q}_{\lettersolid\rotMortarName}}
\newcommand{\qcbs}{\mat{q}_{\letterbeam\lettersolid}}
\newcommand{\qcbb}{\mat{q}_{\letterbeam\letterbeam}}
\newcommand{\qcbl}{\mat{q}_{\letterbeam\rotMortarName}}
\newcommand{\Qcbs}{\mat{Q}_{\letterbeam\lettersolid}}
\newcommand{\Qcbb}{\mat{Q}_{\letterbeam\letterbeam}}
\newcommand{\Qcbl}{\mat{Q}_{\letterbeam\rotMortarName}}
\newcommand{\qcls}{\mat{q}_{\rotMortarName\lettersolid}}
\newcommand{\qclb}{\mat{q}_{\rotMortarName\letterbeam}}
\newcommand{\Qcls}{\mat{Q}_{\rotMortarName\lettersolid}}
\newcommand{\Qclb}{\mat{Q}_{\rotMortarName\letterbeam}}

\newcommand{\scalingMatrixRot}{\mat{\kappa}_{\rotMortarName}}
\newcommand{\ScalingMatrixRot}{\mat{V}_{\rotMortarName}}

\newcommand{\lagrangeRot}{\tns{\lambda}_{\letterbeamrot}}
\newcommand{\lagrangeRothe}{\encapsulate{\tns{\lambda}_{\letterbeamrot,h}^{\ebeam}}}
\newcommand{\dlagrangeRot}{\delta\lagrangeRot}

\newcommand{\lagrangeRotTtT}{\tns{\lambda}_{\twotothree}}
\newcommand{\dlagrangeRotTtT}{\delta\lagrangeRotTtT}

\newcommand{\QlagrangeRot}{\vv{\Lambda}_{\letterbeamrot}}
\newcommand{\qlagrangeRot}{\vv{\lambda}_{\letterbeamrot}}
\newcommand{\qlagrangeRote}{\encapsulate{\qlagrangeRot^\ebeam}}
\newcommand{\qlagrangeRoten}{\vv{\lambda}^\ebeam_{\letterbeamrot,i}}
\newcommand{\dqlagrangeRot}{\delta\qlagrangeRot}

\newcommand{\NlagrangeRot}{\mat{\Phi}}
\newcommand{\NlagrangeRote}{\encapsulate{\NlagrangeRot^\ebeam}}
\newcommand{\NlagrangeRoten}{\Phi^\ebeam}
\newcommand{\NlagrangeRotn}{\Phi}

\newcommand{\dWsolid}{\delta W^\lettersolid}
\newcommand{\Xsolid}{\tns{X}_\lettersolid}
\newcommand{\xsolid}{\tns{x}_\lettersolid}
\newcommand{\Xsolidr}{\tns{X}_{\lettersolid,r}}
\newcommand{\xsolidr}{\tns{x}_{\lettersolid,r}}
\newcommand{\dxsolid}{\delta\tns{x}_\lettersolid}
\newcommand{\usolid}{\tns{u}_\lettersolid}
\newcommand{\dusolid}{\delta\usolid}
\newcommand{\F}{\tnss{F}}
\newcommand{\FNormal}{\tnss{F}^{\tns{n}}}
\newcommand{\FtwoD}{\tnss{F}^{\text{2D}}}
\newcommand{\Spk}{\tnss{S}}
\newcommand{\E}{\tnss{E}}
\newcommand{\dE}{\delta \E}
\newcommand{\loadSolidBody}{\hat{\tns{b}}}
\newcommand{\loadSolidSurface}{\hat{\tns{t}}}

\newcommand{\usolidh}{\tns{u}^\lettersolid_h}
\newcommand{\usolidhref}{{\tns{u}^\lettersolid_{\text{ref}}}}
\newcommand{\Xsolidhe}{\tns{X}^{\lettersolid \esolid}_h}
\newcommand{\usolidhe}{\tns{u}^{\lettersolid \esolid}_h}
\newcommand{\dusolidhe}{\delta\tns{u}^{\lettersolid \esolid}_h}

\newcommand{\Qsolid}{\encapsulate{{\vv{D}^\lettersolid}}}

\newcommand{\DQsolid}{\Delta\Qsolid}
\newcommand{\qsolid}{\encapsulate{{\vv{d}^\lettersolid}}}
\newcommand{\qsolide}{\encapsulate{\vv{d}^{\lettersolid\esolid}}}
\newcommand{\qxsolide}{\encapsulate{\vv{X}^{\lettersolid\esolid}}}
\newcommand{\dqsolid}{\encapsulate{\delta\qsolid}}
\newcommand{\dqsolide}{\encapsulate{\delta\qsolide}}
\newcommand{\Dqsolid}{\encapsulate{\Delta\qsolid}}

\newcommand{\hsolid}{h_{\text{solid}}}

\newcommand{\Nsolid}{\mat{N}}
\newcommand{\Nsolide}{\mat{N}^\esolid}

\newcommand{\Kss}{\mat{K}^\lettersolid}
\newcommand{\Rsolid}{\vv{R}^\lettersolid}

\newcommand{\xisolid}{\xi^\lettersolid}
\newcommand{\etasolid}{\eta^\lettersolid}
\newcommand{\zetasolid}{\zeta^\lettersolid}

\newcommand{\Pbeam}{\Pi_{\nameinternal,\letterbeam}}
\newcommand{\dPbeam}{\delta \Pbeam}
\newcommand{\dWbeam}{\delta W^\letterbeam}
\newcommand{\dWbeamext}{\dWbeam_{\nameexternal}}

\newcommand{\rbeamO}{\tns{r}_0}
\newcommand{\rbeam}{\tns{r}}
\newcommand{\drbeam}{\delta\rbeam}
\newcommand{\ubeam}{\tns{u}_\letterbeam}
\newcommand{\csa}{\alpha}
\newcommand{\csb}{\beta}

\newcommand{\rbeamOhe}{\tns{r}_{0,h}^\ebeam}
\newcommand{\rbeamhe}{\tns{r}_{h}^\ebeam}
\newcommand{\Qbeam}{\vv{D}^\letterbeam_\letterbeamcenterline}

\newcommand{\DQbeam}{\Delta\Qbeam}
\newcommand{\qbeam}{\vv{d}^\letterbeam}
\newcommand{\dqbeam}{\delta\qbeam}

\newcommand{\qbeame}{\vv{d}^{\letterbeam\ebeam}}

\newcommand{\qxbeame}{\vv{X}^{\letterbeam\ebeam}}

\newcommand{\Nbeam}{\mat{H}}
\newcommand{\Nbeame}{\Nbeam^\ebeam}

\newcommand{\xibeam}{\xi^\letterbeam}
\newcommand{\xibeamGauss}{\tilde{\xi}^\letterbeam_{i}}

\newcommand{\Krr}{\mat{K}^\letterbeam_{rr}}
\newcommand{\Krt}{\mat{K}^\letterbeam_{r\letterbeamrot}}
\newcommand{\Ktr}{\mat{K}^\letterbeam_{\letterbeamrot r}}
\newcommand{\Ktt}{\mat{K}^\letterbeam_{\letterbeamrot\letterbeamrot}}
\newcommand{\Rbeam}{\vv{R}^\letterbeam_{r}}
\newcommand{\RbeamRot}{\vv{R}^\letterbeam_{\letterbeamrot}}

\newcommand{\srTension}{\tns{\Gamma}}
\newcommand{\srTensionC}{\tnss{C}_F}
\newcommand{\srBending}{\tns{\Omega}}
\newcommand{\srBendingC}{\tnss{C}_M}

\newcommand{\penPos}{\epsilon_\letterbeamcenterline}

\newcommand{\M}{\mat{M}}
\newcommand{\D}{\mat{D}}

\newcommand{\ScalingMatrix}{\mat{V}_{r}}
\newcommand{\Rc}{\vv{R}_{c,r}}

\newcommand{\Qlagrange}{\vv{\Lambda}_{r}}

\journalname{}

\begin{document}

%\title{A consistent coupling approach including rotations for embedding 1D Cosserat beams into 3D solid volumes}
\title{Consistent coupling of positions and rotations for embedding\\1D Cosserat beams into 3D solid volumes}

\author{
    Ivo Steinbrecher\and
    Alexander Popp\and
    Christoph Meier
}

\institute{
    I. Steinbrecher, A.Popp\at
    Institute for Mathematics and Computer-Based Simulation,\\
    University of the Bundeswehr Munich,\\
    Werner-Heisenberg-Weg 39, D-85577 Neubiberg, Germany\\
    \email{ivo.steinbrecher@unibw.de}
    \and
    C. Meier\at
    Institute for Computational Mechanics,\\
    Technical University of Munich,\\
    Boltzmannstrasse 15, D-85748 Garching b. München, Germany
}

\date{Received: date / Accepted: date}

\maketitle

\begin{abstract}
The present article proposes a mortar-type finite element formulation for consistently embedding curved, slender beams into 3D solid volumes. Following the fundamental kinematic assumption of undeformable \cs{s}, the beams are identified as 1D Cosserat continua with pointwise six (translational and rotational) degrees of freedom describing the \cs (centroid) position and orientation.
A consistent 1D-3D coupling scheme for this problem type is proposed, requiring to enforce both positional  and rotational constraints.
Since Boltzmann continua exhibit no inherent rotational degrees of freedom, suitable definitions of orthonormal triads are investigated that are representative for the orientation of material directions within the 3D solid.
While the rotation tensor defined by the polar decomposition of the deformation gradient appears as a natural choice and will even be demonstrated to represent these material directions in a $L_2$-optimal manner, several alternative triad definitions are investigated.
Such alternatives potentially allow for a more efficient numerical evaluation.
Moreover, objective (\ie frame-invariant) rotational coupling constraints between beam and solid orientations are formulated and enforced in a variationally consistent manner based on either a penalty potential or a Lagrange multiplier potential.
Eventually, finite element discretization of the solid domain, the embedded beams, which are modeled on basis of the geometrically exact beam theory, and the Lagrange multiplier field associated with the coupling constraints results in an embedded mortar-type formulation for rotational and translational constraint enforcement denoted as \textit{\btsvrcFull} (\btsvrc) scheme.
Based on elementary numerical test cases, it is demonstrated that a consistent spatial convergence behavior can be achieved and potential locking effects can be avoided, if the proposed \btsvrc scheme is combined with a suitable solid triad definition.
Eventually, real-life engineering applications are considered to illustrate the importance of consistently coupling both translational \textit{and} rotational degrees of freedom as well as the upscaling potential of the proposed formulation.
This allows the investigation of complex mechanical systems such as fiber-reinforced composite materials, containing a large number of curved, slender fibers with arbitrary orientation embedded in a matrix material.
\end{abstract}

% !TEX root = ../beam-to-volume-rotation.tex

\section{Introduction}
\label{sec:introduction}
Embedding fibers or beams, \ie solid bodies that can mechanically be modeled as dimensionally reduced 1D structures since one spatial dimension is much larger than the other two, into a 3D matrix material is a common approach to enhance the mechanical properties of a structure.
Fiber-reinforced structures can be found in many different fields, \eg in form of steel reinforcements within concrete structures, lightweight fiber-reinforced composite materials based on carbon, glass or polymer fibers in a plastic matrix, or additively manufactured components allowing for a very flexible and locally controlled reinforcement of plastic, metal, ceramic or concrete matrix materials~\cite{Pattinson2019, Mattheij1998, Mattheij2000}.
At a different length scale, fiber embeddings play a key role for essential processes in countless biological systems, \eg in the form of embedded networks (\eg cytoskeleton, extracellular matrix, mucus) or bundles (\eg muscle, tendon, ligament)~\cite{Mueller2015, Grill2021, Lieleg2010, Alberts2008}.
Most of these applications are characterized by geometrically complex embeddings of arbitrarily oriented, slender and potentially curved fibers.
Computational models predicting the response of such reinforced structures are essential for a time- and cost-efficient design and development of technical products, but also to gain fundamental understanding of biological systems at length scales that are not accessible via experiments.
In the context of computational modeling, as considered in the following, the embedded 1D structures will be referred to as fibers or beams, respectively, and the 3D matrix as solid.

One common modeling approach for this physical \btsvcFull problem is based on homogenized, anisotropic material models for the combined fiber-matrix structure~\cite{Agarwal2017, Wiedemann2007}.
This widely used approach is appealing since, \eg no additional degrees of freedom are required to model individual fibers, and existing simulation tools can be used as long as they support anisotropic material laws.
However, such models cannot give detailed information about the interactions between fibers and surrounding matrix as, \eg required to study mechanisms of failure.
Moreover, the fiber distribution in the solid has to be sufficiently homogeneous and a separation of scales is required, \ie the fiber size has to be sufficiently small as compared to the smallest dimension of the overall structure. Eventually, when modeling new fiber arrangements, the homogenization step inherent to these continuum models requires sub-scale information, \eg provided by a model with resolved fiber geometries.

Another modeling approach consists of fully describing the fibers and surrounding solid material as 3D continua.
This leads to a \sutsu coupling problem at the 2D interface between fiber surface and surrounding solid.
In the context of the finite element method, these surfaces can be tied together by either applying fiber and solid discretizations that are conforming at the shared interface or via interface coupling schemes accounting for non-matching meshes, such as the mortar method~\cite{Puso2004, Puso2008, Popp2009, Popp2012a}.
Alternatively, extended finite element methods (\xfem) \cite{Moes2003} or immersed finite element methods \cite{Leichner2019b, Rueberg2016} can be used to represent 2D fiber surfaces embedded in an entirely independent background solid mesh.
While such fully resolved modeling approaches allow to study local effects with high spatial resolution, the significant computational effort associated with these models prohibits their usage for large-scale systems with a large number of slender fibers.

The class of applications considered here typically involves very slender fibers.
In this regime it is well justified, and highly efficient from a computational point of view, to model individual fibers as beams, \eg based on the geometrically exact beam theory~\cite{Reissner1972, Simo1986a, Simo1986b, Simo1988, Cardona1988, Ibrahimbegovic1995, Crisfield1999, Romero2004, Meier2014, Meier2019, Betsch2002, Leyendecker2006}, which is known to combine high model accuracy and computational efficiency~\cite{Romero2008, Bauchau2014}.
Based on the fundamental kinematic assumption of undeformable \cs{s}, such beam models can be identified as 1D \cosserat continua with six degrees of freedom defined at each centerline point to describe the \cs position (three translational degrees of freedom) and orientation (three rotational degrees of freedom).
Thus, the problem of beams embedded in a 3D solid volume can be classified as mixed-dimensional 1D-3D coupling problem between 1D \cosserat continua and a 3D Boltzmann continuum.
A variety of 1D-3D coupling approaches exist in the literature, however most of them involve truss / string models, \ie 1D structural models account only for internal elastic energy contributions from axial tension, \eg \cite{Phillips1976, Chang1987, Elwi1989, Ranjbaran1996, Gomes2001, Kang2014, Kerfriden2020}.
Work on the 1D-3D coupling between beams, \ie full \cosserat continua, and solids is much rarer.
In~\cite{Durville2007}, collocation along the beam centerline is applied to couple beams with a surrounding solid material.
A mortar-type coupling approach is proposed in the authors' previous work \cite{Steinbrecher2020}, where a Lagrange multiplier field is defined along the beam centerline to weakly enforce the coupling constraints.
The 1D-3D coupling between beams and a surrounding fluid field, as relevant for fluid-structure-interaction (FSI) problems, has been considered in some recent contributions~\cite{Hagmeyer2021, Tschisgale2020}.

All the aforementioned 1D-3D \btsFull coupling schemes have in common that only the beam centerline positions, but not the \cs orientations, are coupled to the solid, which will be denoted as \textit{translational 1D-3D coupling}. 
%From a mechanical point of view this can be interpreted as embedding fibers with circular \cs within the solid while assuming frictionless contact interaction at the 2D interface between fiber surface and surrounding solid. 
In such models, an embedded fiber can still perform local twist/torsional rotations, \ie \cs rotations with respect to its centerline tangent vector, relative to the solid.
While this simplified coupling procedure can reasonably describe the mechanics of certain problem classes where such relative rotations will rarely influence the global system response, \eg embedding of straight fibers with circular \cs shape, for most practical applications a more realistic description of the physical problem requires to also couple the rotations of beam and solid.

In a very recent approach by~\cite{Khristenko2021} the \textit{full 1D-3D coupling} problem involving positions and rotations has been addressed for the first time.
The coupling of the two directors spanning the (undeformable) beam \cs with the underlying solid continuum together with the coupling of the \cs centroids results in a total of nine coupling constraints.
One specific focus of this interesting contribution lies on a static condensation strategy, which allows to eliminate the associated Lagrange multipliers and the beam balance equations from the final, discrete system of equations.
The requirement of a $C^1$-continuous spatial discretization of the solid domain, as resulting from the proposed condensation strategy, is satisfied by employing NURBS-based test and trial functions.

The present work proposes a \textit{full 1D-3D coupling} approach based on only six, \ie three translational and three rotational, coupling constraints between the \cs{s} of 1D beams, modeled according to the geometrically exact beam theory, and a 3D solid.
The finite element method is employed for spatial discretization of all relevant fields.
Consistently deriving the \textit{full 1D-3D coupling} on the beam centerline from a \textit{2D-3D coupling} formulation on the beam surface via a first-order Taylor series expansion of the solid displacement field would require to fully couple the two orthonormal directors spanning the (undeformable) beam \cs with the (in-plane projection of the) solid deformation gradient evaluated at the \cs centroid position.
It is demonstrated that such an approach, which suppresses all in-plane deformation modes of the solid at the coupling point, might result in severe locking effects in the practically relevant regime of coarse solid mesh sizes.
Therefore, as main scientific contribution of this work, different definitions of orthonormal triads are proposed that are representative for the orientation of material directions of the 3D continuum in an average sense, without additionally constraining in-plane deformation modes when coupled to the beam \cs.
It is shown that the rotation tensor defined by the polar decomposition of the (in-plane projection of the) deformation gradient appears as a natural choice for this purpose, which even represents the average orientation of material directions of the 3D continuum in a $L_2$-optimal manner.
Moreover, several alternative solid triad definitions are investigated that potentially allow for a more efficient numerical evaluation.

Once these solid triads have been defined, objective (\ie frame-invariant) rotational coupling constraints in the form of relative rotations are formulated for each pair of triads representing the beam and solid orientation.
Their variationally consistent enforcement either based on a penalty potential or a Lagrange multiplier potential, with an associated Lagrange multiplier field representing a distributed coupling moment along the beam centerline, is shown.
Eventually, finite element discretization of the Lagrange multiplier and relative rotation vector field along the beam centerline results in an embedded mortar-type formulation for rotational constraint enforcement.
In combination with a previously developed mortar-type formulation (\btsvc) for \textit{translational 1D-3D coupling}~\cite{Steinbrecher2020}, this results in a \textit{full 1D-3D coupling} approach denoted as \btsvrcFull scheme (\btsvrc).
Finite element discretization of the solid and the embedded (potentially curved) beams inevitably results in non-matching meshes, which underlines the importance of a consistently embedded mortar-type formulation as proposed in this work.
Based on elementary numerical test cases, it is demonstrated that a consistent spatial convergence behavior can be achieved and potential locking effects can be avoided if the proposed \btsvrc scheme is combined with a suitable solid triad definition.
Eventually, real-life engineering applications are considered to illustrate the importance of consistently coupling both translational \textit{and} rotational degrees of freedom as well as the upscaling potential of the proposed formulation to study complex mechanical systems such as fiber-reinforced composite materials, containing a large number of curved, slender fibers with arbitrary orientation embedded in a matrix material.

The remainder of this work is organized as follows:
In \Cref{sec:modeling_assumptions}, we state the fundamental modeling assumptions of the proposed \btsvrc scheme. Specifically, the importance of enforcing both rotational and translational coupling conditions is demonstrated, and the general implications of a 1D-3D coupling approach are discussed. 
In \Cref{sec:large_rotations}, we give a short summary of the theory of large rotations as required to formulate rotational coupling conditions.
In \Cref{sec:problem_formulation}, the governing equations for the solid and beam domains are presented, and objective rotational coupling constraints are defined and  enforced in a variationally consistent manner, either based on a penalty or a Lagrange multiplier potential.
In \Cref{sec:solid_triads}, we propose different definitions of orthonormal triads that are suitable to represent the orientation of solid material directions in an average sense.
In \Cref{sec:discretization}, discretization of the coupling conditions based on the finite element method is considered, once in a \gptslong manner and once as mortar-type approach along with a weighted penalty regularization.
Finally, numerical examples, carefully selected to verify different aspects of the proposed formulation, are presented in \Cref{sec:examples}.

\section{Motivation and modeling assumptions}
\label{sec:modeling_assumptions}

In Section~\ref{sec:modeling_assumptions_1D-3D}, the main modeling assumptions generally underlying 1D-3D coupling schemes will be discussed. Subsequently, in Section~\ref{sec:motivation}, the importance of a full position and rotation coupling (\btsvrc) will be motivated for general application scenarios, and special cases will be discussed, where also a purely translational coupling (\btsvc) can be considered as reasonable approximation.

\subsection{Modeling assumptions underlying the 1D-3D coupling}
\label{sec:modeling_assumptions_1D-3D}

The considered class of 1D-3D coupling schemes is based on the assumption that the fiber material is stiff compared to the solid material, and local fiber \cs dimensions are small compared to the global solid dimensions.
Thus, the solid may be discretized without subtracting the fiber volume, formally resulting in overlapping solid and fiber domains.
While consistent \textit{2D-3D coupling} on the fiber surface would allow for high-resolution stress field predictions in the direct vicinity of the 2D fiber-solid interface, such approaches require an evaluation of coupling constraints on a 2D interface and a sufficient discretization resolution of the solid with mesh sizes smaller than the fiber \cs dimensions, thus in large parts deteriorating the advantages provided by a reduced dimensional description of the fibers.

In truly 1D-3D coupling approaches, the coupling conditions are exclusively defined along the beam centerline, thus preserving the computational advantages of the 
dimensionally reduced beam models.
Of course, such approaches inevitably introduce a modeling error as compared to the 2D-3D coupling, \ie the \textit{surface} tractions on the 2D beam-solid interface are approximated by localized resultant \textit{line} forces and moments acting on the beam centerline.
This has a significant impact on the analytical solution of the problem, as line loads acting on a 3D continuum result in singular stress and displacement fields, \cf \cite{Kelvin1848, Podio-Guidugli2014, Favata2012}.
Thus, convergence of the 1D-3D solution towards the 2D-3D solution is not expected.
However, in the realm of the envisioned applications, we are rather interested in global system responses than in local stress distributions in the direct vicinity of the fibers.
Thus, practically relevant solid element sizes are considered that are larger than the fiber \cs dimension.
In this regime of mesh resolutions, this inherent modeling error of 1D-3D approaches can typically be neglected.

\begin{figure}
\centering
\includegraphics[scale=1]{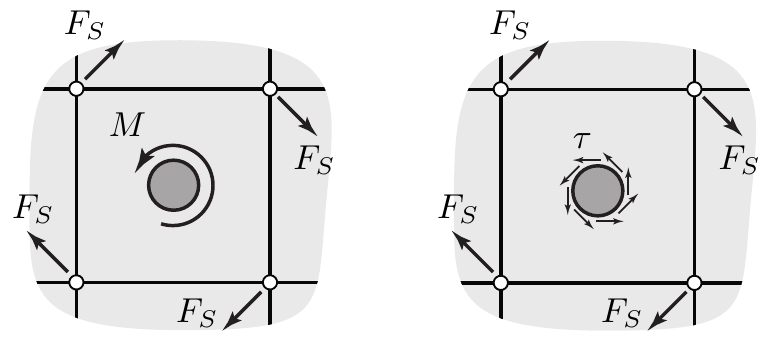}
\caption{Plane coupling problem of a single fiber \cs with a solid finite element mesh -- full 1D-3D coupling (left) vs. 2D-3D coupling (right).}
\label{fig:modeling_assumptions:moment_nodal_forces}
\end{figure}

To verify this statement, consider a plane problem of a beam \cs, loaded with a moment, that is coupled to a solid finite element as depicted in Figure~\ref{fig:modeling_assumptions:moment_nodal_forces}.
As long as the \cs diameter is smaller than the solid finite element mesh size, the resulting discrete nodal forces $F_S$ acting on the solid are independent of the employed coupling approach, \ie either 1D-3D coupling with associated coupling moment $M$ (Figure~\ref{fig:modeling_assumptions:moment_nodal_forces}, left) or 2D-3D coupling with associated coupling surface load $\tau$ (Figure~\ref{fig:modeling_assumptions:moment_nodal_forces}, right).
Obviously, this is an idealized setting, but it illustrates that 1D-3D coupling approaches can be considered as valid models for solid mesh sizes larger than the \cs diameter, which will also be verified in \Cref{sec:examples}.
For a more detailed discussion of this topic the interested reader is referred to our previous publication~\cite{Steinbrecher2020}, specifically to Figure 15 in~\cite{Steinbrecher2020}, which depicts an analogous scenario for the coupling of translational degrees of freedom.

\subsection{Motivation for full translational and rotational coupling}
\label{sec:motivation}

To differentiate the scope of validity of the proposed \btsvrc scheme (coupling of positions and rotations) and of existing \btsvc schemes (coupling of positions only), two application scenarios are discussed.

As first scenario, systems are considered (i) that contain only transversely isotropic fibers (\eg circular \cs shape \textit{and} initially straight) and (ii) whose global system response is dominated by the axial and bending stiffness of the fibers, \ie the torsional contribution is negligible.
As demonstrated in~\cite{Steinbrecher2020}, \btsvc schemes can be considered as a reasonable mechanical model in this case, since local (twist/torsional) rotations of the fibers with respect to their straight axes will rarely influence the global system response.
Torsion-free beam models~\cite{Meier2015} represent an elegant mechanical description of the fibers for such applications. 

As second scenario, systems are considered that contain transversely anisotropic fibers (\eg non-circular \cs shape \textit{or} initially curved).
First, it is clear that twist rotations of the fiber \cs{s} with respect to the centerline tangent (even if not possible in their simplest form as rigid body rotations) will change the global system response, since such fibers exhibit distinct directions of maximal/minimal bending stiffness or initial curvature.
Second, due to the inherent two-way coupling of bending and torsion in initially curved beams~\cite{Meier2015}, bending deformation will inevitably induce torsion in such application scenarios, \ie the global system stiffness is approximated as \textit{too soft} if these torsional rotations are not transferred to the matrix by a proper coupling scheme.
Thus, a unique and consistent mechanical solution for this scenario can only be guaranteed by \btsvrc schemes.

\begin{remark}
In fact, both aforementioned application scenarios might lead to non-unique static solutions if neglecting the rotational coupling. However, for transversely isotropic fibers the non-uniqueness only occurs at the local fiber level, \ie the twist orientation of the fibers is not uniquely defined, which does not influence the global system response. The locally non-unique fiber orientation is typically only an issue from a numerical point of view (\eg linear solvers), and can be effectively circumvented by employing, \eg torsion-free beam models not exhibiting the relevant rotational degrees of freedom. For transversely anisotropic fibers, such local twist rotations will change the global system response. This gives rise to non-unique static solutions on the global level and, thus, has significant implications from a physical point of view.
\end{remark}

\section{Large rotations}
\label{sec:large_rotations}

This section gives a brief overview on the mathematical treatment of finite rotations as required for the formulation of rotational coupling constraints.
For a more comprehesive treatment of this topic, the interested reader is referred to \cite{Simo1986b, Cardona1988, Ibrahimbegovic1995, Romero2004, Meier2019, Betsch1998}.
Let us consider a rotation tensor
\begin{equation}
\label{eq:large_rotations:rottensor}
\triad = \matrix{\gtriad{1}, \gtriad{2}, \gtriad{3}} \in \SO,
\end{equation}
where $\SO$ is the special orthogonal group and the base vectors $\gtriad{i}$ form an orthonormal triad, that maps the Cartesian basis vectors $\tns{e}_{i}$ onto $\gtriad{i}$.
In the following, a rotation pseudo-vector $\rotvec$ is used for its parametrization, \ie $\triad = \triad(\rotvec)$.
The rotation vector describes a rotation by an angle $\rotvecnorm = \norm{\rotvec}$ around the rotation axis $\rotvecaxis = \rotvec / \norm{\rotvec}$.
The parametrization can be given by the well-known Rodrigues formula~\cite{Argyris1982}
\begin{equation}
\label{eq:large_rotations:rodrigues}
\begin{split}
\triad(\rotvec) &= \exp\br{\Sskew[\psi]} \\
&= \tnssI + \sin\rotvecnorm \Sskew[\rotvecaxis] + \br{1-\cos\rotvecnorm}\Sskew^2\br{\rotvecaxis},
\end{split}
\end{equation}
where $\exp\placeholder$ is the exponential map.
%$\exp\br{\tnss{A}} = \tnssI + \tnss{A} + \tnss{A}^2 / 2! + \tnss{A}^3 / 3! + \dots\ $.
Furthermore, $\Sskew \in \so$ is a skew-symmetric tensor, where $\so$ represents the set of skew-symmetric tensors with $\Sskew[\tns{a}] \tns{b} = \tns{a} \times \tns{b} \ \forall \ \tns{a}, \tns{b} \in \R{3}$.
The inverse of the Rodrigues formula \eqref{eq:large_rotations:rodrigues}, \ie the rotation vector as a function of the rotation tensor, will be denoted as $\rotvec(\triad) = \rv(\triad)$ in the remainder of this work.
In practice, Spurrier's algorithm \cite{Spurrier1978} can be used for the extraction of the rotation vector.

Two triads $\triad_1(\rotvec_{1})$ and $\triad_2(\rotvec_{2})$, with their respective rotation vectors $\rotvec_{1}$ and $\rotvec_{2}$, can be related by the relative rotation $\triad_{21}(\rotvec_{21})$.
The relative rotation is given by
\begin{equation}
\label{left_multi}
\begin{split}
\triad_{2}(\rotvec_{2}) &= \triad_{21}(\rotvec_{21}) \triad_1(\rotvec_{1})\\
& \Updownarrow\\
 \triad_{21}(\rotvec_{21}) &= \triad_2(\rotvec_{2}) \triad_1(\rotvec_{1})\tr,
\end{split}
\end{equation}
with the identity $\triad\tr=\triad\inv$ for all elements of $\SO$.
Thus, the (non-additive) rotation vector $\rotvec_{21}=\rv\br{\triad_{21}} \ne \rotvec_{2} - \rotvec_{1}$ describes the relative rotation between $\triad_1$ and $\triad_2$.

In a next step, the variation of the rotation tensor shall be considered, which can be expressed either by an (infinitesimal) additive variation $\drotvec$ of the rotation vector
\begin{equation}
\delta \triad = \at{\ptfrac{}{\epsilon}}_{\epsilon=0} \triad\br{{\rotvec + \epsilon \drotvec}}
= \pfrac{\triad\br{\rotvec}}{\rotvec} \drotvec
,
\end{equation}
or by a (infinitesimal) multiplicative rotation variation $\drotmult$, which is also denoted as spin vector:
\begin{equation}
\delta \triad = \at{\ptfrac{}{\epsilon}}_{\epsilon=0} \triad\br{\epsilon \drotmult} \triad\br{\rotvec}
= \Sskew[\drotmult] \triad\br{\rotvec}
.
\end{equation}
With the relation above and the definition of $\Sskew$, the variations of the triad basis vectors $\dgtriad{i}$ read
\begin{equation}
\dgtriad{i} = \drotmult \times \gtriad{i}.
\end{equation}
The (infinitesimal) additive and multiplicative rotation vector variations can be related according to
\begin{equation}
\label{eq:large_rotations:rot_mult_variation}
\drotvec = \Ttrans(\rotvec) \drotmult,
\end{equation}
where the transformation matrix $\Ttrans(\rotvec)$~\cite{Simo1988} is defined as
\begin{equation}
\begin{split}
\label{eq:large_rotations:transformation_matrix}
\Ttrans(\rotvec) =
& \frac{1}{\rotvecnorm^2} \rotvec \rotvec\tr - \frac{1}{2} \Sskew[\rotvec] \\
+ & \frac{\rotvecnorm}{2 \tan\br{\frac{\rotvecnorm}{2}}} \br{\tnssI - \frac{1}{\rotvecnorm^2} \rotvec \rotvec\tr}
.
\end{split}
\end{equation}
In~\cite{Meier2021}, the objective variation $\objectiveVariation$ of a spatial quantity defined in a moving frame $\triad_1$ is defined as the difference between the total variation and the variation of the base vectors of the moving frame.
In the context of rotational coupling constraints this will be required when expressing the objective variation of a relative rotation vector $\rotvec_{21}$:
\begin{equation}
\label{eq:large_rotations:objective_variation}
\objectiveVariation\rotvec_{21}
=
\drotvec_{21} - \drotmult_1 \times \rotvec_{21}
=
\Ttrans(\rotvec_{21}) (\drotmult_{2} - \drotmult_{1}).
\end{equation}
For a detailed derivation of this expression for the objective variation the interested reader is referred to~\cite{Meier2021}.

\begin{remark}
\label{rem:large_rotations:transformation_matrix_eigenvalue}
Via right-multiplication of \eqref{eq:large_rotations:transformation_matrix} with the rotation vector $\rotvec$ it can easily be shown that $\rotvec$ is an eigenvector (with eigenvalue 1) of $\Ttrans$ and also of $\Ttrans\tr$, \ie $\Ttrans \rotvec = \rotvec$ and $\Ttrans\tr \rotvec = \rotvec$.
This property will be beneficial for derivations presented in subsequent sections.
Every vector parallel to $\rotvec$ is also an eigenvector of $\Ttrans$.
This can be interpreted in a geometrical way:
If the additive increment $\delta \rotvec$ to a rotation vector $\rotvec$ is parallel to the rotation vector, \ie $\delta \rotvec = \delta \rotvecnorm \rotvecaxis$ and $\rotvec = \rotvecnorm \rotvecaxis$, the resulting compound rotation $\rotvec + \delta \rotvec = \br{\rotvecnorm + \delta \rotvecnorm} \rotvecaxis$ is still defined around the rotation axis $\rotvecaxis$.
In this case, the rotation increment is a plane rotation relative to $\triad(\rotvec)$, and the multiplicative and additive rotational increments are equal to each other, $\delta \rotvec = \drotmult$.
\end{remark}

\begin{remark}
In addition to $\triad$, also the symbol $\tnss{R}$ will be used in the following to represent rotation tensors.
\end{remark}

% !TEX root = ../beam-to-volume-rotation.tex

\section{Problem formulation}
\label{sec:problem_formulation}
\begin{figure*}
\centering
\includegraphics[scale=1]{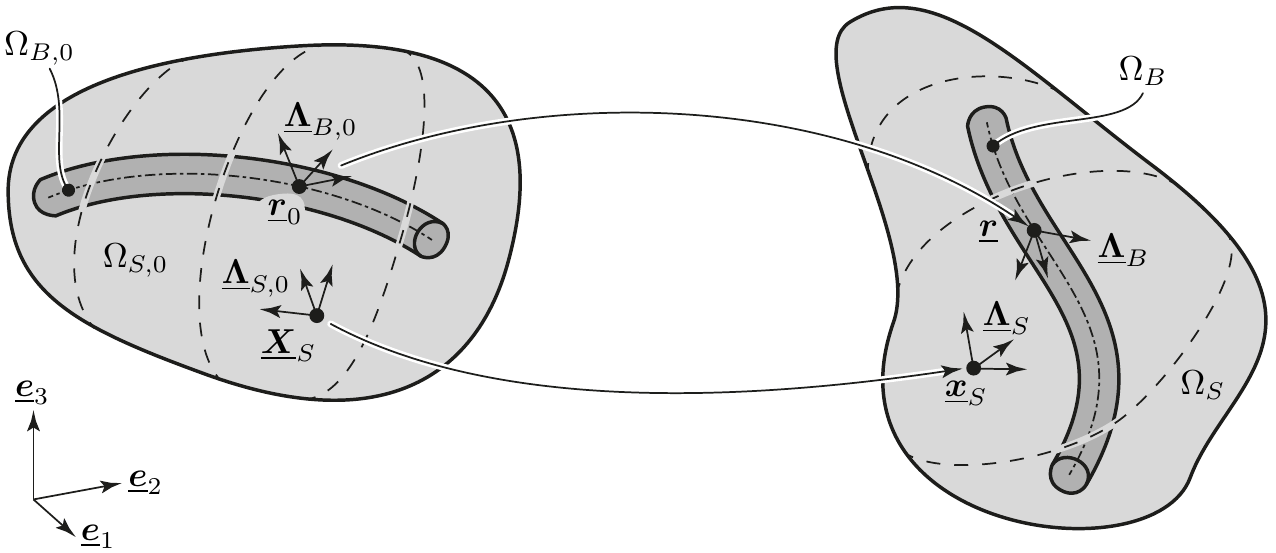}
\caption{Employed notations and relevant kinematic quantities defining the 3D finite deformation \btsvrc problem.}
\label{fig:problem_formulation:btsvrc_problem}
\end{figure*}

We consider a 3D finite deformation \btsvrcFull problem (\btsvrc) as shown in Figure~\ref{fig:problem_formulation:btsvrc_problem}.
All quantities are refereed to a Cartesian frame $\ex, \ey, \ez$.
For simplicity, we focus on quasi-static problems in this work, while the presented \btsvrc method is directly applicable to dynamic problems as well.
The principle of virtual work serves as basis for the proposed finite element formulation.
Contributions to the total virtual work of the system can be split into solid, beam and coupling terms, where the solid and beam terms are independent of the coupling constraints, \ie well-established modeling and discretization techniques can be used for these single fields, \cf \cite{Steinbrecher2020}.

\subsection{Solid formulation}
The solid body is modeled as a 3D \boltz continuum, defined by its domain $\domainSolidref \subset \R3$ in the reference configuration, with boundary $\domainSolidSurfaceref$.
Throughout this work, the subscript $\placeholder_0$ indicates a quantity in the reference configuration.
A solid material point can be identified by its reference position $\Xsolid \in \R3$.
The current position $\xsolid \in \R3$ relates to $\Xsolid$ via the displacement field $\usolid \in \R3$, \ie
\begin{equation}
\xsolid\br{\Xsolid} = \Xsolid + \usolid\br{\Xsolid}.
\end{equation}
The domain and surface of the solid in the deformed configuration are $\domainSolid$ and $\domainSolidSurface$, respectively.
Virtual work contributions $\dWsolid$ of the solid are given by
\begin{equation}
\begin{split}
\dWsolid
=&
\intSolid{\Spk : \dE} \\
&-
\intSolid{\loadSolidBody \cdot \dusolid}
-
\intSolidNeumann{\loadSolidSurface \cdot \dusolid}
,
\end{split}
\end{equation}
where $\delta$ denotes the (total) variation of a quantity, $\Spk \in \R{3\times 3}$ is the second Piola--Kirchhoff stress tensor, $\E\in \R{3\times 3}$ is the work-conjugated Green--Lagrange strain tensor, $\loadSolidBody \in \R3$ is the body load vector and $\loadSolidSurface \in \R3$ are surface tractions on the Neumann boundary $\domainSolidNeumann \subset \domainSolidSurfaceref$.
The Green-Lagrange strain tensor is defined as $\E = \frac{1}{2}\br{\F\tr \F - \tnssI}$, where the deformation gradient $\F\in \R{3\times 3}$ is defined according to
\begin{equation}
\label{eq:problem_formulation:solid_defgrad}
\F = \frac{\partial \xsolid}{\partial \Xsolid}.
\end{equation}
For the compressible or nearly incompressible solid material, we assume existence of a hyperelastic strain energy function $\Psi(\E)$, which allows to determine the second Piola--Kirchhoff stress tensor according to $\Spk=\frac{\partial \Psi(\E)}{\partial \E}$.

\subsection{\Gexact beam theory}
The beams are modeled as 1D \cosserat continua embedded in 3D space based on the \gexact \sr beam theory.
Thus, each beam \cs is described by six degrees of freedom, namely three positional and three rotational degrees of freedom.
This results in six deformation modes of the beam: axial tension, bending ($2\times$), shear ($2\times$) and torsion.
%In the context of the \btsvrc method, employing a beam theory with six degrees of freedom per beam \cs is a natural choice, since in Section~\ref{sec:problem_formulation:btsvrc} six constraint equations per beam \cs will be imposed.

The \cs centroids are connected by the centerline curve $\rbeam(\sbeam) \in \R3$, where $\sbeam \in [0,L] =: \domainBeamref \subset \R{}$ is the arc-length coordinate along the beam centerline $\domainBeamref$ in the reference configuration, and $L$ the corresponding reference length.
The displacement of the beam centerline $\ubeam(\sbeam)\in\R3$ relates the reference position $\rbeamO$ to the current position $\rbeam$ via
\begin{equation}
\rbeam(\sbeam) = \rbeamO(\sbeam) + \ubeam(\sbeam).
\end{equation}
The orientation of the beam \cs field is described by the following field of right-handed orthonormal triads $\triadbeam(\sbeam) := [\gtriadbeam{1}(\sbeam), \gtriadbeam{2}(\sbeam), \gtriadbeam{3}(\sbeam)]=\triadbeam(\rotvecbeam(\sbeam)) \in \SO$, which maps the global Cartesian basis vectors $\ei$ onto the local \cs basis vectors $\gtriadbeam{i}(\sbeam) = \triadbeam \ei$ for $i=1,2,3$. Therein, $\rotvecbeam \in \R{3}$ is the rotation pseudo-vector chosen as parametrization for the triad. Moreover, the triad field in the reference configuration is denoted as $\triadbeamO(\sbeam) := [\gtriadbeam{1,0}(\sbeam), \gtriadbeam{2,0}(\sbeam), \gtriadbeam{3,0}(\sbeam)]=\triadbeamO(\tns{\psi}_{B,0}(\sbeam))$, and the relative rotation between the triads in reference and current configuration is denoted as $\tnss{R}_B:= \triadbeam \triadbeamO\tr$. According to the fundamental kinematic assumption of undeformable \cs{s}, the position of an arbitrary material point within the beam \cs either in the reference or in the current configuration can be expressed as follows:
\begin{align}
\label{eq:problem_formulation:beam_kinematic_assumption}
\tns{X}_B(\sbeam, \csa, \csb) &= \rbeamO(\sbeam) + \csa \gtriadbeam{2,0}(\sbeam) + \csb \gtriadbeam{3,0}(\sbeam),\\
\label{eq:problem_formulation:beam_kinematic_assumption2}
\tns{x}_B(\sbeam, \csa, \csb) &= \rbeam(\sbeam) + \csa \gtriadbeam{2}(\sbeam) + \csb \gtriadbeam{3}(\sbeam),
\end{align}
where $\csa$ and $\csb$ represent in-plane coordinates. Based on a hyperelastic stored-energy function according to
\begin{equation}
\begin{split}
\Pbeam &= \intBeamCenterline{ 
\tilde{\Pi}_{\nameinternal,\letterbeam}} \\
\text{with} \quad \tilde{\Pi}_{\nameinternal,\letterbeam}&=\frac{1}{2}(\srTension\tr \srTensionC \srTension
+
\srBending\tr \srBendingC \srBending)
\end{split}
\end{equation}
the material force stress resultants $\tns{F}=\frac{\partial \tilde{\Pi}_{\nameinternal,\letterbeam}}{\partial \srTension}$ and moment stress resultants $\tns{M}=\frac{\partial \tilde{\Pi}_{\nameinternal,\letterbeam}}{\partial \srBending}$ can be derived. Here, $\srTension \in \R{3}$ is a material deformation measure representing axial tension and shear, $\srBending \in \R3$ is a material deformation measure representing torsion and bending, and $\srTensionC \in \R{3\times3}$ and $\srBendingC \in \R{3\times3}$ are \cs constitutive matrices.
Eventually, the beam contributions to the weak form are given by
\begin{equation}
\dWbeam = \dPbeam + \dWbeamext,
\end{equation}
with the virtual work $\dWbeamext$ of external forces and moments.

\subsection{\BtsvrcFull (\btsvrc)}
\label{sec:problem_formulation:btsvrc}

In the proposed \btsvrc method, the pointwise six degrees of freedom associated with the beam centerline positions and \cs triads are coupled to the surrounding solid, \ie
\begin{align}
\label{eq:problem_formulation:position_coupling}
\rbeam - \xsolid &= \tnsO \quad \text{on} \quad \domainCoupling
\\
\label{eq:problem_formulation:rotation_coupling}
\rotvecbeamsolid &= \tnsO \quad \text{on} \quad \domainCoupling.
\end{align}
Herein, $\domainCoupling = \domainSolidref \cap \domainBeamref$ is the one-dimensional coupling domain between the beam centerline and the solid volume, \ie the part of the beam centerline that lies within the solid.
The rotational coupling between beam \cs and solid as presented in this section is in close analogy to the generalized \cs interaction laws proposed in~\cite{Meier2021}.
The rotation vector $\rotvecbeamsolid$ describes the relative rotation between a beam \cs triad $\triadbeam$ and a corresponding triad $\triadsolid$ associated with the current solid configuration,
\begin{equation}
\label{eq:problem_formulation:relative_rotation_vector}
\rotvecbeamsolid = \rv\br{\triadsolid \triadbeam\tr}.
\end{equation}
Opposite to $\triadbeam$, which is well defined along the beam centerline, there is no obvious or unique definition for $\triadsolid$ in the solid domain.
In Section~\ref{sec:solid_triads}, different definitions of the solid triad $\triadsolid$ are presented and investigated.
However, for the derivation of the coupling equations, it is sufficient to assume the general form $\triadsolid = \triadsolid(\F)$, \ie formulating the solid triad as a general function of the solid deformation gradient in the current configuration.

The formulation of the constraint equations along the beam centerline brings about an advantageous property of the \btsvrc method:
the translational \eqref{eq:problem_formulation:position_coupling} and rotational \eqref{eq:problem_formulation:rotation_coupling} coupling constraints are completely decoupled.
Therefore, the rotational coupling equations \eqref{eq:problem_formulation:rotation_coupling} can be interpreted as a direct extension to the \btsvc method, which only couples the beam centerline positions to the solid as derived and thoroughly discussed in~\cite{Steinbrecher2020}.
In what follows, two different constraint enforcement strategies for the rotational coupling conditions will be presented.

\begin{remark}
In Section~\ref{sec:examples}, we compare the \btsvrc method to a full 2D-3D coupling approach that enforces constraints at the 2D beam-solid interface.
The governing equations, as well as the discretized coupling terms for this 2D-3D coupling scheme are stated in \Cref{sec:appendix:full_2d_3d,sec:appendix:full_2d_3d_discret}.
\end{remark}

\subsubsection{Penalty potential}

We consider a quadratic space-continuous penalty potential between beam \cs triads and solid triads defined along the beam centerline:
\begin{equation}
\label{eq:problem_formulation:total_gp_potential}
\couplingPotentialRotGP
=
\intCoupling{\couplingpotentialRotGP}
= \intCoupling{\frac{1}{2} \rotvecbeamsolid\tr \rotPenaltyTns \rotvecbeamsolid},
\end{equation}
with the \cs coupling potential $\couplingpotentialRotGP = \couplingpotentialRotGP(\sbeam)$ and the symmetric penalty tensor $\rotPenaltyTns \in \R{3\times3}$.
Variation of the penalty potential leads to the following contribution to the weak form:
\begin{equation}
\begin{split}
\dcouplingPotentialRotGP
& =
\intCoupling{\pfrac{\couplingpotentialRotGP}{\rotvecbeamsolid} \objectiveVariation\rotvecbeamsolid} \\
& =
\intCoupling{\br{\objectiveVariation\rotvecbeamsolid}\tr \rotPenaltyTns \rotvecbeamsolid}.
\end{split}
\end{equation}
Therein, $\objectiveVariation\rotvecbeamsolid$ is the objective variation of the rotation vector $\rotvecbeamsolid$.
Making use of \eqref{eq:large_rotations:objective_variation}, the variation of the total potential becomes, \cf  \cite{Meier2021},
\begin{equation}
\dcouplingPotentialRotGP
=
\intCoupling{\br{\drotmultsolid - \drotmultbeam}\tr \Ttrans\tr(\rotvecbeamsolid) \rotPenaltyTns \rotvecbeamsolid},
\end{equation}
where $\drotmultsolid$ and $\drotmultbeam$ are multiplicative variations associated with the solid and beam triad, respectively.
Here, we consider penalty tensors of the form $\rotPenaltyTns = \penRot \tnssI$ with a scalar penalty parameter $\penRot\in\R{+}$ with physical unit $\unit{Nm/m}$.
With this definition and the identity $\Ttrans\tr(\rotvec) \rotvec = \rotvec$ (\cf Remark~\ref{rem:large_rotations:transformation_matrix_eigenvalue}) the variation of the penalty potential simplifies to
\begin{equation}
\label{eq:problem_formulation:total_gp_potential_variation}
\dcouplingPotentialRotGP
=
\penRot \intCoupling{\br{\drotmultsolid - \drotmultbeam}\tr \rotvecbeamsolid}.
\end{equation}
It is well-known from the \gexact beam theory that the (multiplicative) virtual rotations $\drotmultbeam$ are work-conjugated to the moment stress resultants.
Therefore, $\penRot \rotvecbeamsolid$ can be directly interpreted as the (negative) coupling moment acting on the beam \cs.

\subsubsection{Lagrange multiplier potential}
\label{sec:problem_formulation:btsvrc:mortar}
Alternatively, the Lagrange multiplier method can be employed to impose the rotational coupling constraints.
A Lagrange multiplier field $\lagrangeRot = \lagrangeRot(\sbeam) \in \R3$ is therefore defined on the coupling curve $\domainCoupling$.
For now, this field is a purely mathematical construct in the sense of generalized coupling forces associated with the coupling conditions \eqref{eq:problem_formulation:rotation_coupling}.
The Lagrange multiplier potential for the rotational coupling is
\begin{equation}
\label{eq:problem_formulation:total_mortar_potential}
\couplingPotentialRotMortar = \intCoupling{\lagrangeRot\tr \rotvecbeamsolid}.
\end{equation}
Variation of the Lagrange multiplier potential again leads to a constraint contribution to the  weak form, \ie
\begin{equation}
\label{eq:problem_formulation:mortar_variation_potential}
\dcouplingPotentialRotMortar =
\underbrace{\intCoupling{\dlagrangeRot\tr \rotvecbeamsolid}}_{\dWrotLambda}
+
\underbrace{\intCoupling{\lagrangeRot\tr \objectiveVariation\rotvecbeamsolid}}_{-\dWrotC}.
\end{equation}
Therein, $\dWrotLambda$ and $\dWrotC$ are the variational form of the coupling constraints and the virtual work of the generalized coupling forces $\lagrangeRot$, respectively.
With \eqref{eq:large_rotations:objective_variation} the virtual work of the generalized coupling forces becomes
\begin{equation}
\label{eq:problem_formulation:mortar_variation_coupling_constraints}
-\dWrotC
=
\intCoupling{ (\drotmultsolid - \drotmultbeam)\tr \Ttrans\tr(\rotvecbeamsolid) \lagrangeRot}.
\end{equation}
Since the multiplicative rotation variations $\drotmultbeam$ are work-conjugated to the moment stress resultants of the beam,
the term $-\Ttrans\tr(\rotvecbeamsolid) \lagrangeRot$ can be interpreted as a distributed coupling moment acting along the beam centerline.

\begin{remark}
\label{rem:problem_formulation:interpretation_lagrange_rot}
For a vanishing relative rotation $\rotvec_{21} = \tnsO$, as enforced in the space-continuous problem setting according to~\eqref{eq:problem_formulation:rotation_coupling}, the identity $-\Ttrans\tr(\rotvecbeamsolid) = \tnssI$ holds true and the rotational Lagrange multipliers exactly represent the coupling moments along the beam centerline.
However, for the discretized problem this is only an approximation.
\end{remark}

\subsubsection{Objectivity of \btsvrcFull}

As indicated above, the solid triad field depends on the solid deformation gradient $\F$.
It can easily be shown, that the presented solid triad definitions \solidTriadPolar, \solidTriadAverage and \solidTriadFixTriad, in \Cref{sec:solid_triads} are objective with respect to an arbitrary rigid body rotation $\rotMatArbitrary \in \SO$, \ie
\begin{equation}
\label{eq:problem_formulation:rotated_solid}
\triadsolid^* = \triadsolid(\rotMatArbitrary \F)=\rotMatArbitrary\triadsolid(\F).
\end{equation}
The \gexact beam model employed in this contribution is also objective \cite{Meier2014, Meier2019}, \ie
\begin{equation}
\label{eq:problem_formulation:rotated_beam}
\triadbeam^* = \rotMatArbitrary \triadbeam.
\end{equation}
Equations~\Cref{eq:problem_formulation:rotated_solid,eq:problem_formulation:rotated_beam} inserted into the definition of the relative beam-to-solid rotation vector according to~\eqref{eq:problem_formulation:relative_rotation_vector} gives the rotated relative rotation vector,
\begin{equation}
\rotvecbeamsolid^* = \rv(\rotMatArbitrary \triadsolid \triadbeam\tr {\rotMatArbitrary}\tr) = \rotMatArbitrary \rotvecbeamsolid,
\end{equation}
where the identity $\rv(\rotMatArbitrary \triad {\rotMatArbitrary}\tr) = \rotMatArbitrary \rv(\triad)$ has been used. Thus, the rotational coupling conditions~\eqref{eq:problem_formulation:rotation_coupling} in combination with the proposed solid triad definitions and the employed \gexact beam models are objective.
As shown in~\cite{Meier2021}, in this case also an associated penalty potential of type \eqref{eq:problem_formulation:total_gp_potential} or an associated Lagrange multiplier potential of type~\eqref{eq:problem_formulation:total_mortar_potential} is objective.

The previous considerations show objectivity of the proposed (space-continuous) 1D-3D coupling approaches.
However, in the realm of the finite element method, \cf \Cref{sec:discretization}, it is important to demonstrate that objectivity is preserved also in the discrete problem setting.
It is well known that the discretized deformation gradient, as required for the definition of solid triads, is objective as long as standard discretization schemes (\eg via Lagrange polynomials) are applied to the displacement field of the solid.
Also the employed beam finite element formulation based on the geometrically exact beam theory is objective, even though this topic is not trivial and the interested reader is referred to \cite{Meier2014, Meier2019}.
Therefore, it can be concluded that the proposed 1D-3D coupling schemes are objective for the space-continuous as well as for the spatially discretized problem setting.

\begin{remark}
\label{rem:objectivity}
Objectivity is the main reason for formulating the rotational coupling constraints~\eqref{eq:problem_formulation:rotation_coupling} based on the relative rotation vector, \ie $\rotvecbeamsolid = \tnsO$, \cf \cite{Meier2021}.
As alternative choice for the rotational coupling constraints the difference between the beam and solid triad rotation vectors, \ie $\rotvecbeam - \rotvecsolid = \tnsO$, could be considered.
However, such coupling constraints would result in a non-objective coupling formulation~\cite{Meier2021}.
\end{remark}

% !TEX root = ../beam-to-volume-rotation.tex

\section{Definition of solid triad field}
\label{sec:solid_triads}

One of the main aspects of the present work is the definition of a suitable right-handed orthonormal triad field $\triadsolid$ in the solid, which is required for the coupling constraint~\eqref{eq:problem_formulation:rotation_coupling}.
This is by no means a straightforward choice, and different triad definitions will lead to different properties of the resulting numerical coupling scheme.
In the following, a brief motivation will be given for the concept of solid triads before different solid triad definitions will be proposed.

\subsection{Motivation of the solid triad concept}
\label{sec:solid_triads_motivation}

If the embedded beam is considered as a 3D body, a consistent 2D-3D coupling constraint between the 2D beam surface and the surrounding 3D solid can be formulated as 
\begin{equation}
\label{eq:solid_triads_2d-3d_coupling}
\tns{x}_B - \xsolid = \tnsO \quad \text{on} \quad \domainCouplingFull.
\end{equation}
Therein, $\domainCouplingFull$ is the 2D-3D coupling surface, \ie the part of the beam surface that lies withing the solid volume.
In the following, let $\Xsolidr$ denote the line of material solid points that coincide with the beam centerline in the reference configuration, \ie $\Xsolidr=\rbeamO$. Furthermore, the orthonormal triad $\triadsolidO= [\gtriadsolid{1,0}, \gtriadsolid{2,0}, \gtriadsolid{3,0}]$ shall represent material directions of the solid that coincide with the beam triad in the reference configuration according to
\begin{equation}
\label{eq:refsol_equal_refbeam}
\triadsolidO = \triadbeamO.
\end{equation}
The corresponding quantities in the deformed configuration are denoted as $\xsolidr$ and $\triadsolid$. Let us now expand the position field in the solid as Taylor series around $\xsolidr$, \ie
\begin{equation}
\label{eq:solid_triads_taylor}
\xsolid = \xsolidr + \F \,\, \Delta \tns{X} + O( \Delta \tns{X}^2),
\end{equation}
where $\tns{F}$ is the deformation gradient of the solid according to~\eqref{eq:problem_formulation:solid_defgrad}. The 1D-3D coupling strategy underlying the proposed \btsvrc scheme relies on the basic assumption of slender beams, \ie $R \ll L$, where $R$ is a characteristic \cs dimension (\eg the radius of circular \cs{s}). This assumption allows to truncate the Taylor series after the linear term as long as small increments $\Delta \tns{X}=\csa \gtriadsolidO{2} + \csb \gtriadsolidO{3}$, with $\csa,\csb \le R$, are considered:
\begin{equation}
\label{eq:solid_triads_taylor2}
\xsolid \approx \xsolidr + \csa \gtriadsolid{2} + \csb \gtriadsolid{3},
\end{equation}
which results in an error of order $\order{R^2}$. Here, the directors ${\tns{g}}_{\lettersolid 2}$ and ${\tns{g}}_{\lettersolid 3}$, which are not orthonormal in general, represent the push-forward of the solid directions $\gtriadsolid{2,0}$ and $\gtriadsolid{3,0}$:
\begin{equation}
\label{eq:solid_triads_taylor2b}
\gtriadsolid{i}:=\F \,\, \gtriadsolidO{i} \quad \text{for} \quad i=2,3.
\end{equation}
It follows from~\eqref{eq:solid_triads_taylor2} and~\eqref{eq:problem_formulation:beam_kinematic_assumption2} that the 2D-3D coupling conditions~\eqref{eq:solid_triads_2d-3d_coupling} between the beam surface and the expanded solid position field are exactly fulfilled if the following 1D-3D coupling constraints are satisfied:
\begin{align}
\label{eq:solid_triads_taylor3}
\xsolidr &=\rbeam, \\
\label{eq:solid_triads_taylor4}
\gtriadsolid{2} &=\gtriadbeam{2}, \quad \gtriadsolid{3}=\gtriadbeam{3}.
\end{align}
Coupling constraints of the form~\eqref{eq:solid_triads_taylor4} enforce that the material fibers $\tns{g}_{\lettersolid 2}$ and $\tns{g}_{\lettersolid 3}$ of the solid remain orthonormal during deformation, thus enforcing vanishing in-plane strains of the solid at the coupling point $\tns{x}_{\lettersolid,r}=\rbeam$.
In Section~\ref{sec:examples}, it will be demonstrated that constraints of this type lead to severe locking effects when applied to finite element discretizations that are relevant for the proposed \btsvrc scheme, \ie solid mesh sizes that are larger than the beam \cs dimensions.
It will be demonstrated that such locking effects can be avoided if the solid triad field is defined in a manner that only captures the purely rotational contributions to the local solid deformation at $\xsolidr=\rbeam$ without additionally constraining the solid directors in the deformed configuration.
As will be demonstrated in the next sections, the rotation tensor defined by the polar decomposition of the deformation gradient is an obvious choice for this purpose, but also alternative solid triad definitions are possible.
Table~\ref{tbl:solid_triads:variants} gives an overview of the solid triad variants proposed in the following.
\begin{table*}
\centering
\caption{Listing of the different solid triad variants presented in this contribution.}
\label{tbl:solid_triads:variants}
\begin{tabular}{ll}
\toprule
solid triad & description \\
\midrule
\solidTriadPolar & obtained from the polar decomposition of the solid deformation gradient\\
\solidTriadFixFiber{2/3} & fix one chosen solid material direction to the solid triad \\
\solidTriadAverage & fix average of two solid material directions to the solid triad \\
\solidTriadFixTriad & orthogonal solid material directions stay orthogonal\\
\bottomrule
\end{tabular}
\end{table*}

All of these solid triad definitions $\triadsolid=[\tilde{\tns{g}}_{\lettersolid 1}, \tilde{\tns{g}}_{\lettersolid 2}, \tilde{\tns{g}}_{\lettersolid 3}]$ will be a function of the solid deformation gradient $\F$, \ie $\triadsolid = \triadsolid(\F)$.
Moreover, all solid triad definitions will be constructed in a manner such that the associated orthonormal base vectors $\tilde{\tns{g}}_{\lettersolid 2}$ and $\tilde{\tns{g}}_{\lettersolid 3}$ represent the effective rotation of the non-orthonormal directors ${\tns{g}}_{\lettersolid 2}$ and ${\tns{g}}_{\lettersolid 3}$ in an average sense. Thus, it will be required that $\tilde{\tns{g}}_{\lettersolid 2}$ and $\tilde{\tns{g}}_{\lettersolid 3}$ lie within a plane defined by the normal vector
\begin{equation}
\label{eq:normal}
\tns{n} = \normalize{{\tns{g}}_{\lettersolid 2} \times {\tns{g}}_{\lettersolid 3}},
\end{equation}
in the following denoted as the $\tns{n}$-plane.
Eventually, in the examples in Section~\ref{sec:examples}, two desirable properties of the solid triad field for the proposed \btsvrc method are identified:
\begin{description}
\item[(i)] The solid triad should be invariant, \ie symmetric/unbiased with respect to the reference in-plane beam \cs basis vectors $\gtriadbeamO{2}$ and $\gtriadbeamO{3}$.
\item[(ii)] The resulting \btsvrc method should not lead to locking effects in the spatially discretized coupled problem.
\end{description}
These properties will be investigated for the following solid triad definitions.

\subsection{Polar decomposition of the deformation gradient (\solidTriadPolar)}
\label{sec:solid_triads:polar_decomposition}
Based on polar decomposition, the deformation gradient of the solid problem can be split into a product $\F = \tnss{v} \tnss{R}_S = \tnss{R}_S \tnss{U} $ consisting of a rotation tensor $\tnss{R}_S \in \SO$ and a (spatial or material) positive definite symmetric tensor $\tnss{v}$ or $\tnss{U}$, respectively, which describes the stretch.
An explicit calculation rule for the rotation tensor, \eg based on $\tnss{v}$, can be stated as:
\begin{align}
\tnss{v}^2 &= \F \F\tr, \\
\label{eq:polar_decomp_total}
 \tnss{R}_S &= \tnss{v}^{-1} \F.
\end{align}
As mentioned above, it is desirable that the orthonormal base vectors $\tilde{\tns{g}}_{\lettersolid 2}$ and $\tilde{\tns{g}}_{\lettersolid 3}$ of the solid triad $\triadsolid$ lie in a plane with normal vector $\tns{n}$ according to~\eqref{eq:normal}. It can easily be verified that the rotation tensor $\tnss{R}_S$ associated with the total deformation gradient $\F$ according to~\eqref{eq:polar_decomp_total} will in general not satisfy this requirement. Thus, a modification will be presented in the following to preserve this property.

\subsubsection{Construction of \solidTriadPolar triad}

Since the sought-after solid triad shall be uniquely defined already by the two in-plane directors $\gtriadsolid{2}$ and $\gtriadsolid{3}$, a modified version of the deformation gradient will be considered,
\begin{equation}
\F^{\tns{n}} = \normal \otimes \gtriadsolidO{1} + \gtriadsolid{2} \otimes \gtriadsolidO{2} + \gtriadsolid{3} \otimes \gtriadsolidO{3},
\end{equation}
which consists of the projection of the total deformation gradient $\F$ into the $\normal$-plane extended by the additional term $\normal \otimes \gtriadsolidO{1}$.
This modified deformation gradient ensures that the two relevant in-plane basis vectors are correctly mapped, \ie $\gtriadsolid{2}=\FNormal \gtriadsolidO{2}$ and $\gtriadsolid{3}=\FNormal \gtriadsolidO{3}$, while the third basis vector, which is not relevant for the proposed coupling procedure, is mapped onto the normal vector of the $\normal$-plane, \ie $\normal = \FNormal \gtriadsolidO{1}$.
This specific definition of a deformation gradient allows for the following multiplicative split:
\begin{equation}
\label{sec:2D_3D_split}
\FNormal = \FtwoD \rotMatNormal,
\end{equation}
where $\rotMatNormal$ describes the (pure) rotation from the initial solid triad $\triadsolidO$ onto a (still to be defined) orthonormal intermediate triad $\bar{\triad}=[\bar{\tns{g}}_{1},\bar{\tns{g}}_{2},\bar{\tns{g}}_{3}]$, whose base vectors $\bar{\tns{g}}_{2}$ and $\bar{\tns{g}}_{3}$ lie within the $\normal$-plane, and $\FtwoD$ represents a (quasi-2D) in-plane deformation between $\bar{\tns{g}}_{2}$ and $\bar{\tns{g}}_{3}$ and the non-orthonormal base vectors ${\tns{g}}_{2}$ and ${\tns{g}}_{3}$. Now, by applying the polar decomposition only to the in-plane deformation, \ie
\begin{equation}
\label{eq:polar_decomp_2D}
\FtwoD = \strechtwoD \rotMatSolidtwoD,
\end{equation}
a solid triad can be defined from the initial triad $\triadsolidO$ as:
\begin{equation}
\label{eq:triad_def_polar}
\triad_{S,\namepolar} = \rotMatSolidtwoD  \rotMatNormal \triadsolidO.
\end{equation}
Once an intermediate triad $\bar{\triad}$ is defined, the required rotation tensors $\rotMatSolidtwoD$ and $\rotMatNormal$ can 
be calculated as follows:
\begin{enumerate}
\item $\rotMatNormal = \bar{\triad}  \triadsolidO\tr$,
\item $\FtwoD = \FNormal  (\rotMatNormal)\tr$,
\item $(\strechtwoD)^2 = \FtwoD (\FtwoD)\tr$,
\item $\rotMatSolidtwoD = (\strechtwoD)^{-1} \FtwoD$.
\end{enumerate}
The last remaining question is the definition of the triad $\bar{\triad}$.
It can be shown that the choice of this triad is arbitrary and does not influence the result, since a corresponding in-plane rotation offset would be automatically considered/compensated (in the sense of a superposed rigid body rotation) via the rotational part $\rotMatSolidtwoD$ of the in-plane polar decomposition~\eqref{eq:polar_decomp_2D}.
For example, a simple choice is given by $\bar{\tns{g}}_{1}\!=\tns{n}$, $\bar{\tns{g}}_{2}\!={\tns{g}}_{\lettersolid 2}/ \|{\tns{g}}_{\lettersolid 2}\|$ and $\bar{\tns{g}}_{3}\!=\tns{n} \times \bar{\tns{g}}_{2}$, which coincides with the solid triad definition later discussed in Section~\ref{sec:solidTriadFixFiber}.

\begin{remark}
It can be verified that $\rotMatSolid =  \rotMatSolidtwoD \rotMatNormal$ is fulfilled for quasi-2D deformation states, \eg for pure torsion load cases where the beam axis remains straight during the entire deformation (see example in Section~\ref{sec:examples:convergence}). In this case, the (simpler) polar decomposition of the total deformation gradient $\F$ according to~\eqref{eq:polar_decomp_total} can exploited.
\end{remark}

\subsubsection{Properties of \solidTriadPolar triad}
In contrast to alternative solid triad definitions that will be investigated in the following sections, the definition according to~\eqref{eq:triad_def_polar}, referred to as \solidTriadPolar or by the subscript $\placeholder_{\namepolar}$, is not biased by an ad-hoc choice of material directors in the solid that are coupled to the beam.
Instead, the rotation tensor $\rotMatSolid$ describes the rotation of material directions coinciding with the principle axes of the deformation (\ie it maps the principle axes from the reference to the spatial configuration), which has two important implications: First, the choice of material directions that are coupled depend on the current deformation state and will in general vary in time.
Second, the principle axes represent an orthonormal triad per definition, and, thus the coupling to the beam triad will not impose any constraints on the local in-plane deformation of the solid.
Consequently, this solid triad variant fulfills both requirements (i) and (ii) as stated above.

Eventually, a further appealing property of the \solidTriadPolar triad shall be highlighted. Let $\theta_0 \in [-\pi, \pi]$ represent the orientation of arbitrary in-plane directors in the reference configuration defined to coincide for solid and beam according to $\tns{g}_{S,0}(\theta_0)=\tns{g}_{B,0}(\theta_0)=\cos{(\theta_0)} \, \tns{g}_{B2,0} +\sin{(\theta_0)} \, \tns{g}_{B3,0}$. Their push-forward is given by $\tns{g}_{S}(\theta_S(\theta_0)) = \FNormal \, \tns{g}_{S,0}(\theta_0)$ for the solid and $\tns{g}_{B}(\theta_B(\theta_0)) = \tnss{R}_B \, \tns{g}_{B,0}(\theta_0)$ for the beam, where the angles  $\theta_S \in [-\pi,\pi]$ and $\theta_B \in [-\pi,\pi]$ represent the corresponding in-plane orientations in the deformed configuration (see Appendix~\ref{sec:appendix:proof_least_squares} for a detailed definition). Since in-plane shear deformation is permissible for the solid but not for the beam, the orientations $\theta_S(\theta_0)$ and $\theta_B(\theta_0)$ cannot be identical for all $\theta_0 \in [-\pi,\pi]$ and arbitrary deformation states. However, as demonstrated in Appendix~\ref{sec:appendix:proof_least_squares}, when coupling the beam triad to the \solidTriadPolar triad according to~\eqref{eq:triad_def_polar}, the beam directors $\tns{g}_{B}(\theta_B(\theta_0)$ represent the orientation of the solid directors $\tns{g}_{S}(\theta_S(\theta_0)$ in an average sense such that the following $L_2$-norm is minimized:
\begin{align}
\label{eq:least_square}
\int \limits_{-\pi}^{\pi} \! (\theta_S(\theta_0)-\theta_B(\theta_0))^2 \mathrm d \theta_0 \rightarrow \text{min.} \,\,\, \text{for} \,\,\, \triadbeam \!=\! \triadsolidpolar.\!
\end{align}
In conclusion, \solidTriadPolar is an obvious choice for the solid triad with many favorable properties, \eg it represents the average orientation of material solid directions in a $L_2$-optimal manner. However, it requires the calculation of the square root of a tensor, and more importantly, for latter variation and linearization procedures also the first and second derivatives of the tensor square root with respect to the solid degrees of freedom.
This results in considerable computational costs, since this operation has to be performed at local \gp level. Therefore, alternative solid triad definitions will be proposed in the following that can be computed more efficiently, while still being able to represent global system responses with sufficient accuracy.

\subsection{Alternative solid triad definitions}
\label{sec:solid_triads:alternative_triads}

All solid triad variants considered in the following rely on the non-orthonormal solid directors $\tns{g}_{\lettersolid 2}$ and $\tns{g}_{\lettersolid 3}$ according to~\eqref{eq:solid_triads_taylor2b}, their normalized counterparts 
\begin{equation}
\label{eq:normalized_solid_directors}
\tns{g}_{S i}':=\normalize{\tns{g}_{\lettersolid i}} \quad \text{for} \quad i=2,3
\end{equation}
and the corresponding normal vector $\tns{n}$ according to~\eqref{eq:normal}. Based on these definitions, three different variants will be exemplified in the following.

\subsubsection{Fixed single solid director (\solidTriadFixFiber{\text{2/3}})}
\label{sec:solidTriadFixFiber}

In the first variant, denoted as \solidTriadFixFiber{\text{2/3}}, the orientation of one single solid director, either $\tns{g}_{S2}'$ or $\tns{g}_{S3}'$, is fixed to the solid triad, \cf \Cref{fig:solid_triads:plane_triads_fix_fiber}.
The choice which solid material direction to couple is arbitrary. Therefore, two variants will be distinguished:
\begin{align}
\label{eq:solid_triads_TriadFixFiber_2}
\triadsolidfixfiber{2}
&=
\matrix{
\tns{n}, \tns{g}_{S2}', \tns{n} \times \tns{g}_{S2}'
}
\\
\label{eq:solid_triads_TriadFixFiber_3}
\triadsolidfixfiber{3}
&=
\matrix{
	\tns{n}, \tns{g}_{S3}' \times \tns{n}, \tns{g}_{S3}'
}
,
\end{align}
Since the variant \solidTriadFixFiber{\text{2/3}} does not fulfill the requirement (i) as stated above, it will only be considered for comparison reasons in the 2D verification examples in Section~\ref{sec:examples}.

\subsubsection{Fixed average solid director (\solidTriadAverage)}
\label{sec:solid_triads:3d_triad}

\begin{figure}
\newcommand{\spacingfigure}{0cm}
\centering
\subfigure[]{
	\label{fig:solid_triads:plane_triads_reference}
	\hspace{\spacingfigure}\includegraphics[page=1, scale=1]{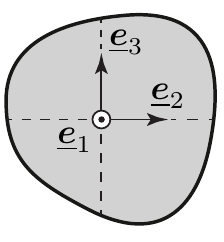}\hspace{\spacingfigure}}
\hfil
\subfigure[]{
	\label{fig:solid_triads:plane_triads_fix_fiber}
	\hspace{\spacingfigure}\includegraphics[page=2, scale=1]{figures/solid_triads_plane_coupling_methods.pdf}\hspace{\spacingfigure}}
\\
\subfigure[]{
	\label{fig:solid_triads:plane_triads_average}
	\hspace{\spacingfigure}\includegraphics[page=4, scale=1]{figures/solid_triads_plane_coupling_methods.pdf}\hspace{\spacingfigure}}
\hfil
\subfigure[]{
	\label{fig:solid_triads:plane_triads_fix_triad}
	\hspace{\spacingfigure}\includegraphics[page=3, scale=1]{figures/solid_triads_plane_coupling_methods.pdf}\hspace{\spacingfigure}}
\caption{
Illustration of \solidTriadFixFiber{2}, \solidTriadAverage and \solidTriadFixTriad solid triad definitions for an exemplary 2D problem setting.
For simplicity it is assumed that the beam reference triad aligns with the Cartesian frame $\ex, \ey, \ez$, \ie $\triadbeamO = \tnssI$.
\subref{fig:solid_triads:plane_triads_reference}~Reference configuration,
\subref{fig:solid_triads:plane_triads_fix_fiber}~\solidTriadFixFiber{2},
\subref{fig:solid_triads:plane_triads_average}~\solidTriadAverage
and \subref{fig:solid_triads:plane_triads_fix_triad}~\solidTriadFixTriad.
}
\label{fig:solid_triads:plane_triads}
\end{figure}

In order to solve this problem, \ie to define a solid triad that is symmetric with respect to the base vectors  $\tns{g}_2'$ and $\tns{g}_3'$, an alternative variant denoted as \solidTriadAverage is proposed, which relies on the average of the directors $\tns{g}_2'$ and $\tns{g}_3'$, \cf \Cref{fig:solid_triads:plane_triads_average}:
\begin{equation}
\label{eq:averaging}
\gtriadsolidaverage = 
\normalize{
	\tns{g}_{S2}' + \tns{g}_{S3}'
}.
\end{equation}
With this average vector the solid triad can be constructed as:
\begin{equation}
\begin{split}
\label{eq:solid_triads:averaged_triad}
\triadsolidaverage &= \tnss{R}\br{-(\pi/4) \tns{n}} \triadsolidaverageRef \\
\text{with} \quad
\triadsolidaverageRef &= \matrix{\tns{n}, \gtriadsolidaverage, \tns{n} \times \gtriadsolidaverage}.
\end{split}
\end{equation}
%Figure~\ref{fig:solid_triads:3D_triads} illustrates the aforementioned construction procedure.
%\begin{figure*}
%\centering
%\includegraphics[scale=1]{figures/solid_triads_3d_triad.pdf}
%\caption{Construction of the three-dimensional solid triad \solidTriadAverage.}
%\label{fig:solid_triads:3D_triads}
%\end{figure*}
The rotation tensor $\tnss{R}\br{-(\pi/4) \tns{n}}$ in~\eqref{eq:solid_triads:averaged_triad} represents a "back-rotation" of the constructed reference triad $\triadsolidaverageRef$ by an angle of $-\pi/4$ to ensure that the resulting solid triad aligns with the beam triad in the reference configuration according to~\eqref{eq:refsol_equal_refbeam}.
In Section~\ref{sec:examples}, it will be shown numerically that the variant \solidTriadAverage, similar to the variant \solidTriadPolar, fulfills both requirements (i) and (ii) stated above.

\begin{remark}
Theoretically, an additive director averaging procedure such as~\eqref{eq:averaging} can result in a singularity if the underlying vectors are anti-parallel, \ie $\tns{g}_{S2}' =- \tns{g}_{S3}'$.
However, since the associated material directors  are orthogonal in the reference configuration, \ie $\tns{g}_{S2,0}\tr \tns{g}_{S3,0}=0$, and shear angles smaller than $\pi/2$ can be assumed, this singularity will not be relevant for practical applications.
\end{remark}

\subsubsection{Fixed orthogonal solid material directions (\solidTriadFixTriad)}
In the last considered solid triad definition, both material directors $\tns{g}_{S2}'$ and $\tns{g}_{S3}'$ are coupled to the solid triad simultaneously.
This variant enforces 
that the directors $\tns{g}_{S2}'$ and $\tns{g}_{S3}'$ remain orthogonal to each other, and thus it is denoted as \solidTriadFixTriad, indicated by a subscript $\placeholder_{\namefixtriad}$.
The \solidTriadFixTriad variant is realized by applying the rotational coupling constraints \eqref{eq:problem_formulation:rotation_coupling} twice, once with $\triadsolidfixfiber{2}$ according to~\eqref{eq:solid_triads_TriadFixFiber_2} and once with $\triadsolidfixfiber{3}$ according to~\eqref{eq:solid_triads_TriadFixFiber_3}.

Opposed to the other triad definitions in this section, this version additionally imposes a constraint on the solid displacement field by enforcing all shear strain components to vanish at the coupling point.
In Section~\ref{sec:examples}, it will be demonstrated that this over-constrained solid triad definition can lead to severe shear locking effects, \ie requirement (ii) from \Cref{sec:solid_triads_motivation} is not satisfied.
Thus, also this variant will only be considered for comparison reasons in the 2D verification examples in Section~\ref{sec:examples}.

\subsection{Variation of the solid rotation vector}
In the coupling contributions to the weak form \cref{eq:problem_formulation:total_gp_potential_variation,eq:problem_formulation:mortar_variation_potential} the multiplicative rotation vector variation $\drotmultsolid$ (spin vector) of a solid rotation vector $\rotvecsolid$ arises.
The spin vector is work-conjugated with the coupling moments, \ie it is required to calculate the virtual work of a moment acting on the solid in a variationally consistent manner.
In contrast to the beam spin vector $\drotmultbeam$, which represents the multiplicative variation of primal degrees of freedom in the finite element discretization of the \gexact \sr beam theory and is discretized directly, no such counterpart exists for the solid field.
Therefore, it is assumed that the solid spin vector can be stated as a function of a set of generalized solid degrees of freedom $\qGeneral$ (which will later be identified as nodal position vectors in the context of a finite element discretization) and their variations $\dqGeneral$.
The additive variation of the solid rotation vector $\rotvecsolid(\qGeneral)$ then reads
\begin{equation}
\drotvecsolid = \pfrac{\rotvecsolid(\qGeneral)}{\qGeneral} \dqGeneral.
\end{equation}
The multiplicative and additive variations are related via \eqref{eq:large_rotations:rot_mult_variation}, which gives the spin vector associated with the solid triad as a function of the generalized solid degrees of freedom:
\begin{equation}
\label{eq:solid_triads:variation}
\drotmultsolid = \Ttrans\inv\br{\rotvecsolid(\qGeneral)} \pfrac{\rotvecsolid(\qGeneral)}{\qGeneral} \dqGeneral.
\end{equation}

\begin{remark}
Alternatively, the solid spin vector can be expressed by the variations of the corresponding solid triad basis vectors $\gtriadsolid{i}$ and their variations $\dgtriadsolid{i}$, \cf \cite{Meier2014, Meier2019}:
\begin{equation*}
\begin{split}
\drotmultbeamsolid
\!=\!&
\br{\dgtriadsolid{2}\tr \gtriadsolid{3}} \gtriadsolid{1}
\!\!+\! \br{\dgtriadsolid{3}\tr \gtriadsolid{1}} \gtriadsolid{2}
\!\!+\! \br{\dgtriadsolid{1}\tr \gtriadsolid{2}} \gtriadsolid{3}
\\
=&
\brackets{(}{.}{
\br{\gtriadsolid{1} \dyad \gtriadsolid{3}} \pfrac{\gtriadsolid{2}}{\qGeneral}
+ \br{\gtriadsolid{2} \dyad \gtriadsolid{1}} \pfrac{\gtriadsolid{3}}{\qGeneral}
}\\
& \brackets{.}{)}{+ \br{\gtriadsolid{3} \dyad \gtriadsolid{2}} \pfrac{\gtriadsolid{1}}{\qGeneral}
} \dqGeneral.
\end{split}
\end{equation*}
This formulation for the solid spin vector is equivalent to the one in \eqref{eq:solid_triads:variation}, but only contains the solid triad basis vectors and their variations.
Therefore, this definition of the solid spin vector is better suited for solid triads constructed via their basis vector.
Especially in the implementation of the finite element formulation, it is advantageous to avoid the computation and inversion of the transformation matrix in \eqref{eq:solid_triads:variation}.
Nonetheless, in the remainder of this contribution, the solid spin vector as defined in \eqref{eq:solid_triads:variation} is used to improve readability of the equations.
\end{remark}

\section{Spatial discretization}
\label{sec:discretization}

In this work, spatial discretization of the beam, solid and coupling problem will exclusively be based on the finite element method.
In the following, a subscript $\placeholder_h$ refers to an interpolated field quantity, superscripts $\esolid$ and $\ebeam$ indicate that the quantity is defined for a solid element $\esolidName$ and a beam element $\ebeamName$, respectively.
Accordingly, $\ebeamsolid$ refers to coupling terms between the solid element $\esolidName$ and beam element $\ebeamName$.
The global element count is made up of $\nesolid$ solid finite elements and $\nebeam$ beam finite elements.

\subsection{Solid and beam problem}
\label{sec:discretization:solid_and_beam}

For the solid domain an isoparametric finite element approach is used to interpolate position, displacement and virtual displacement field within each solid element $\domainSolidhe$:
\begin{align}
\label{eq:discret_fe_solid}
\Xsolidhe &= \Nsolide\br{\xisolid, \etasolid, \zetasolid} \qxsolide
\\
\label{eq:discret_fe_solid_displacement}
\usolidhe &= \Nsolide\br{\xisolid, \etasolid, \zetasolid} \qsolide
\\
\label{eq:discret_fe_solid_virtual_displacement}
\dusolidhe &= \Nsolide\br{\xisolid, \etasolid, \zetasolid} \dqsolide.
\end{align}
Therein, $\Nsolide \in \R{3\times \nedofsolid}$ is the element shape function matrix, which depends on the solid element parameter coordinates $\xisolid, \etasolid,  \zetasolid \in \R{}$.
Furthermore, $\qxsolide \in \R{\nedofsolid}$, $\qsolide \in \R{\nedofsolid}$ and $\dqsolide \in \R{\nedofsolid}$ are the element reference position vector, element displacement vector and element virtual displacement vector, respectively.
Each solid element has $\nedofsolid$ degrees of freedom.

\begin{figure}
\centering
\includegraphics[scale=1]{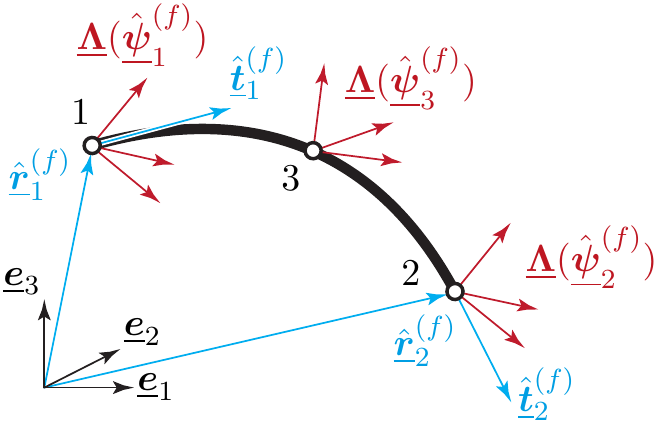}
\caption{
	Degrees of freedom for a single beam element used in this work.
	All quantities related to the beam centerline position are depicted in blue, all \cs orientation related quantities are depicted in red.
}
\label{fig:discretization:beam_element}
\end{figure}

The beam finite elements used in this work are based on the \sr formulation presented in \cite{Meier2018, Meier2019}.
\Cref{fig:discretization:beam_element}. illustrates the degrees of freedom for a single beam finite element.
The beam centerline interpolation is $\C1$-continuous based on third-order Hermite polynomials with two centerline nodes per element.
Each node for the centerline interpolation has 6 degrees of freedom: 3 for the nodal position $\rbeamne_i$ and 3 for the centerline tangent $\rbeampne_i$ at the node, thus resulting in a total of 12 element degrees of freedom describing the beam centerline position.
The interpolated position of the beam centerline is
\begin{equation}
\begin{split}
\rbeamhe &= \Nbeame(\xibeam) \vector{{{}\rbeamne_1}\tr, {{}\rbeampne_1}\tr, {{}\rbeamne_2}\tr, {{}\rbeampne_2}\tr}\tr \\
&= \Nbeame(\xibeam) \br{\qxbeame + \qbeame},
\end{split}
\end{equation}
with the beam position shape function matrix $\Nbeame \in \R{3\times12}$, the beam centerline reference position vector $\qxbeame \in \R{12}$ and the beam centerline displacement vector $\qbeame \in \R{12}$.
Furthermore, $\xibeam \in \R{}$ is the parameter coordinate along the beam centerline.

A triad interpolation scheme based on three element nodal rotation vectors $\rotvecbeamne_1$, $\rotvecbeamne_2$ and $\rotvecbeamne_3$ is utilized \cite{Crisfield1999}.
The third node is placed in the middle of the element and carries no translational degrees of freedom, only rotational ones.
The three local nodal rotation vectors serve as primal degrees of freedom for the interpolated rotation field along the beam centerline.
Each local rotation vector has 3 degrees of freedom, thus resulting in a total of 9 rotational degrees of freedom per beam finite element.
The interpolation of the beam \cs triad along the beam centerline is a non-trivial task and requires an orthonormal interpolation scheme for the interpolated triad $\triadbeamhe(\xibeam)$ to guarantee that the interpolated triad field is still a member of the rotational group $\SO$.
Furthermore, objectivity of the discrete beam deformation measures has to be preserved by the interpolation, which is a challenging task if rotational degrees of freedom are involved.
In this contribution we will refer to the interpolated triad field $\triadbeamhe(\xibeam) = \nl\br{\xibeam, \rotvecbeamne_1, \rotvecbeamne_2, \rotvecbeamne_3}$ as an abstract nonlinear function of the beam parameter coordinate and the nodal rotation vectors.
%Finally, the discretized rotation vector of a beam \cs can be calculated via $\rotvecbeamhe(\xibeam) = rv\br{\triadbeamhe(\xibeam)}$.
The corresponding interpolated field of multiplicative rotation vector increments $\Drotmultbeamhe(\xibeam)$ has been consistently derived in \cite{Crisfield1999} and reads:
\begin{equation}
\label{eq:discretization:multiplicative_rotations}
\Drotmultbeamhe = \sum_{i=1}^{3} \Irotne_i (\xibeam) \Drotvecbeamne_i = \Irote(\xibeam) \Dqbeamrote.
\end{equation}
Therein, $\Irotne_i \in \R{3\times3}$ are generalized shape function matrices for the multiplicative nodal rotation increments $\Drotvecbeamne_i$, and $\Irote \in \R{3 \times 9}$ and $\Dqbeamrote \in \R9$ are the corresponding element-wise assembled quantities.
It should be pointed out that $\Irotne_i$ are nonlinear functions of the beam parameter coordinate and the nodal rotation vectors of the beam element, \ie these rotational shape functions are deformation-dependent.
To avoid this nonlinearity in the discretized spin vector, \ie the virtual rotation field $\drotmultbeamhe$, which would require additional linearization contributions to calculate a consistent tangent, the beam finite elements employed in this work follow a \pg discretization approach.
Therein, standard Lagrange shape functions are used to interpolate the discretized nodal spin vectors:
\begin{equation}
\label{eq:discretization:spin_vector}
\drotmultbeamhe = \sum_{i=1}^{3} \Lrotne_i (\xibeam) \drotvecbeamne_i = \Lrote(\xibeam) \dqbeamrote.
\end{equation}
Here, $\Lrotne_i \in \R{}$ are standard second-order Lagrange polynomials, and $\drotvecbeamne_i$ are the nodal spin vectors. Again, this equation can be assembled element-wise, thus resulting in the shape function matrix $\Lrote \in \R{3 \times 9}$ and the element spin vector $\dqbeamrote \in \R9$.

In what follows, all coupling terms are evaluated on the beam centerline.
This requires the projection of points along the beam centerline parameter space into the solid element parameter space, which in turn is achieved by solving the set of nonlinear equations $\Xsolidhe\br{\xisolid, \etasolid, \zetasolid} = \rbeamOhe\br{\xibeam}$, for a given $\xibeam$.
To improve readability, the superscripts indicating the beam and solid elements will be omitted from now on.
They will however be stated in the virtual work contributions and the integration domains in order to clearly indicate pair-wise values.
Additionally, any dependency on element parameter coordinates will not be stated explicitly.

\begin{remark}
While the $C^1$-continuous centerline representation of the employed beam elements~\cite{Meier2018} is not mandatory for the 
considered \btsvcFull problem, it offers significant advantages in problems additionally involving beam-to-beam~\cite{Meier2017} or beam-to-solid contact interaction, which 
will be addressed in our future research.
\end{remark}

\subsection{\gptslong coupling of \cs rotations}
\label{sec:discretization:gpts_coupling}

Evaluating the variation of the total coupling potential \eqref{eq:problem_formulation:total_gp_potential_variation} based on the discretized solid position field and beam \cs rotation field as presented in the last section yields the discrete variation of the coupling potential:
\begin{equation}
\label{eq:discretization:penalty_potential_integral}
\begin{split}
\dcouplingPotentialRotGPhe = \penRot \intCouplingheOpen{\brackets{(}{.}{ \Ttrans\inv(\rotvecsolidh) \pfrac{\rotvecsolidh}{\qsolid} \dqsolid}}
\\
\intCouplingheClose{\brackets[4]{.}{)}{- \Lrot \dqbeamrot}\tr \rotvecbeamsolidh}.
\end{split}
\end{equation}
Therein, $\domainCouplinghe = \domainBeamhe \cap \domainSolidhe$ is the discretized coupling domain between beam element $\ebeam$ and solid element $\esolid$.
The integral in \eqref{eq:discretization:penalty_potential_integral} is evaluated numerically via a Gauss--Legendre quadrature, resulting in a \gptslong (\gpts) coupling scheme.
From a mechanical point of view this can be interpreted a weighted enforcement of the rotational constraints at each integration point along the beam, \ie a \gptslong type coupling:
\begin{equation}
\label{eq:discretization:penalty_potential_integral_gauss}
\begin{split}
&\dcouplingPotentialRotGPhe \approx \penRot \intGaussLegendre{\brackets{[}{.}{\brackets{(}{.}{ \Ttrans\inv(\rotvecsolidh) \pfrac{\rotvecsolidh}{\qsolid} \dqsolid }}}
\\
&\qquad \qquad \qquad\qquad \brackets[4]{.}{]}{\brackets[4]{.}{)}{- \Lrot \dqbeamrot}\tr \rotvecbeamsolidh}_{\xibeam = \xibeamGauss} \weightGauss,
\end{split}
\end{equation}
where $\numberGauss$ is the number of Gauss--Legendre points, $\xibeamGauss$ is the beam element parameter coordinate for Gauss--Legendre point $\indexGauss$ with the corresponding weight $w_\indexGauss$.
Again, in order to improve the readability of the remaining equations in this subsection, the explicit indication of the evaluation at the Gauss--Legendre points will be omitted in the following.
The previous equation can now be stated in matrix form as
\begin{equation}
\label{eq:discretization:penalty_potential_integral_gauss_matrix}
\begin{split}
\dcouplingPotentialRotGPhe
&\approx
\matrix{\dqbeamrot\tr & \dqsolid\tr}
\intGaussLegendre{\weightGauss
\matrix{\fcbeamRotGP \\ \fcsolidRotGP}}
\\
&=
\matrix{\dqbeamrot\tr & \dqsolid\tr} \matrix{\rcbeamRotGP \\ \rcsolidRotGP}.
\end{split}
\end{equation}
Therein, the abbreviations $\fcbeamRotGP \in \R{9}$ and $\fcsolidRotGP \in \R{\nedofsolid}$ for the generalized \gp coupling forces on the rotational beam degrees of freedom and the generalized solid element degrees of freedom, respectively, have been introduced:
\begin{equation}
\begin{split}
\fcbeamRotGP &= - \penRot \Lrot \tr \rotvecbeamsolidh
\\
\fcsolidRotGP &= \penRot \br{\pfrac{\rotvecsolidh}{\qsolid}}\tr \Ttrans\trinv(\rotvecsolidh) \rotvecbeamsolidh.
\end{split}
\end{equation}
Furthermore, $\rcbeamRotGP \in \R{9}$ and $\rcsolidRotGP \in \R{\nedofsolid}$ are the beam and solid coupling residual vectors.
Employing a \nr algorithm to solve the global system of nonlinear equations, a linearization of the residual vectors with respect to the element degrees of freedom is required, which reads:
\begin{equation}
\matrix{\Delta \rcbeamRotGP \\ \Delta \rcsolidRotGP}
=
\intGaussLegendre{
\weightGauss
\matrix{
\pfrac{\fcbeamRotGP}{\rotvecbeamh} \Ttrans(\rotvecbeamh) \Irot &
\pfrac{\fcbeamRotGP}{\qsolid} \\
\pfrac{\fcsolidRotGP}{\rotvecbeamh} \Ttrans(\rotvecbeamh) \Irot &
\pfrac{\fcsolidRotGP}{\qsolid}
}
\matrix{\Dqbeamrot \\ \Dqsolid}
}.
\end{equation}
Therein, the transformation matrix $\Ttrans(\rotvecbeamh)$ appears, since the linearization is performed with respect to the multiplicative rotation increments $\Drotmultbeamh$.
Furthermore, the generalized shape function matrix $\Irot$ follows from the interpolation of the multiplicative rotation increments, \cf \eqref{eq:discretization:multiplicative_rotations}.
The previously derived matrices and vectors are all defined on \btsFull element pair level.
Since no additional degrees of freedom are introduced, the pair-wise contributions can simply be assembled and added to the global linear system of equations.
The \gptslong coupling approach is presented here to illustrate how the rotational coupling conditions can be enforced in a point-wise manner.
However, in \cite{Steinbrecher2020} it has been shown that a \gptslong coupling approach leads to spurious contact locking for embedded one-dimensional beams in three-dimensional solid volumes.
Therefore, this approach will not be investigated further in the remainder of this contribution, but a mortar-type coupling is proposed instead.

\subsection{Mortar-type coupling of \cs rotations}
\label{sec:discretization:mortar_coupling}
Employing a mortar-type coupling approach, the rotational Lagrange multiplier field $\lagrangeRot$ introduced in Section~\ref{sec:problem_formulation:btsvrc:mortar} is also approximated with a finite element interpolation, \cf \cite{BenBelgacem1999,Popp2010,Wohlmuth2000}.
The rotational Lagrange multiplier field is defined along the beam centerline and accordingly its finite element approximation is defined along the beam finite element and reads as follows:
\begin{equation}
\lagrangeRothe = \sum_{i=1}^{\nenlagrange} \NlagrangeRoten_i(\xibeam) \qlagrangeRoten = \NlagrangeRote(\xibeam) \qlagrangeRote,
\end{equation}
where $\nenlagrange$ is the number of Lagrange multiplier nodes on beam element $\ebeam$, $\NlagrangeRoten_i$ is the shape function for the local node $i$ and $\qlagrangeRoten \in \R{3}$ is the rotational Lagrange multiplier at node $i$.
Furthermore, $\NlagrangeRote \in \R{3 \times 3 \nenlagrange}$ is the element-wise assembled Lagrange multiplier shape function matrix for a beam element and $\qlagrangeRote \in \R{3 \nenlagrange}$ is the vector with all corresponding discrete rotational Lagrange multiplier values per beam element.
As indicated by the dependency on beam parameter coordinate $\xibeam$, the Lagrange multiplier field is defined along the beam centerline.
However, there is no requirement that the Lagrange multiplier shape functions are identical to the beam centerline shape functions, or even that the number of beam nodes matches the number of Lagrange multiplier nodes.
A more thorough discussion on the choice of Lagrange multiplier shape functions is given at the end of this section.

When inserting the finite element interpolations, the discretized variation of the coupling constraints \eqref{eq:problem_formulation:mortar_variation_potential} reads
\begin{equation}
\label{eq:discretization:rotational_constraint_equations}
\dWrotLambdahe = %\intCouplinghe{\dlagrangeRoth\tr \rotvecbeamsolidh}
%=
\dqlagrangeRot\tr \intCouplinghe{
\underbrace{
	\NlagrangeRot\tr \rotvecbeamsolidh
}_{\gcRot}
}
=
\dqlagrangeRot\tr \rcRot.
\end{equation}
Therein, the abbreviations $\gcRot \in \R{\nedoflagrange}$ and $\rcRot \in \R{\nedoflagrange}$ represent the integrand of the pair constraint equations and the residual of the pair constraints equations, respectively.
The discretized virtual work of the coupling forces \eqref{eq:problem_formulation:mortar_variation_coupling_constraints} reads
\begin{equation}
\label{eq:discretization:bts-full-local-residuum}
\begin{split}
-\dWrotChe 
&=
\matrix{\dqbeamrot\tr & \dqsolid\tr}
\matrix{\intCouplinghe{\fcbeamRot} \\ \intCouplinghe{\fcsolidRot}}
\\
&=
\matrix{\dqbeamrot\tr & \dqsolid\tr}
\matrix{\rcbeamRot \\ \rcsolidRot}.
\end{split}
\end{equation}
Therein, the abbreviations $\fcbeamRot \in \R{9}$ and $\fcsolidRot \in \R{\nedofsolid}$ represent the integrand of the beam and solid element coupling forces, \ie
\begin{equation}
\begin{split}
\fcbeamRot &=
	\Lrot\tr \Ttrans\tr(\rotvecbeamsolidh) \NlagrangeRot \qlagrangeRot,
\\
\fcsolidRot &= 
	\pfrac{\rotvecsolidh}{\qsolid}\tr \Ttrans\trinv(\rotvecsolidh) \Ttrans\tr(\rotvecbeamsolidh) \NlagrangeRot \qlagrangeRot.
\end{split}
\end{equation}
Furthermore, $\rcbeamRot \in \R{9}$ and $\rcsolidRot \in \R{\nedofsolid}$ are the beam and solid coupling residual vectors, respectively.
Again, a linearization of the residual contributions with respect to the discrete \btsFull pair degrees of freedom is required for the \nr algorithm.
The linearization is:
\begin{equation}
\label{eq:discretization:bts-full-local-linearized}
\matrix{\Delta\rcbeamRot \\ \Delta\rcsolidRot \\ \Delta\rcRot}
= \matrix{
\qcss & \qcsb & \qcsl \\
\qcbs & \qcbb & \qcbl \\
\qcls & \qclb & \matO
}
\matrix{\Dqbeamrot \\ \Dqsolid \\ \qlagrangeRot}.
\end{equation}
Therein, the abbreviations $\qcxx$ for the stiffness matrices of the pair-wise coupling terms have been introduced, \ie
\begin{equation}
\begin{split}
&\matrix{
\qcss & \qcsb & \qcsl \\
\qcbs & \qcbb & \qcbl \\
\qcls & \qclb & \matO
}
\\& \qquad=
\intCouplinghe{
	\matrix{
		\pfrac{\fcbeamRot}{\rotvecbeamh} \Ttrans(\rotvecbeamh) \Irot &
		\pfrac{\fcbeamRot}{\qsolid} &
		\pfrac{\fcbeamRot}{\qlagrangeRot} \\
		\pfrac{\fcsolidRot}{\rotvecbeamh} \Ttrans(\rotvecbeamh) \Irot &
		\pfrac{\fcsolidRot}{\qsolid} &
		\pfrac{\fcsolidRot}{\qlagrangeRot} \\
		\pfrac{\gcRot}{\rotvecbeamh} \Ttrans(\rotvecbeamh) \Irot &
		\pfrac{\gcRot}{\qsolid} &
		\matO
}}.
\end{split}
\end{equation}
As in the \gpts case, the previously derived vectors and matrices are all defined on \btsFull element pair level.
However, in this case additional unknowns have been introduced, \ie the rotational Lagrange multipliers $\qlagrangeRot$ on pair level.
In practice, all derivatives explicitly stated in \cref{eq:discretization:bts-full-local-residuum,eq:discretization:bts-full-local-linearized} are evaluated using forward automatic differentiation (\fad), \cf \cite{Korelc2016}, using the Sacado software package \cite{SacadoWebsite}, which is part of the Trilinos project \cite{TrilinosWebsite}.

At this point it should be pointed out that all coupling integrals are evaluated numerically using so-called segment-based integration, \cf \cite{Steinbrecher2020,Farah2015}.
Therein, the beam finite element parameter space is divided into subsegments at points where the beam crosses a solid finite element face.
Each subsegment is subsequently integrated using a \gale quadrature with a fixed number of integration points. 
This leads to a highly accurate numerical integration procedure and allows for the resulting finite element coupling method to pass classical patch tests in \sutsu problems as well as constant stress transfer tests in \btsFull problems, \cf \cite{Steinbrecher2020,Farah2015}.

The choice of proper Lagrange multiplier basis functions is important for the mathematical properties of the resulting finite element discretization.
The Lagrange multiplier shape functions must fulfill an $\mathrm{inf}\text{-}\mathrm{sup}$ condition to guarantee stability of the mixed finite element method \cite{Boffi2013}.
This is a well-studied topic in the context of classical \sutsu mesh tying or contact.
However, as pointed out in \cite{Steinbrecher2020}, \btsFull coupling problems diverge from the standard \sutsu case in some aspects.
First, the discretization along the beam centerline with Hermite polynomials is unusual compared to standard (\ie Lagrange polynomial-based) finite element discretizations.
Also, the 1D-3D coupling can be classified as a mixed-dimensional embedded mesh problem, since there is no explicit curve in the solid domain to match the beam centerline, which can lead to stability issues \cite{Sanders2012a}.
Additionally, in this contribution we deal with rotational coupling, which also differs from the standard \sutsu case.
A deep mathematical analysis of these properties is beyond the scope of the present contribution.
However, we will build upon the extensive studies and findings from \cite{Steinbrecher2020}, where it has been shown that a linear interpolation of the Lagrange multipliers combined with a penalty regularization leads to a stable finite element formulation of the coupled problem.
Instabilities might only occur if the beam finite elements become shorter than the solid finite elements.
However, this is not a mesh size relation that is within the envisioned applications for the \btsvrc method.
Alternative approaches to avoid the mentioned instabilities are available in the literature, such as Nitsche's method \cite{Dolbow2009,Hautefeuille2012,Sanders2012},  or discontinuous Galerkin formulations \cite{Hautefeuille2012,Sanders2012a}.
Another appealing approach is the so-called vital vertex method introduced in \cite{Bechet2009}, where discrete Lagrange multipliers are inserted at intersection points between the coupled meshes, \ie the intersections between a beam finite element centerline and the solid elements in the presented \btsFull case.

\subsection{\BtsvcFull (\btsvc)}
In \Cref{sec:discretization:gpts_coupling,sec:discretization:mortar_coupling}, \gpts and mortar-type coupling discretizations for the rotational coupling between the beam \cs and the solid volume \eqref{eq:problem_formulation:rotation_coupling} have been presented.
The translational coupling of the beam centerline \eqref{eq:problem_formulation:position_coupling}, however, is entirely based on the mortar-type \btsvcFull (\btsvc) method previously introduced in \cite{Steinbrecher2020}.
The resulting linearized system of equations for the centerline translation coupling with saddle point structure reads:
\begin{equation}
\label{eq:discretization:btsvc_discret_global_system}
\matrix{
\Kss & \matO & \matO & -\M\tr \\
\matO & \Krr & \Krt & \D\tr \\
\matO & \Ktr & \Ktt & \matO \\
-\M & \D & \matO & \matO
}
\vector{\DQsolid \\ \DQbeam \\ \DQbeamrot \\ \Qlagrange}
=
\begin{bmatrix}
-\Rsolid \\
-\Rbeam \\
-\RbeamRot \\
\M \Qsolid - \D \Qbeam
\end{bmatrix}.
\end{equation}
Therein, $\Kss$ is the solid tangent stiffness matrix, $\Qsolid$ and $\DQsolid$ are the global solid displacement vector and its increment, respectively, and $\Rsolid$ is the residual of the solid degrees of freedom.
In \eqref{eq:discretization:btsvc_discret_global_system}, the beam terms are split into centerline and rotation contributions indicated by $\placeholder_{\letterbeamcenterline}$ and $\placeholder_{\letterbeamrot}$, respectively.
At this point it should be noted that due to the employed \pg method the beam stiffness matrices are non-symmetric, as will be the case for the rotational coupling contributions to the global stiffness matrix.
Furthermore, $\Qbeam$ and $\DQbeam$ are the global beam centerline displacement vector and its increment, and $\DQbeamrot$ represents the global vector collecting the multiplicative rotation increments associated with the nodal triads of the beam finite elements.
The update of the rotation state has to be preformed according to \cite{Meier2019}.
The \btsvc coupling is represented by the discrete mortar matrices $\D$ and $\M$ and the centerline Lagrange multipliers $\Qlagrange$.
The structure of the global stiffness matrix in \eqref{eq:discretization:btsvc_discret_global_system} illustrates that only the beam centerline degrees of freedom (and not the rotational degrees of freedom) are coupled to the solid degrees of freedom in the previously proposed \btsvc coupling scheme.

\subsection{Combined mortar-type coupling of translations and rotations (\btsvrc)}

The global system of equations for the mortar-type \btsvrc method is the combination of the mortar-type coupling for the centerline positions \btsvc \eqref{eq:discretization:btsvc_discret_global_system} and the mortar-type coupling of the beam \cs rotations \eqref{eq:discretization:bts-full-local-linearized}.
The resulting global system of equations becomes:
\begin{equation}
\label{eq:discretization:global_system_mixed_formulation}
\begin{split}
\matrix{
	\Kss + \Qcss & \matO & \Qcsb & -\M\tr & \Qcsl \\
	\matO & \Krr & \Krt & \D\tr & \matO \\
	\Qcbs & \Ktr & \Ktt + \Qcbb & \matO & \Qcbl \\
	-\M & \D & \matO & \matO & \matO \\
	\Qcls & \matO & \Qclb & \matO & \matO
}
\vector{\DQsolid \\ \DQbeam \\ \DQbeamrot \\ \Qlagrange \\ \QlagrangeRot}
\\
= \begin{bmatrix}
-\Rsolid - \RcsolidRot\\
-\Rbeam\\
-\RbeamRot - \RcbeamRot \\
\M \Qsolid - \D \Qbeam \\
-\RcRot
\end{bmatrix},
\end{split}
\end{equation}
where $\Qcxx$ are the globally assembled pair-wise stiffness matrices $\qcxx$, and $\Rcx$ are the globally assembled residual vectors of the local rotational coupling contributions $\rcx$.
Additionally, $\QlagrangeRot$ are the globally assembled rotational Lagrange multipliers $\qlagrangeRot$.
Therefore, the size of the global system of equations of the uncoupled system is extended by the total number of translational and rotational Lagrange multipliers.

\begin{remark}
The structure of the global system of equations for \btsvrc \eqref{eq:discretization:global_system_mixed_formulation} illustrates the direct coupling of the rotational degrees of freedom of the beam with the solid degrees of freedom, \ie $\Qcbl$ and $\Qclb$.
Disregarding all other advantages of our \btsvrc method, this motivates its from a pure numerical point of view, as possible rigid body rotations of straight embedded fibers around their centerline are constrained, which is not the case for the \btsvc method \eqref{eq:discretization:btsvc_discret_global_system}.
\end{remark}

\begin{remark}
In the \btsvc method, the mortar-type coupling matrices $\D$ and $\M$ only depend on the reference configuration, \ie they only have to be calculated once and can be stored for the entire simulation.
In the \btsvrc method, the (rotational) coupling terms $\Qcxx$ depend on the current configuration, \ie the coupling terms have to be re-evaluated in each Newton-Raphson step.
%From a purely computational point of view, this is a drawback of \btsvrc compared to \btsvc.
However, this should not be viewed as a drawback of \btsvrc scheme, rather as a simplification of the \btsvc variant, which results from neglecting the rotational coupling terms.
\end{remark}

\subsection{Penalty regularization}

In the present mortar-type coupling case (\btsvrc) the constraint equations are enforced with the Lagrange multiplier method, thus resulting in a mixed formulation.
However, a direct solution of the global system \eqref{eq:discretization:global_system_mixed_formulation} might introduce certain drawbacks, such as an increased system size compared to the uncoupled system and a generalized saddle point structure.
In \cite{Steinbrecher2020} the constraint equations have therefore been enforced using a well-known penalty regularization, which means that a relaxation of the translational coupling constraints $-\M \Qsolid + \D \Qbeam = \Rc = \vectO$ in the form of $\Qlagrange = \penPos \ScalingMatrix\inv \Rc$ is introduced.
Therein, $\penPos \in \R{+}$ is a scalar penalty parameter and $\ScalingMatrix$ is a scaling matrix to account for non-uniform weighting of the constraint equations \cite{Steinbrecher2020, Yang2005}.
The numerical examples in \cite{Steinbrecher2020} show that for reasonably chosen penalty parameters the resulting violation of the constraint equations due to their relaxation does not have any impact on the accuracy of the \btsvc method.
Therefore, the constraint enforcement of the new rotational coupling equations \eqref{eq:discretization:rotational_constraint_equations} is also carried out with the penalty method.
The constraint relaxation is achieved through
\begin{equation}
\label{eq:discretization:constraint_relaxation}
\QlagrangeRot = \penRot \ScalingMatrixRot\inv \RcRot,
\end{equation}
again with a scalar penalty parameter $\penRot \in \R{+}$ and a global scaling matrix for the rotational Lagrange multipliers $\ScalingMatrixRot$.
The global scaling matrix is assembled from the nodal scaling matrices $\scalingMatrixRot^{(i,i)}$ for the Lagrange multiplier node $i$, \ie
\begin{equation}
\scalingMatrixRot^{(i,i)} = \intCouplingh{\NlagrangeRotn_i} \matI^{3\times 3}.
\end{equation}
With the introduction of the constraint relaxation \eqref{eq:discretization:constraint_relaxation}, the Lagrange multipliers $\QlagrangeRot$ are no longer independent degrees of freedom of the system, but a function of the beam rotations and solid displacements.
Therefore, they can be eliminated from the global system of equations \eqref{eq:discretization:global_system_mixed_formulation}, which results in the condensed linear system of equations
\begin{equation}
\matrix{
\mat{A}_{\lettersolid\lettersolid} & \mat{A}_{\lettersolid\letterbeamcenterline} & \mat{A}_{\lettersolid\letterbeamrot} & \\
\mat{A}_{\letterbeamcenterline\lettersolid} & \mat{A}_{\letterbeamcenterline\letterbeamcenterline} & \mat{A}_{\letterbeamcenterline\letterbeamrot} & \\
\mat{A}_{\letterbeamrot\lettersolid} & \mat{A}_{\letterbeamrot\letterbeamcenterline} & \mat{A}_{\letterbeamrot\letterbeamrot}
}
\vector{\DQsolid \\ \DQbeam \\ \DQbeamrot}
=
\begin{bmatrix}
-\Rsolid - \RcsolidRot\\
-\Rbeam\\
-\RbeamRot - \RcbeamRot
\end{bmatrix}.
\end{equation}
Therein, the following abbreviations have been introduced for improved readability:
\begin{equation}
\begin{split}
\mat{A}_{\lettersolid\lettersolid} &= \penRot \M \ScalingMatrix\inv \M\tr + \Qcss + \penRot \Qcsl \ScalingMatrixRot\inv \Qcls\\ 
\mat{A}_{\lettersolid\letterbeamcenterline} &= -\penRot \M \ScalingMatrix\inv \D \tr\\
\mat{A}_{\lettersolid\letterbeamrot} &= \Qcsb + \penRot \Qcsl \ScalingMatrixRot\inv \Qclb\\
\mat{A}_{\letterbeamcenterline\lettersolid} &= -\penRot \D \ScalingMatrixRot\inv \M \tr\\
\mat{A}_{\letterbeamcenterline\letterbeamcenterline} &= \Krr + \penRot \D \ScalingMatrix\inv \D \tr\\
\mat{A}_{\letterbeamcenterline\letterbeamrot} &= \Krt\\
\mat{A}_{\letterbeamrot\lettersolid} &= \Qcbs + \penRot \Qcbl \ScalingMatrixRot\inv \Qcls\\
\mat{A}_{\letterbeamrot\letterbeamcenterline} &= \Ktr\\
\mat{A}_{\letterbeamrot\letterbeamrot} &= \Ktt + \Qcbb + \penRot \Qcbl \ScalingMatrixRot\inv \Qclb.
\end{split}
\end{equation}

\section{Examples}
\label{sec:examples}

The following numerical examples are set up using the beam finite element pre-processor MeshPy \cite{MeshPyWebsite} and are simulated with our in-house parallel multi-physics research code BACI \cite{BaciWebsite}.

\subsection{Single element moment test}
\begin{figure}
\centering
\includegraphics[scale=1]{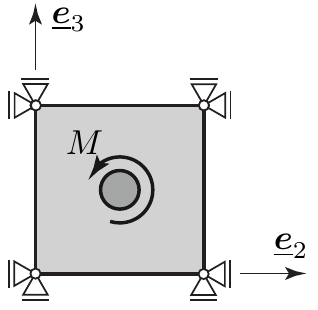}
\caption{Moment test problem.}
\label{fig:examples:moment_test_problem}
\end{figure}

\begin{figure}
\centering
\subfigure[\solidTriadPolar]{\label{fig:examples:moment_test_results:polar}\includegraphics[resolution=300]{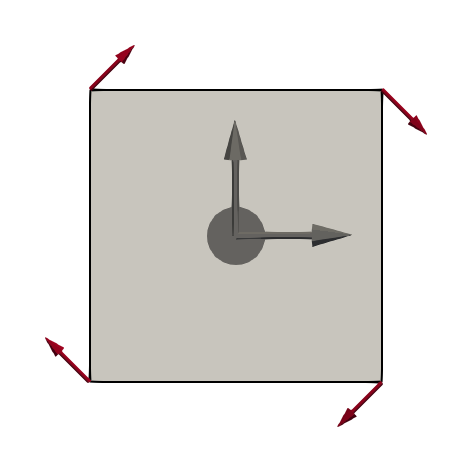}}
\subfigure[\solidTriadFixFiber{2}]{\label{fig:examples:moment_test_results:fix2}\includegraphics[resolution=300]{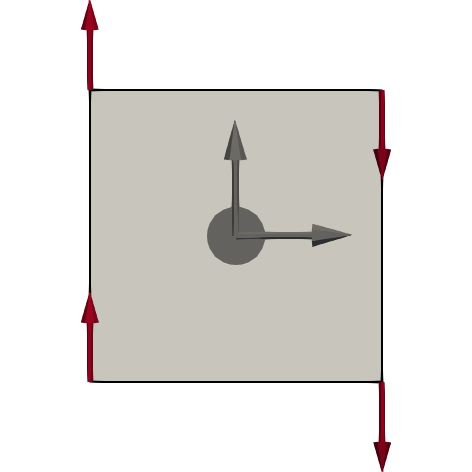}}
\subfigure[\solidTriadFixFiber{3}]{\label{fig:examples:moment_test_results:fix3}\includegraphics[resolution=300]{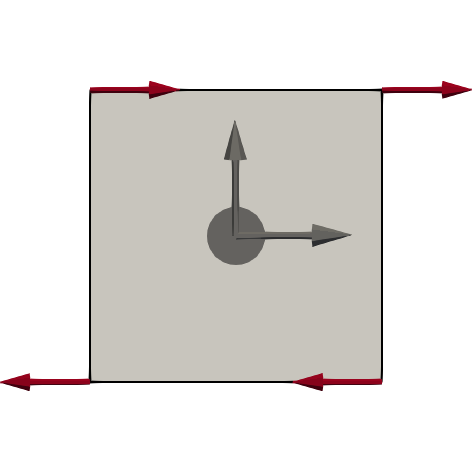}}
\subfigure[\solidTriadAverage]{\label{fig:examples:moment_test_results:average}\includegraphics[resolution=300]{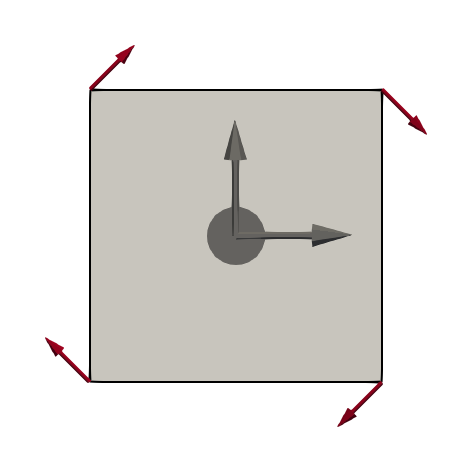}}
\subfigure[\solidTriadFixTriad]{\label{fig:examples:moment_test_results:fixtriad}\includegraphics[resolution=300]{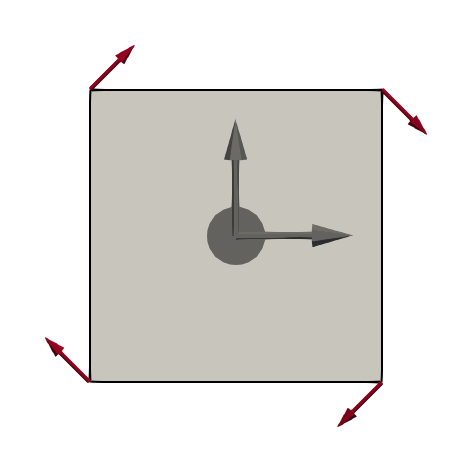}}
\caption{Results for the moment test problem -- the nodal reaction forces are shown for different solid triads.}
\label{fig:examples:moment_test_results}
\end{figure}

The first problem setup is depicted in \Cref{fig:examples:moment_test_problem}.
A straight beam is embedded inside a solid cube ($E=\unit[1]{N/m^2}$, $\nu=0$) and the beam is loaded with a distributed torsion moment in $\ex$ direction, which is constant along the beam centerline.
This example is used to investigate how a moment on a beam is transferred to solid nodal forces.
The cube is modeled with a single eight-noded hexahedral element and all solid degrees of freedom are fixed.
A single \sr beam finite element is used to discretize the beam.
No Dirichlet boundary conditions are applied on the beam and the coupling between the beam and the solid is realized with our novel mortar-type \btsvrc method.
Thus, the only interaction between the beam and the solid is the transfer of the external moment.
The resulting nodal reaction forces for the different solid triad definitions introduced in \Cref{sec:solid_triads} are depicted in \Cref{fig:examples:moment_test_results}.
Therein, the results for the \solidTriadPolar, \solidTriadAverage and \solidTriadFixTriad variants, \cf \Cref{fig:examples:moment_test_results:polar,fig:examples:moment_test_results:fixtriad,fig:examples:moment_test_results:average}, match up to machine precision.
In general, however, the solid coupling reaction forces may differ for different definitions of the solid triad, as visible for the variants \solidTriadFixFiber{2/3} in \Cref{fig:examples:moment_test_results:fix2,fig:examples:moment_test_results:fix3}.
This observation can be explained by the fact that the representation of a moment via nodal forces is non-unique, \ie there is an infinite number of possible force pair combinations to achieve this.
However, from a mechanical point of view, the force pairs resulting from the  \solidTriadPolar, \solidTriadAverage and \solidTriadFixTriad variants seem more natural than the ones for the \solidTriadFixFiber{2/3} variants.
Moreover, the former three variants result in the (unique) force pair solution if the moment is applied as a constant shear stress on the beam surface, \cf \Cref{sec:modeling_assumptions}.
Additionally, it can be observed for the \solidTriadFixFiber{2/3} variants that the choice which local solid direction is coupled to the solid triad drastically affects the result for the nodal forces.

\subsection{Shear test}
\label{sec:examples:single_element_shear_test}
\begin{figure}
\centering
\includegraphics[scale=1]{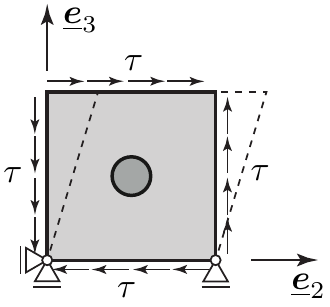}
\caption{Shear test problem.}
\label{fig:examples:shear_test_problem}
\end{figure}

The next elementary test case is illustrated in Figure~\ref{fig:examples:shear_test_problem}.
The problem geometry is the same as in the previous example.
The cube (side length $h = \unit[1]{m}$) is fixed at two bottom corner points to constrain all rigid body modes.
A constant surface load $\tau = \unit[0.001]{N/mm^2}$ is applied to the surfaces of the cube, as depicted in Figure~\ref{fig:examples:shear_test_problem}.
No boundary conditions are applied to the beam.
This problem illustrates how the specific solid triads affect the shear stiffness of the solid element and will be studied in two steps.

\begin{figure}
\centering
\includegraphics[scale=0.2]{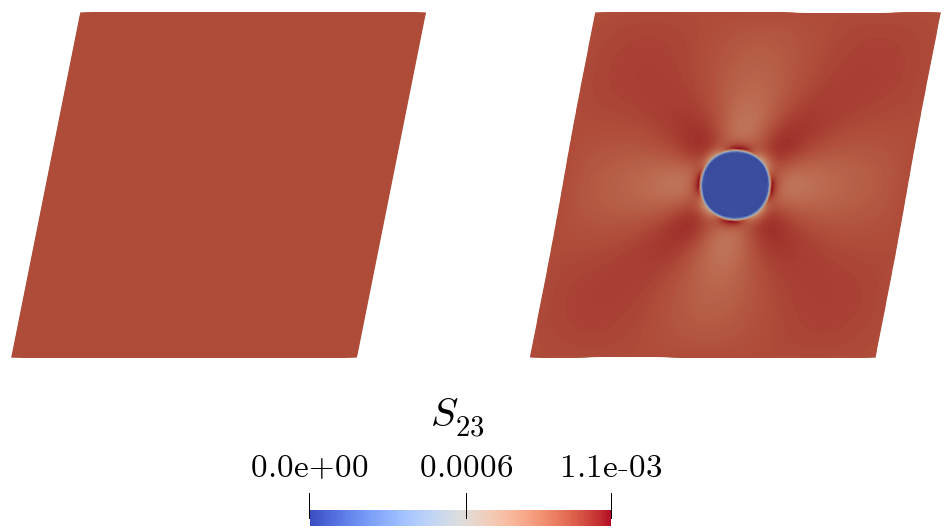}
\caption{
Shear test problem -- reference solution for solid without a beam (left) and solid with an embedded beam (2D-3D coupling), $r=\unit[0.1]{m}$ (right).
The second Piola-Kirchhoff stress $S_{23}$ is shown in the solid.
Displacements are scaled by a factor of $100$.
}
\label{fig:examples:shear_test_2D-3D_stress}
\end{figure}

In a first step, we want to investigate the impact of the local stiffening effect the beam \cs has on the surrounding solid material.
To do so, a reference solution is created by applying a full 2D-3D beam-to-solid coupling scheme, \ie the coupling conditions are enforced on the beam surface, \cf~\Cref{sec:appendix:full_2d_3d,sec:appendix:full_2d_3d_discret}.
For comparison purposes, a variant of this problem is simulated without the embedded beam, \ie the pure solid shear problem.
Figure~\ref{fig:examples:shear_test_2D-3D_stress} illustrates the shear stress in the solid, with and without the embedded beam, for an exemplary beam radius $r=\unit[0.1]{m}$.
As expected, the solution is uniform in the entire solid volume for the pure solid variant.
In the 2D-3D beam-to-solid coupling variant, the embedded beam affects the solid stress and displacement fields.
The overall displacement of the solid is smaller than for the variant without a beam, thus demonstrating the stiffening effects of the beam \cs.
In agreement with our fundamental modeling assumption of overlapping beam and solid domains (see Section~\ref{sec:modeling_assumptions}), the solid shear stress inside the beam domain is zero.
Outside of the beam domain, the solid shear stress field shows slight fluctuations due to the local constraints enforcing the 2D-3D coupling at the beam surface.
However, close to the boundaries of the cube, these fluctuations become negligible and the shear stress field is quite homogeneous and therefore very similar to the pure solid shear problem.

\begin{figure}
	\centering
	\subfigure[no beam]{\includegraphics[resolution=300]{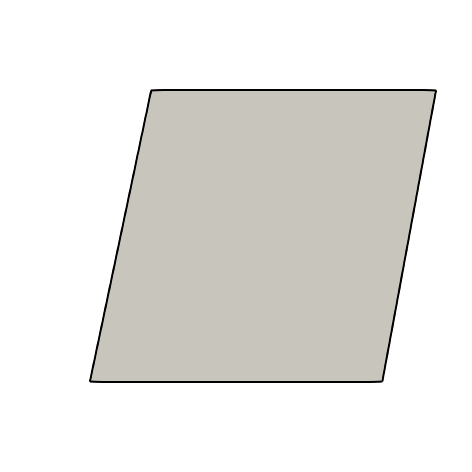}}
	\subfigure[\solidTriadPolar]{\includegraphics[resolution=300]{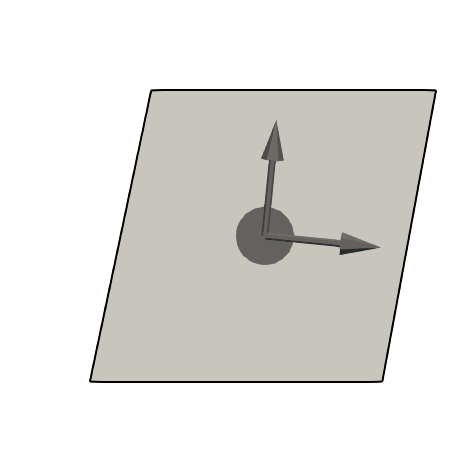}}
	\subfigure[\solidTriadFixFiber{2}]{\includegraphics[resolution=300]{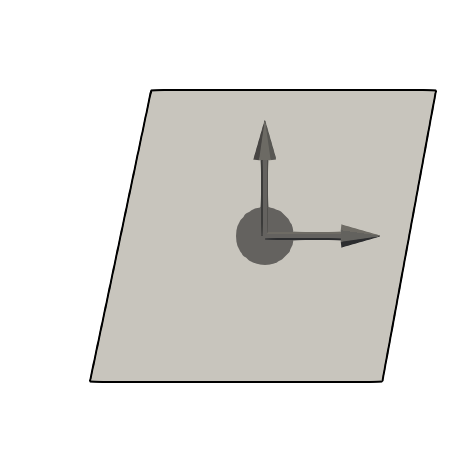}}
	\subfigure[\solidTriadFixFiber{3}]{\includegraphics[resolution=300]{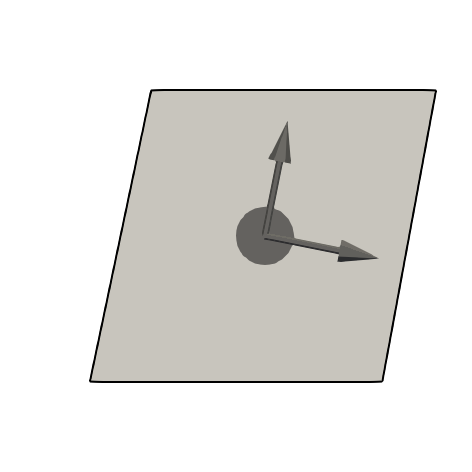}}
	\subfigure[\solidTriadAverage]{\includegraphics[resolution=300]{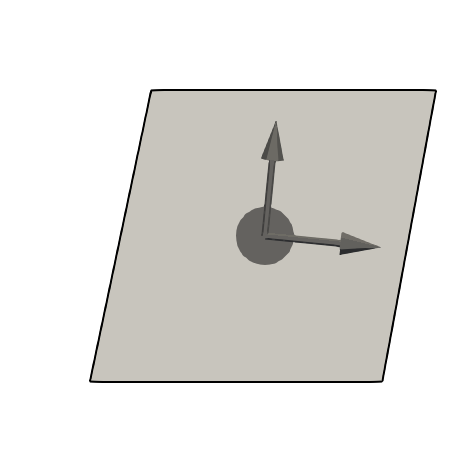}}
	\subfigure[\solidTriadFixTriad]{\includegraphics[resolution=300]{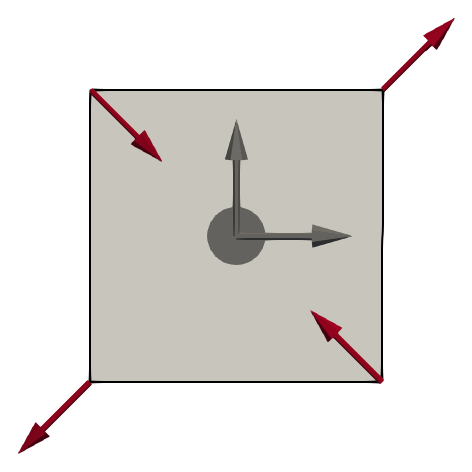}}
	\caption{Results for the shear test problem -- the nodal reaction forces are shown for different solid triads.
	Displacements are scaled with a factor of $100$.}
	\label{fig:examples:shear_test_results}
\end{figure}

In a second step, this problem is simulated with one single solid finite element to investigate potential shear locking effects.
The coupling between beam and solid is now realized with our novel \btsvrc method and a rotational penalty parameter of $\penRot = \unit[100]{Nm/m}$.
In Figure~\ref{fig:examples:shear_test_results}, the deformed solid element and the resulting coupling reaction forces on the solid nodes are depicted for the different solid triad definitions and again for the problem without embedded beam.
Due to the orthogonality constraints in the \solidTriadFixTriad variant, no shear mode remains in the solid finite element, \ie it is rigid with respect to shear deformations (in fact, small deformations can be observed due to the penalty regularization).
In this example, all other solid triad definitions result in a solid displacement field matching the variant without embedded beam up to machine precision.
Table~\ref{tbl:examples:shear_test_results} states the rigid body rotation angle $\rotvecbeamScalar$ of the beam for the different solid triad variants.
The rotation angle $\rotvecbeamScalar$ of the beam depends on the employed solid triad variant.
With the \solidTriadFixFiber{2} variant the beam does not rotate at all since the orientation of the local solid material fiber does not change.
The \solidTriadFixFiber{3} variant, on the other hand, results in the largest rotation of the beam, since the solid triad is coupled to the solid material fiber which undergoes the largest orientation change.
Although they are not identical up to machine precision, the \solidTriadPolar and \solidTriadAverage variants lead to very similar results for the rotation of the beam, \ie roughly an average of the \solidTriadFixFiber{2} and \solidTriadFixFiber{3} variants.
\begin{table}
	\centering
	\caption{Numerical results for the shear test problem.}
	\label{tbl:examples:shear_test_results}
	\begin{tabular}{lr}
	\toprule
		solid triad            & $\rotvecbeamScalar$ \\
	\midrule
		\solidTriadPolar       &         -0.09899932 \\
		\solidTriadFixFiber{2} &         -0.00000000 \\
		\solidTriadFixFiber{3} &         -0.19485464 \\
		\solidTriadAverage     &         -0.09742732 \\
		\solidTriadFixTriad    &         -0.00099010 \\
	\bottomrule
	\end{tabular}
\end{table}

\begin{figure}
\centering
\includegraphics[scale=1]{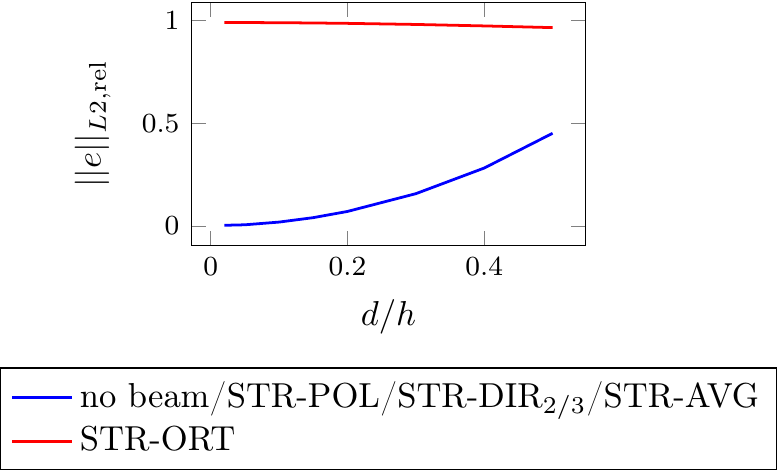}
\caption{
Relative displacement error $\errorRel$ for different beam diameter to solid cube length ratios $d/h$.
The relative error is computed with respect to the 2D-3D reference solution.
The curves represent different solid triad variants.
}
\label{fig:examples:shear_test_2D-3D_error}
\end{figure}

The results show that the presented solid triads lead to either no shear stiffening effects in the solid (\solidTriadPolar, \solidTriadFixFiber{2,3} and \solidTriadAverage) or to severe locking resulting in a complete constraining of all shear modes (\solidTriadFixTriad).
To assess which variant resembles best the resolved 2D-3D coupling scheme, the relative $L_2$-displacement error
\begin{equation}
\label{eq:examples:l2_error_relative}
\errorRel =
\frac{\sqrt{\intSolid{ \norm{ \usolidh - \usolidhref }^2 }}}{\sqrt{\intSolid{ \norm{ \usolidhref }^2 }}}
\end{equation}
is compared.
In the results presented in the following, the reference solution is the solution obtained with a fine solid mesh and a 2D-3D coupling.
Figure~\ref{fig:examples:shear_test_2D-3D_error} illustrates $\errorRel$ for different beam diameter to solid cube length ratios $d/h$.
The relative error for the \solidTriadFixTriad variant is almost constant $1$ for all beam diameters ratios, \ie even for beam \cs sizes similar to the cube dimensions a full constraining of all shear modes does not accurately describe the physical coupling.
For all other variants the behavior of the relative error is the same, since none of them constrain the shear deformation mode in the solid, \ie the beam \cs{s} rotate with the solid without constraining it.
For small ratios of beam radius to solid cube length the error is close to zero.
For larger ratios of beam radius to solid cube length, the error increases as there is a real physical stiffening effect due to the embedded beam \cs in the 2D-3D problem that is not captured by the 1D-3D coupling schemes.
However, in the entire range of practically relevant solid mesh sizes (relative to the beam \cs size) as illustrated in Figure~\ref{fig:examples:shear_test_2D-3D_error}, the solid triad variants that do not constrain the in-plane deformation of the solid result in a better approximation of the physical system behavior as compared to the \solidTriadFixTriad triad.

\subsection{Fiber composite under shear loading}
\label{sec:examples:shear_composite}
\begin{figure}
\centering
\includegraphics[scale=1]{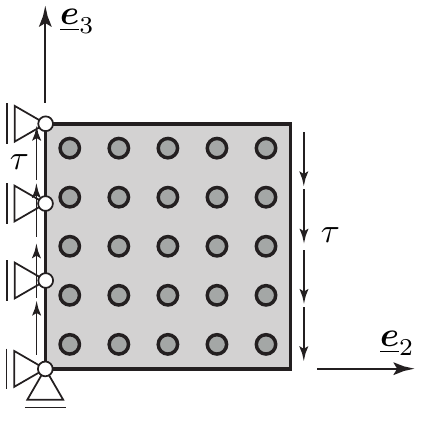}
\caption{
Fiber composite under shear loading -- Problem setup of 5$\times$5 embedded fibers inside a solid cube.
}
\label{fig:examples:shear_convergence_problem}
\end{figure}

In this example, multiple fibers are placed inside a solid cube, \cf Figure~\ref{fig:examples:shear_convergence_problem}.
The solid cube has the dimensions $\unit[1]{m}\times\unit[1]{m}\times\unit[1]{m}$ and consists of a hyperelastic \svk material model ($E=\unit[1]{N/m^2}$, $\nu=0.0$).
Embedded inside the solid cube are 5$\times$5 fibers with a radius of $\unit[0.0125]{m}$.
All fibers point in $\ez$ direction.
The solid is fixed in $\ey$ direction at the left boundary, and loaded with two equilibrating shear loads ($\tau=\unit[0.01]{N/m^2}$) at the left and right boundary.
To constrain the remaining rigid body mode, the lower left corner point is fixed in $\ez$ axis.
The fibers are coupled to the solid via the \btsvrc method and no additional boundary conditions act on the fibers.
The cube is meshed with $7\times 7\times1$ solid \hex{8} elements, and each fiber is represented by a single \sr beam finite element.
The penalty parameters for the \btsvrc method are $\penPos = \unit[100]{N/m^2}$ and $\penRot = \unit[100]{Nm/m}$.
In this example, the results obtained with the \btsvrc method and different solid triads will be compared with a spatially converged reference solution, where the coupling between the beam surfaces and solid volume is discretized in a surface-to-volume (2D-3D) manner, \ie the beam surface instead of the centerline is fixed to the solid, \cf \Cref{sec:appendix:full_2d_3d,sec:appendix:full_2d_3d_discret}.

\begin{figure*}
\centering
\includegraphics[resolution=300]{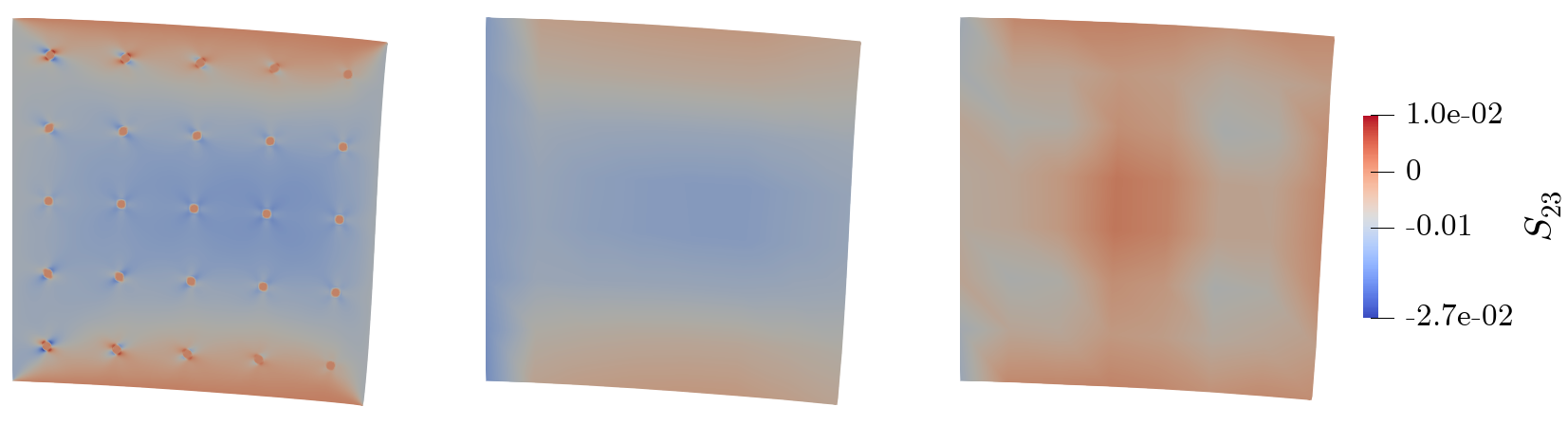}
\caption{
Fiber composite under shear loading -- Full 2D-3D coupling (left), \solidTriadPolar, \solidTriadFixFiber{2/3} and \solidTriadAverage (middle), and \solidTriadFixTriad (right).
The second Piola-Kirchhoff stress $S_{23}$ is shown in the solid.
}
\label{fig:examples:shear_convergence_results}
\end{figure*}

The resulting shear stresses are visualized in Figure~\ref{fig:examples:shear_convergence_results}.
In the full 2D-3D model, there are stress concentrations at the interface between the beam surfaces and the solid.
It is important to point out that the \btsvrc method (1D-3D), is not able to capture these stress concentrations, regardless of the employed solid triad.
However, this has not been the intention of the \btsvrc method in the first place, but instead we want to make sure that the far field stress in the solid is represented accurately.
\Cref{fig:examples:shear_convergence_results} illustrates the shear stress results obtained with the \solidTriadPolar, \solidTriadFixFiber{2/3} and \solidTriadAverage solid triads.
In the reference solution the in-plane shear stress is positive at the top and bottom of the cube and negative in the middle.
The results obtained with the \solidTriadPolar, \solidTriadFixFiber{2/3} and \solidTriadAverage solid triads are similar to the ones obtained with 2D-3D coupling.
However, the results with the \solidTriadFixTriad solid triads clearly exhibit drastic shear locking effects due to the \mbox{(over-)} constraining of orthogonal solid directions.
Table~\ref{tbl:examples:shear_convergence_results} provides the displacement at the top right corner of the cube for the 2D-3D reference solution and different types of solid triad fields, as well as the relative error.
The error for the \solidTriadFixTriad solid triad is six times larger than for all other solid triads.
This again illustrates the unwanted locking effects introduced by the \solidTriadFixTriad solid triads variant.
\begin{table*}
\centering
\caption{
Numerical results for the fiber composite under shear loading -- the displacement $\tns{u}$ at the top right corner of the cube for the 2D-3D reference solution and different types of solid triad fields, and the relative error.}
\label{tbl:examples:shear_convergence_results}
\begin{tabular}{llrr}
\toprule
	coupling type     & solid triad                                                    &      $\tns{u}$ in $\unit{m}$ & $\frac{\norm{\tns{u} - \tns{u}_{\text{ref}}}}{\norm{\tns{u}_{\text{ref}}}}$ \\
\midrule
	2D-3D (reference) & --                                                             & $[0, 0.0311342, -0.0706488]$ &            -- \\
\midrule
	1D-3D             & \solidTriadPolar, \solidTriadFixFiber{2/3}, \solidTriadAverage & $[0, 0.0299373, -0.0681153]$ &    $3.6282\%$ \\
	                  & \solidTriadFixTriad                                            & $[0, 0.0293469, -0.0547324]$ &    $19.559\%$ \\
\bottomrule
\end{tabular}
\end{table*}
%			norm	rel error
%ref	0,0311342	-0,0706488	0,07720486611011	0
%triad	0,0293469	-0,0547324	0,062103753102128	0,195597943093965
%polar	0,029937	-0,0681153	0,074403750329469	0,036281596248683

At this point a short recap of the first three examples for each of the investigated solid triad constructions is given to summarize their applicability in the context of our \btsvrc method:
\begin{description}
\item[\solidTriadPolar] All basic consistency tests are fulfilled by this variant.
However, due to the computational complexity of the polar decomposition in 3D, \cf Section~\ref{sec:solid_triads:polar_decomposition}, this variant is not used in the remaining examples presented in this contribution.
\item[\solidTriadFixFiber{2/3}] The examples show that the (arbitrary) choice of the solid material direction for the construction of the solid triad can have a considerable effect on the results.
Therefore, these variants will not be employed in the following.
However, for comparison purposes they will be included in the spatial convergence example, \cf Section~\ref{sec:examples:convergence}.
\item[\solidTriadAverage] All basic consistency tests are fulfilled by the averaged solid triad and the results are very close to the ones obtained via the \solidTriadPolar variant, while being less expensive from a computational point of view.
This variant is used in the remaining examples of this contribution.
\item[\solidTriadFixTriad] This variant leads to considerable shear locking in the range of coarse solid mesh resolutions, which is exactly the range of interest for the proposed 1D-3D coupling schemes.
Therefore, this variant will not be used in the remainder of this contribution.
\end{description}

\subsection{Transfer of constant torque}
\label{sec:examples:patch_test}
\begin{figure}
\centering
\includegraphics[scale=1]{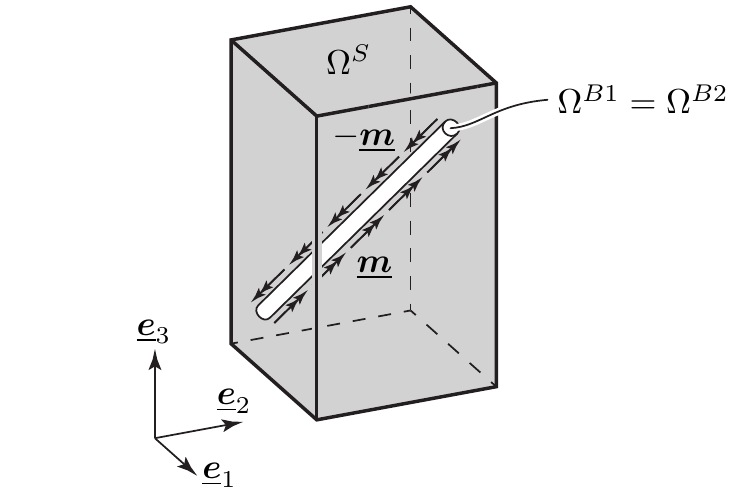}
\caption{
Problem setup for the transfer of constant torque example.
Both beams $\Omega^{B1}$ and $\Omega^{B2}$ occupy the same spatial position.
}
\label{fig:examples:patch_test_problem}
\end{figure}

This example serves as a consistency test for the \btsvrc method and its ability to transfer a constant torque.
It is an extension of the \emph{constant stress transfer problem} for the \btsvc method previously presented in \cite{Steinbrecher2020}.
The example is inspired by classical patch tests, which are well-established tools to investigate the consistency of finite element formulations \cite{Taylor1986}.
The constant torque test is depicted in Figure~\ref{fig:examples:patch_test_problem}.
It consists of a solid block $\domainSolid$ with two embedded beams $\domainBeam[1]$ and $\domainBeam[2]$.
The two beams occupy the same spatial position.
The solid is fixed at the lower surface and no external loads are applied.
One beam is loaded with a torsion load $\tns{m}$, and the other beam with a torsion load $-\tns{m}$, both acting along their axial direction.
The magnitude of the torsion load is $\unit[10]{Nm/m}$.
Based on the space-continuous problem description, the opposing loads on the two beams cancel out each other, and in sum the two beams transfer no loads to the solid.
This gives the trivial solution $\usolid = \tnsO$ for the displacement field in the solid, \cf \cite{Steinbrecher2020}, and a constant solution for the beam rotations along their axis.
In this test it shall be verified that this solution can also be represented in the spatially discretized setting using an arbitrarily coarse discretization.
Both beams are coupled to the solid via the \btsvrc method.
There is no direct interaction between the two beams, but all interactions are transferred through the solid domain.

The geometry and material parameters are taken from \cite{Steinbrecher2020}.
The dimensions of the solid block are $\unit[1]{m}\times\unit[1]{m}\times\unit[2]{m}$ and a \svk material model ($E = \unit[10]{N/m^2}$,
$\nu = 0.3$) is employed.
The block is discretized with $4\times 4 \times 7$ eight-noded, first-order hexahedral elements.
The circular \cs{s} of the two beams have a radius of $\unit[0.05]{m}$, and the beam material parameters are $E=\unit[100]{N/m^2}$ and $\nu=0$.
The beams $B1$ and $B2$ are discretized with $5$ and $7$ \sr beam finite elements, respectively.
This results in a non-matching discretization between the two beams as well as between the beams and the solid.
Coupling between the beams and the solid is realized with a linear interpolation of both the translational and rotational Lagrange multipliers.
The \solidTriadAverage solid triads are employed in this example, \cf Section~\ref{sec:solid_triads:3d_triad}.
The penalty parameters are $\penPos = \unit[100]{N/m^2}$ and $\penRot = \unit[100]{Nm/m}$.

\begin{figure}
\centering
\includegraphics[resolution=300]{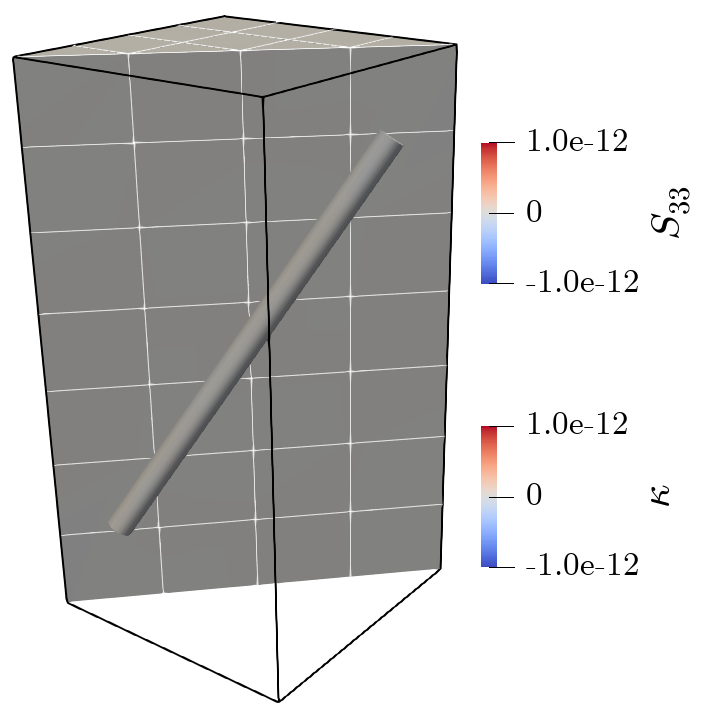}
\caption{
Results for the constant torque transfer test.
The second \pk stress $S_{33}$ is shown in the solid and the curvature $\kappa$ in the middle of each beam element is shown in the beams.
}
\label{fig:examples:patch_test_result}
\end{figure}

Figure \ref{fig:examples:patch_test_result} illustrates the results of this test.
The stress in the solid and the curvature in the beam are indeed zero up to machine precision, thus matching the expected analytical solution.

This example illustrates the ability of the \btsvrc method to exactly represent a constant torsion state along the beam and the consistency of the coupling terms despite the fact that arbitrary non-matching meshes are involved.

\subsection{Spatial convergence}
\label{sec:examples:convergence}
\begin{figure}
\centering
\includegraphics[scale=1]{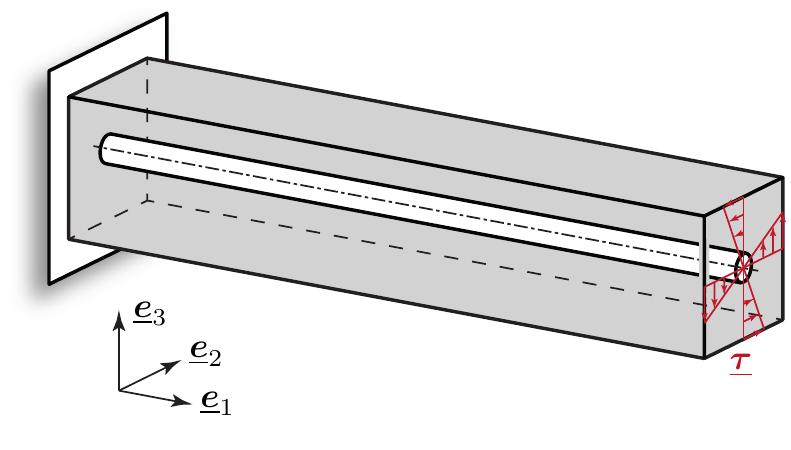}
\caption{
Convergence test case for the \btsvrc method -- Problem setup of a coupled beam and solid structure.
}
\label{fig:examples:convergence_problem}
\end{figure}

This numerical example investigates the spatial convergence properties of the \btsvrc method under uniform mesh refinement.
The problem is depicted in Figure~\ref{fig:examples:convergence_problem}.
It consists of a solid block with the dimensions $\unit[5]{m}\times\unit[1]{m}\times\unit[1]{m}$ and a \svk material model ($E = \unit[10]{N/m^2}$,
$\nu = 0$).
A beam (\cs radius $\unit[0.125]{m}$, $L=\unit[5]{m}$, $E=\unit[300]{N/m^2}$, $\nu=0$) is embedded inside the solid block.
No external loads or Dirichlet boundary conditions are applied to the beam, \ie homogeneous Neumann boundary conditions at both ends.
The right end of of the block is loaded with a shear stress $\tns{\tau}$.
The shear stress at point $\tns{p} = L \ex + y \ey + z \ez$ reads
\begin{equation}
\tns{\tau}=\br{-z \ey + y \ez}\unit[0.05]{N/m^3},
\end{equation} 
thus resulting in a total torque of $\unit[1.65885\cdot 10^{-2}]{Nm} $.
This example can be interpreted as an adapted version of the spatial convergence problem in \cite{Steinbrecher2020} to verify the scenario of rotational coupling.
A similar problem is also investigated in \cite{Khristenko2021}.
The spatial convergence behavior of the \btsvrc method will be analyzed with respect to a spatially converged reference solution obtained with a 2D-3D coupling discretization, as described in \Cref{sec:appendix:full_2d_3d}.
To compare the results, the $L_2$ displacement error in the solid is calculated via
\begin{equation}
\label{eq:examples:l2_error}
\error =
\frac{1}{V_0} \sqrt{\intSolid{ \norm{ \usolidh - \usolidhref }^2 }}.
\end{equation}
Here, $V_0=\unit[1]{m^3}$ is the solid volume in the reference configuration.
It should be pointed out that the 2D-3D coupling problem does not have the same analytical solution as the \btsvrc problem, because the 1D-3D coupling results in a singularity in the analytical solution, \cf \Cref{sec:modeling_assumptions_1D-3D}.
Therefore, spatial convergence of the \btsvrc method towards the reference solution is not expected all the way towards the asymptotic limit of arbitrarily small solid element sizes, but only in the practically relevant regime of solid mesh sizes that are larger than the beam \cs radius.
In this regime, the singularity, \ie the difference between the 1D-3D and 2D-3D models can not be fully resolved by the finite element solution space.
This fact can be exploited to obtain reasonably accurate results with our \btsvrc (\ie 1D-3D) method for the envisioned applications and practically relevant mesh resolutions.

\begin{figure}
\centering
\includegraphics[scale=1]{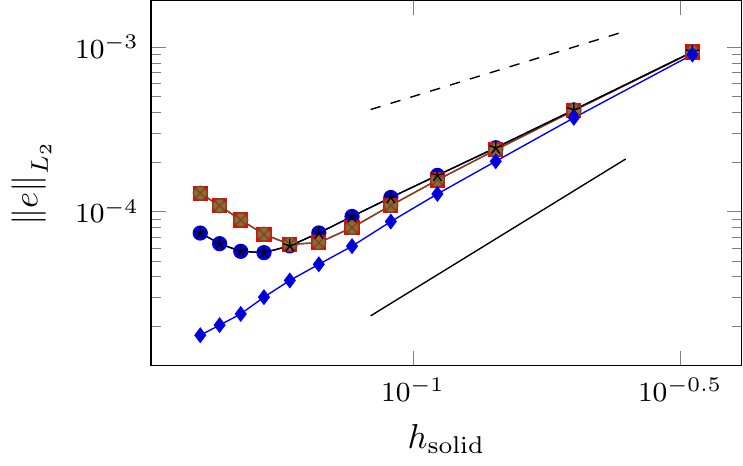}\hphantom{\parbox{1.7cm}{\ }}
\\[2mm]
\includegraphics[scale=1]{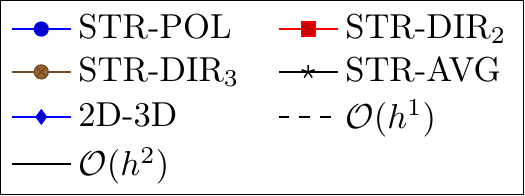}
\caption{
Spatial convergence plot for different solid triads and the 2D-3D reference solution.
}
\label{fig:examples:convergence_result}
\end{figure}

The solution to the presented problem has a point symmetry around the $\ex$ axis.
Therefore, the \solidTriadPolar and \solidTriadAverage solid triad variants coincide and give the same numerical results up to machine precision.
Similarly, the results obtained with the \solidTriadFixFiber{1/2} variants match up to machine precision.
\Cref{fig:examples:convergence_result} shows the convergence plot for different types of solid triads as well as for the 2D-3D coupling approach.
For coarse discretizations an excellent convergence behavior can be observed for the \btsvrc variants, slightly below the convergence rate of the reference 2D-3D method, but with a significantly reduced computational cost.
All \btsvrc convergence plots exhibit a kink at a certain solid mesh resolution: the \solidTriadFixFiber{1/2} variants at around $\hsolid = \unit[0.07]{m}$, and the \solidTriadPolar and \solidTriadAverage variants at around $\hsolid = \unit[0.06]{m}$.
Figuratively speaking, the difference between the 1D-3D and 2D-3D coupling model becomes dominant at the kink position, since the solid element size to beam \cs diameter ratio becomes smaller.
Nevertheless, the kink only occurs when the solid element size is already smaller than the \cs radius, which is far away from the envisioned geometric relations for the \btsvrc method anyways.
The results confirm that for solid element sizes larger than the beam \cs diameter, \ie our desired and practically relevant discretization case, the results obtained with the \btsvrc method (1D-3D) exhibit excellent spatial convergence properties and thus give a very good approximation of the 2D-3D coupling problem.

\subsection{Plane cantilever bending}

\begin{figure}
\centering
\subfigure[]{\label{fig:examples:plane_cantilever_bending_problem}\includegraphics[scale=1]{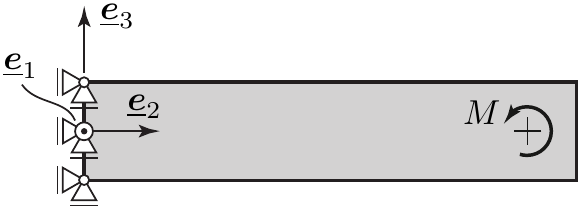}}
\hfill
\subfigure[]{\label{fig:examples:plane_cantilever_bending_result}\includegraphics[resolution=300]{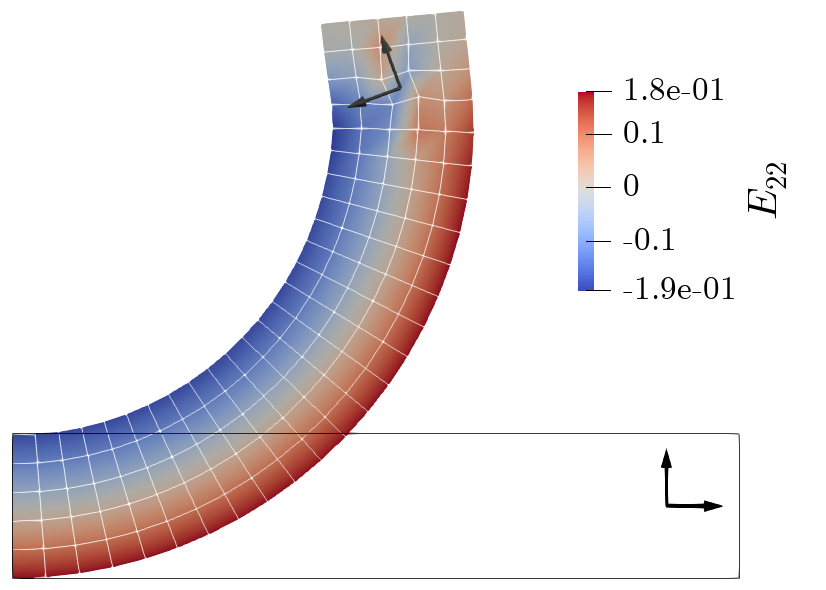}}
\caption{
Plane cantilever bending problem.
\subref{fig:examples:plane_cantilever_bending_problem} Problem setup of the cantilever beam with a moment load. \subref{fig:examples:plane_cantilever_bending_result} Deformed cantilever with the Green--Lagrange strains $E_{22}$.
}
\end{figure}

In this example we consider a cantilever structure modeled as a solid continuum subject to a moment load.
The problem is illustrated in Figure~\ref{fig:examples:plane_cantilever_bending_problem}.
The cantilever has the dimensions $\unit[5]{m}\times\unit[1]{m}$ and consists of a Saint Venant--Kirchhoff material ($E = \unit[10]{N/m^2}$, $\nu = 0$).
At the left boundary all displacement components are fully constrained.
A moment load $M=\unit[0.0290888]{Nm}$ acts on the cantilever at the material point $\tns{X}_M=\unit[{[4.5, 0]\tr}]{m}$.
This moment is chosen such that, if the cantilever were modeled using 1D beam theory, it should bend exactly to a quarter circle, due to a pure bending deformation in the region between the Dirichlet boundary and the applied moment.
Directly imposing a conservative moment load on a solid, \ie a \boltz continuum, which exhibits no rotational degrees of freedom is a non-trivial task.
Standard approaches would require to model the moment as a (deformation-dependent) load/traction field distributed across an arbitrarily chosen sub-volume of the solid.
In this example we impose the external moment on the solid structure by defining a solid triad (\solidTriadAverage) at the application point of the moment.
The nodal external forces effectively acting on the solid are obtained by projecting the moment to the solid finite element space via the discrete version of \eqref{eq:solid_triads:variation}.
The cantilever is discretized with $20\times 4$ plane four-noded, first-order quadrilateral elements.
In Figure~\ref{fig:examples:plane_cantilever_bending_result} the deformed cantilever is illustrated.
The global displacement behavior is as expected, \ie the cantilever bends to a quarter circle.
Of course, the local strain state close to the point where the external moment is applied is not meaningful in a continuum mechanics sense, since we impose a singular moment at that point.
However, according to Saint Venant's principle, a linear stress distribution across the beam height, as expected for the pure bending of a slender beam-like structure, can be observed at a sufficient distance from the point where the moment is induced.
This example illustrates that the presented rotational coupling approach is not limited to the coupling of beam \cs orientations, but can also be used as a stand-alone feature to impose moments onto a solid domain in a variationally consistently manner.
It should be pointed out that this example has only been carried out in 2D for reasons of simplicity, while the illustrated capability is available in 3D problems, too.
%If a suitable 3D solid triad is used, \eg \solidTriadAverage, the same procedure can be performed for 3D problems.

\subsection{Plate with embedded beam}
\begin{figure}
\centering
\includegraphics[scale=1]{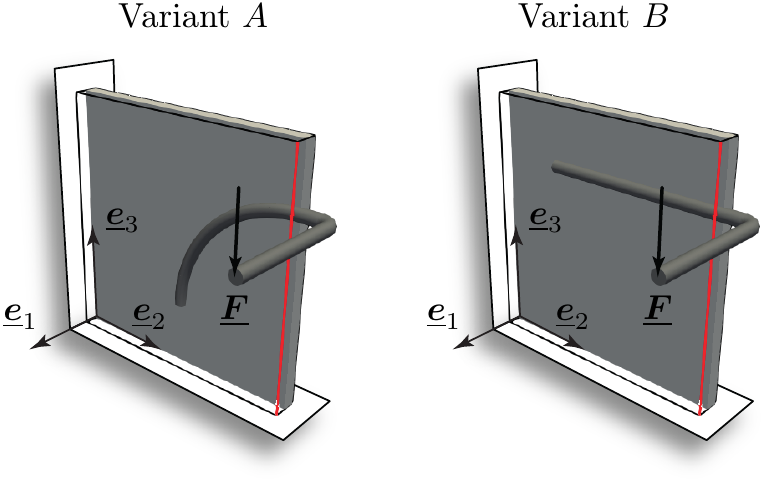}
\caption{
Problem setup for the plate with embedded beam examples.
In variant $A$ the embedded part of the beam has the shape of a quarter circle, in variant $B$ it is straight.
In both cases the red line indicates along which edge the results are plotted in Figure~\ref{fig:examples:plate_embedded_beam_results_plot}.
}
\label{fig:examples:plate_embedded_beam_problem}
\end{figure}

In this example a beam is only partially embedded inside a solid plate and loaded with a tip force.
Two different geometry variants of the embedded beam are considered, \cf Figure~\ref{fig:examples:plate_embedded_beam_problem}.
In variant $A$ the embedded part of the beam has the shape of a quarter circle, while it is straight in variant $B$.
The plate has the dimensions $\unit[1]{m}\times\unit[1]{m}\times\unit[0.1]{m}$ and consists of a Saint Venant--Kirchhoff material ($E = \unit[1]{N/m^2}$, $\nu = 0.3$).
The embedded \sr beam has a \cs radius $r=\unit[0.025]{m}$ and the material parameters are $E = \unit[100]{N/m^2}$ and $\nu = 0$.
In both variants the beam is loaded with a tip load $\tns{F} = -0.0001\ez$ at the end that sticks out of the solid domain.
The solid plate is fully clamped at the left and at the bottom.

The coupling of beam and solid is realized with our novel \btsvrc method and compared to the \btsvc method from \cite{Steinbrecher2020}, \ie the one without rotational coupling.
First-order Lagrange polynomials are employed to  discretize the translational and rotational Lagrange multipliers.
The penalty parameters are $\penPos = \unit[100]{N/m^2}$ and $\penRot = \unit[100]{Nm/m}$.
The solid plate is modeled with $1\times10\times10$ eight-noded solid-shell elements \cite{Bischoff1997,Vu-Quoc2003a}, while the entire beam is discretized with six \sr beam finite elements.
The resulting global finite element model has $807$ degrees of freedom.
A full 3D model, also resolving the beam with three-dimensional solid finite elements and consisting of $90{,}190$ second-order tetrahedra (\tet{10}) elements, serves as a comparison.
The discretization of the full 3D model has been chosen such that mesh convergence is guaranteed.
Consequently, the full 3D model consists of $270{,}570$ degrees of freedom.

\begin{figure*}
\centering
\includegraphics[resolution=300]{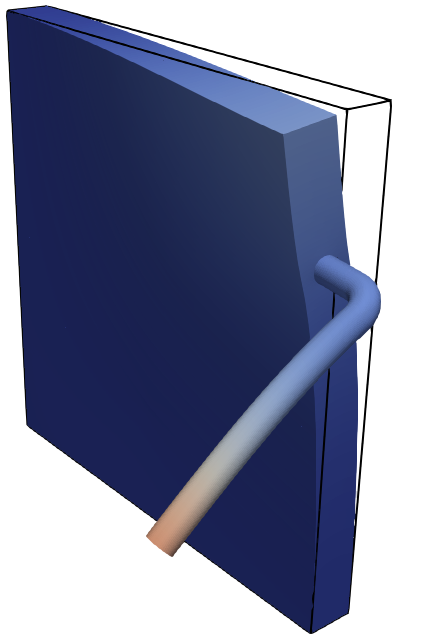}
\hfil
\includegraphics[resolution=300]{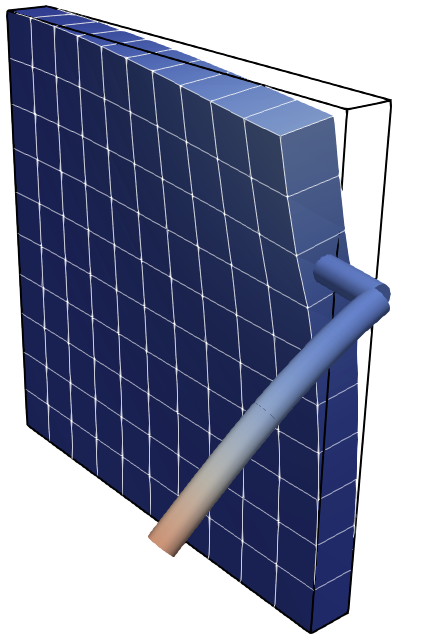}
\hfil
\includegraphics[resolution=300]{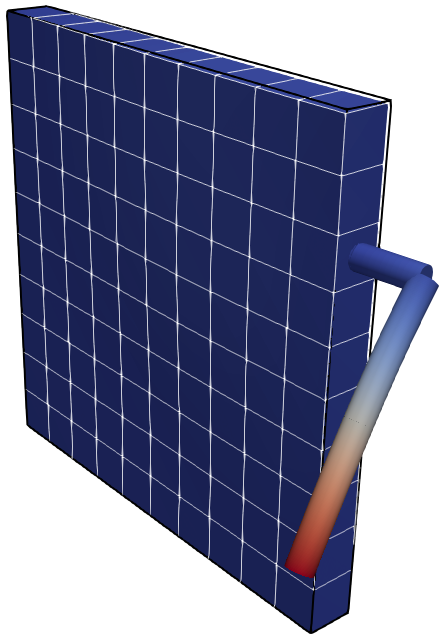}
\\[3mm]
\includegraphics[resolution=300]{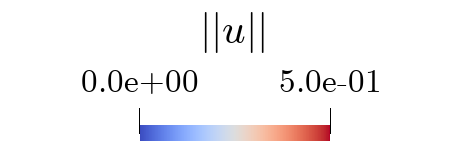}
\caption{
Deformed plate with embedded beam, variant $A$ -- for the full model (left), \btsvrc model (middle) and \btsvc model (right).
The contour plots visualize the displacement magnitude.
}
\label{fig:examples:plate_embedded_beam_results_A}
\end{figure*}

\begin{figure*}
\centering
\includegraphics[resolution=300]{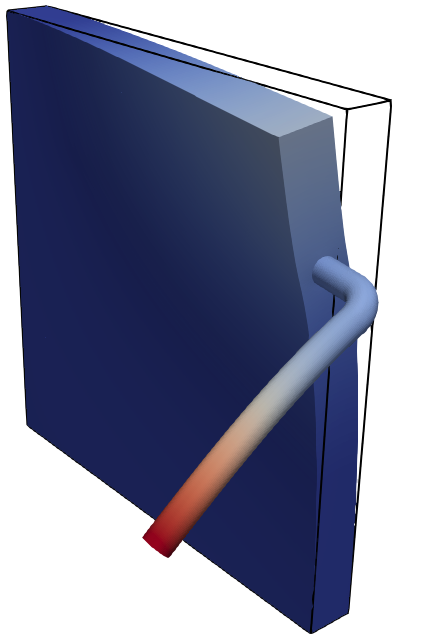}
\hfil
\includegraphics[resolution=300]{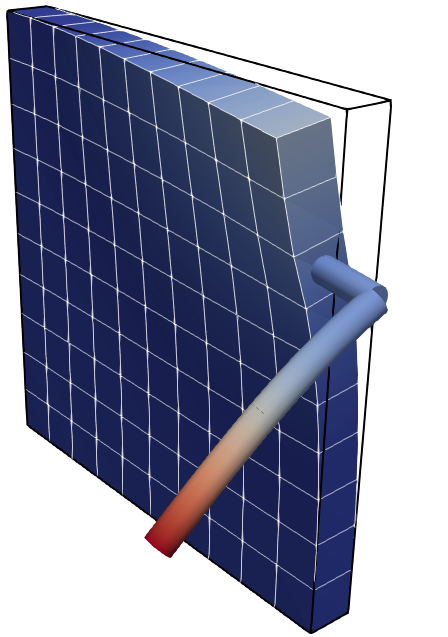}
\\[3mm]
\includegraphics[resolution=300]{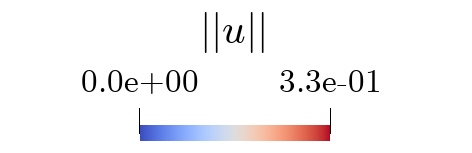}
\caption{
Deformed plate with embedded beam, variant $B$ -- for the full 3D model (left), \btsvrc model (right).
The contour plots visualize the displacement magnitude.
}
\label{fig:examples:plate_embedded_beam_results_B}
\end{figure*}

\begin{figure*}
\centering
\includegraphics[scale=1]{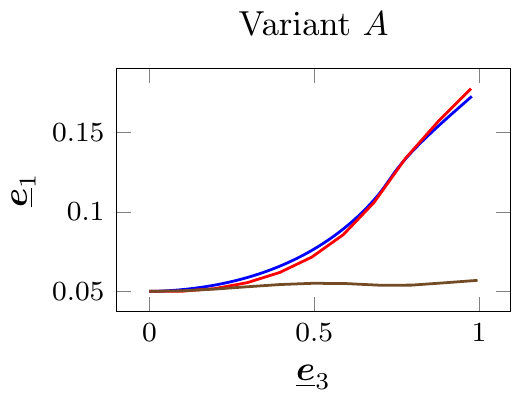}
\hfil
\includegraphics[scale=1]{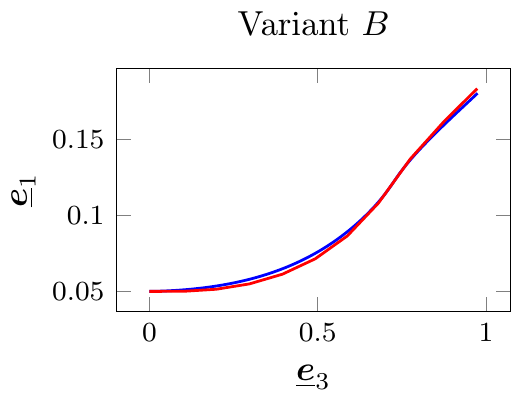}
\\
\includegraphics[scale=1]{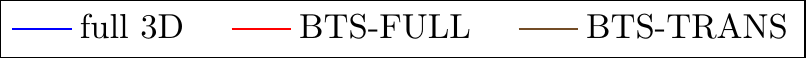}
\caption{
Deformed configuration of the edge indicated in Figure~\ref{fig:examples:plate_embedded_beam_problem} for variants $A$ and $B$, and for different modeling techniques.
}
\label{fig:examples:plate_embedded_beam_results_plot}
\end{figure*}

The results for variant $A$ are shown in Figure~\ref{fig:examples:plate_embedded_beam_results_A}.
It can be seen that the the full 3D model and the new \btsvrc method exhibit the same overall behavior, while the beam experiences much larger deformations and the solid smaller ones in the \btsvc model without rotational coupling.
This is due to the fact that in the full 3D problem a considerable portion of the external load is transferred from the beam to the solid via shear stresses on the beam surface, which are represented by moments in the reduced-dimensional model.
Only the new \btsvrc method is able to capture these coupling moments.
Figure~\ref{fig:examples:plate_embedded_beam_results_B} shows the results for variant $B$.
In this case a solution for the purely translational \btsvc method (\ie only centerline position coupling) does not even exist within a quasi-static framework, since the beam has an unconstrained rigid body rotation mode around its axis of the embedded part.
Again, the displacement results of the full 3D problem and the \btsvrc model are very close to each other.
A more detailed comparison of the different variants in given in Figure~\ref{fig:examples:plate_embedded_beam_results_plot}.
Therein, the displacements along the curve indicated in Figure~\ref{fig:examples:plate_embedded_beam_problem} are visualized.
Now it also becomes clear quantitatively that the displacement results obtained with the \btsvrc method are very close to the ones obtained with the full 3D problem.
Considering that the former reduces the number of degrees of freedom by a factor of about $330$ as compared with the latter, this is a remarkable result and showcases the efficiency of the new \btsvrc method for challenging applications.

\subsection{Twisted plate}
\label{sec:examples:plate_twist}
\begin{figure*}
\centering
\includegraphics[scale=1]{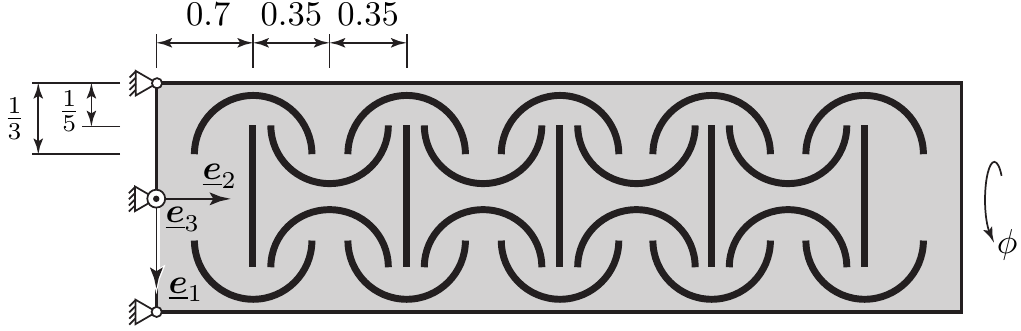}
\hfill
\includegraphics[scale=1]{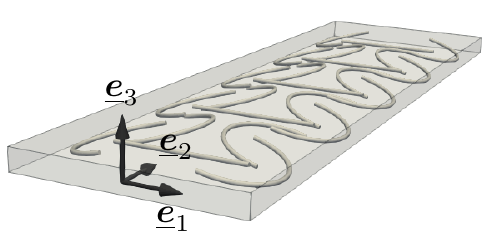}
\caption{
Twisted plate problem -- Fiber placement in the pate, all dimensions are in $\unit{m}$ (left) and 3D view illustrating the rotation of the curved fibers around the $\ey$ axis (right).
}
\label{fig:examples:plate_twist_problem}
\end{figure*}

In this final example we consider a plate, with complex, spatially distributed fiber reinforcements in 3D, \cf Figure~\ref{fig:examples:plate_twist_problem}.
The plate has the dimensions $\unit[1]{m}\times\unit[3.5]{m}\times\unit[0.1]{m}$ and consists of a Neo-Hookean material ($E = \unit[1]{N/m^2}$, $\nu = 0.3$).
The plate is fully clamped at the left face.
The right face of the plate is rotated around the $\ey$ axis with the rotation angle $\phi = [0, 2 \pi]$, \ie the plate is twisted along the $\ey$ axis.
Different shapes of fibers are embedded in the plate: semicircles with a radius of $\unit[0.25]{m}$ and straight lines with a length of $\unit[0.6]{m}$.
The fiber semicircles are rotated by $\pm\unit[15]{^{\circ}}$ with respect to the $\ey$ axis to make the example more challenging and represent general 3D fiber-solid element intersection scenarios.
The embedded fibers have a \cs radius $r=\unit[0.01]{m}$ and the material parameters are $E = \unit[400]{N/m^2}$ and $\nu = 0$.
The coupling of fibers and solid is realized with the \btsvrc method (\solidTriadAverage solid triad, $\penPos = \unit[100]{N/m^2}$ and $\penRot = \unit[100]{Nm/m}$).
First-order Lagrange polynomials are employed to discretize the translational and rotational Lagrange multipliers.
The solid plate is modeled with $10\times35\times2$ eight-noded solid-shell elements, while each fiber is discretized with four \sr beam finite elements, thus resulting in a total of 92 beam finite elements.
The displacement controlled twisting deformation of the plate is applied within 100 quasi-static load steps.
At this point it should be mentioned that this example could not be solved with the \btsvc method, since the rigid body rotation modes of the straight fibers lead to a non-converging \nr algorithm in the very first load step.
This underlines the advantages of the mechanically consistent coupling provided by the \btsvrc method.

\begin{figure*}
\begin{minipage}{\textwidth}
\centering
\setlength{\tabcolsep}{12pt}
\begin{tabular}{ccccc}
load step 50 & load step 75 & load step 85 & load step 95 & load step 100 \\
($\phi = \pi$) & ($\phi = \frac{7}{2}\pi$) & ($\phi = \frac{17}{10}\pi$) & ($\phi = \frac{19}{10}\pi$) & ($\phi = 2\pi$) \\[2mm]
\includegraphics[resolution=300]{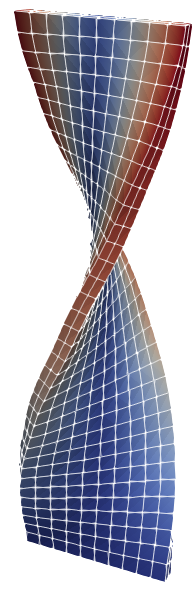} &
\includegraphics[resolution=300]{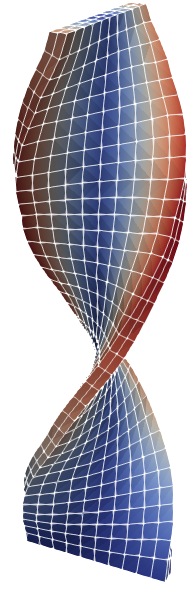} &
\includegraphics[resolution=300]{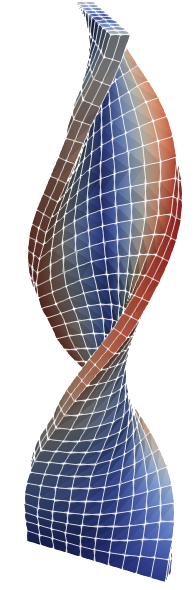} &
\includegraphics[resolution=300]{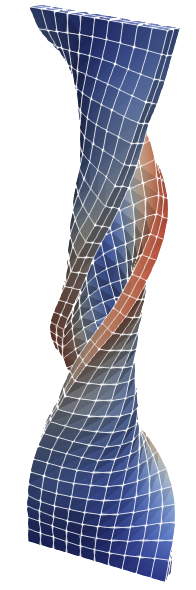} &
\includegraphics[resolution=300]{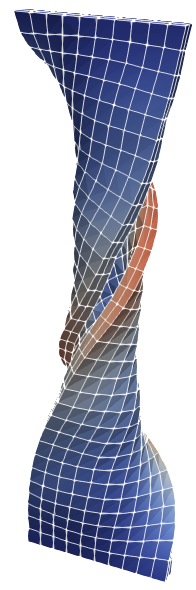} \\
\end{tabular}
\hspace{2.5cm}\includegraphics[resolution=300]{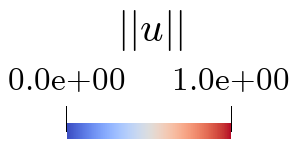}
\end{minipage}
\caption{
Deformed twisted plate problem at different load steps.
The magnitude of the displacements is shown in the solid.
}
\label{fig:examples:plate_twist_result_time_step}
\end{figure*}

\begin{figure}
\centering
\alignVcenter{\includegraphics[scale=1]{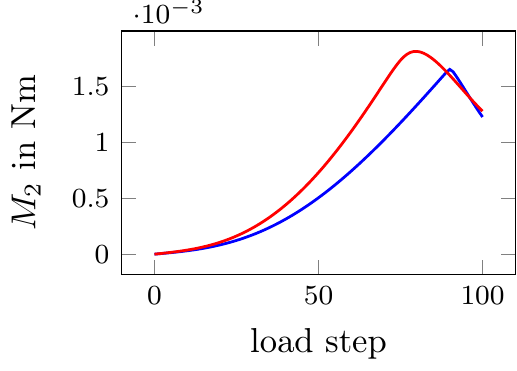}}
\includegraphics[scale=1]{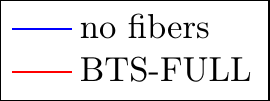}
\caption{
Reaction moment $M_2$ around the $\ey$ axis at the fully clamped surface of the plate over the course of the simulation -- for the plate with and without fiber-reinforcements.
}
\label{fig:examples:plate_twist_result_plot}
\end{figure}

Figure~\ref{fig:examples:plate_twist_result_time_step} illustrates the deformed structure at different load steps.
Until load step $75$, the reinforced plate exhibits a more or less homogeneous twist along the $\ey$ axis.
From load step $75$ to load step $100$, the reinforced plate folds around the $\ey$ axis.
To assess the non-linear behavior of this structure and evaluate the global impact of the fiber-reinforcements, the fiber-reinforced plate is compared to a simple plate (same material) without any fibers.
Figure~\ref{fig:examples:plate_twist_result_plot} depicts the reaction moment $M_2$ around the $\ey$ axis at the fully clamped surface of the plate with and without fiber-reinforcements.
Until load step $70$, the structures behave similarly.
However, as expected the fiber-reinforcements lead to an increased reaction moment for the same twist angle $\phi$, \ie to a stiffer structural response.
Both structures exhibit a limit point with an unstable post-critical solution, \ie the structures would collapse if the twist is applied in a load-controlled manner.
The fiber-reinforcements affect the critical point of the structure such that the instability occurs at a smaller twist angle and  the critical moment is increased.
This illustrates the complex influences that fiber-reinforcements may have on the global non-linear behavior of a structure.

\begin{figure*}
\centering
\includegraphics[resolution=300]{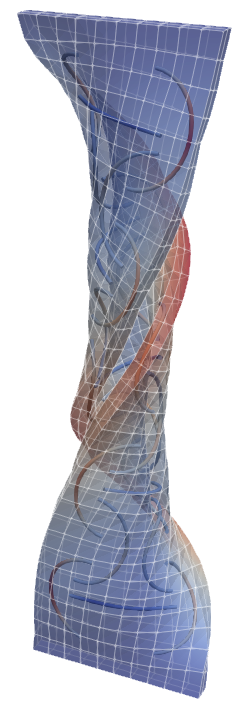}
\hspace{5mm}
\includegraphics[resolution=300]{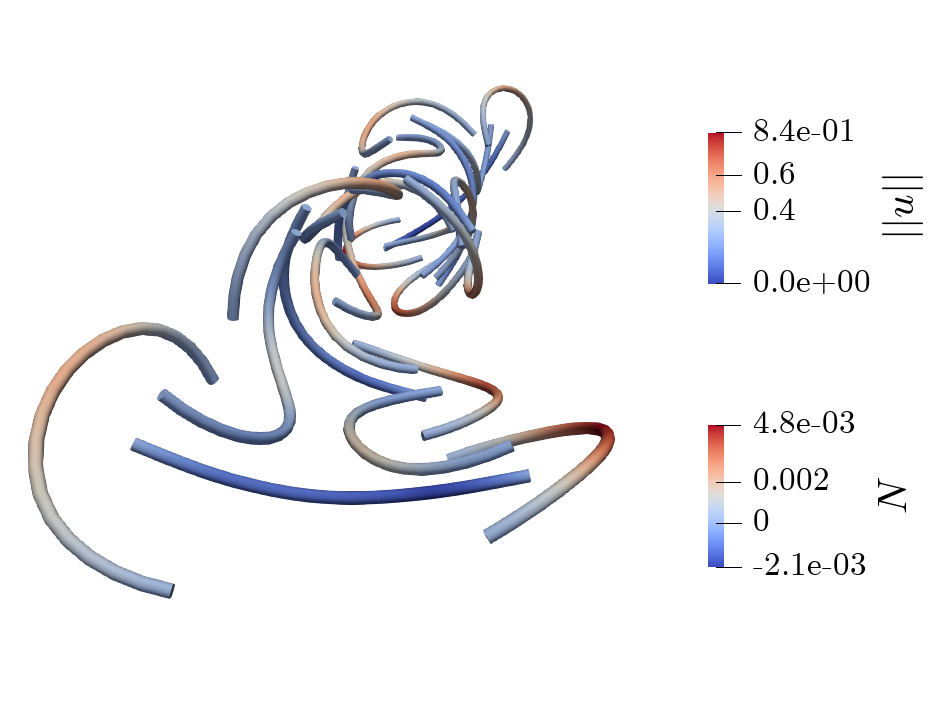}
\caption{
Deformed twisted plate problem -- full plate (right) and closeup of the embedded fibers (right).
Magnitude of the displacements is visualized in the solid and the axial force $N$ is shown in the beams.
}
\label{fig:examples:examples_plate_twist_results_deformed}
\end{figure*}

\begin{figure*}
\centering
\includegraphics[scale=1]{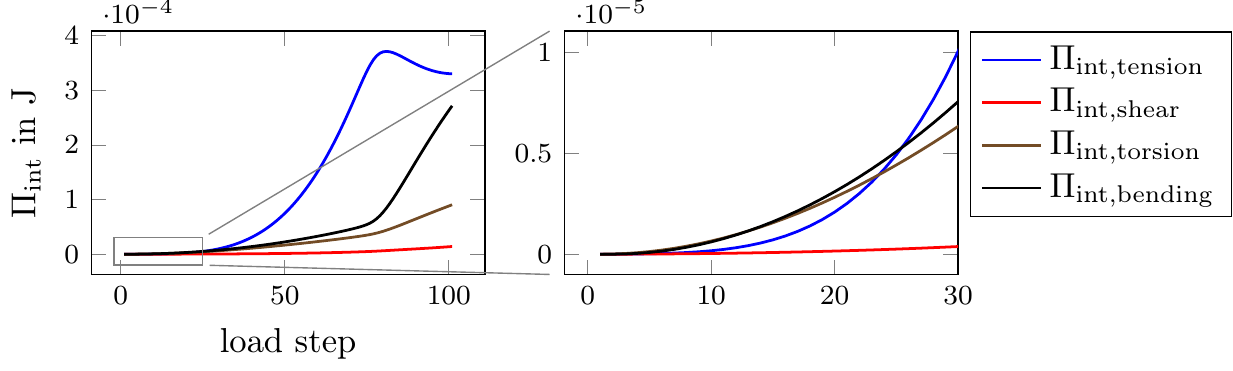}
\caption{
Twisted plate problem -- internal elastic energies in the fibers, split up in tension, shear, torsion and bending contributions.
}
\label{fig:examples:plate_twist_results_plot_energy}
\end{figure*}

Figure~\ref{fig:examples:examples_plate_twist_results_deformed} illustrates the final configuration and shows a close-up view of the deformed embedded fibers.
The maximum normal stresses in the fibers resulting from axial and bending deformations can be estimated for this example as $\approx \unit[15]{N/m^2}$ and $\approx \unit[26]{N/m^2}$ (not visualized in the figure), respectively. 
In the solid the maximum principal Cauchy stress is $\unit[0.578]{N/m^2}$ (not visualized in the figure).
As expected, the stresses in the stiff fibers are much larger than in the relatively soft solid matrix.
To further investigate the influences of the different deformation modes of the fibers, Figure~\ref{fig:examples:plate_twist_results_plot_energy} depicts the tension, shear, torsion and bending contributions to the total internal elastic energy of the fibers over the course of the simulation.
In the first few load steps, the main contributors to the internal elastic energy of the system are bending and torsion deformations, \cf right part of Figure~\ref{fig:examples:plate_twist_results_plot_energy}.
This can be attributed to the fact that in the beginning of the simulation the deformations of the plate mainly take place in in $\ez$ direction, which predominantly causes bending and torsion deformations of the fibers.
As the plate is twisted further, geometrically non-linear effects materialize especially on the outer edges of the plate.
The edges form helix like curves.
Due to the constrained displacements in $\ez$ direction at the clamped surfaces, the outside edges of the plate are stretched in $\ez$ direction.
This causes axial tension in the fiber semicircles at the outside.
Starting at approximately load step $25$, the main contribution to the internal elastic fiber energy comes from axial deformations.
In the post-buckling state, the bending deformation of the plate, and therefore also of the fibers increases.
This causes an increase in the internal elastic bending and torsion energy of the fibers.
Moreover, shear deformations only have a minor contribution to the total internal energy of the fibers, which is expected due to the slenderness of the embedded fibers, thus motivating a future use of the \btsvrc method in combination with shear stiff \kl beam theories \cite{Meier2014, Meier2015}.

The considerable contributions of bending and torsional energy to the internal elastic energy of the fibers demonstrates the importance of consistently representing these modes and coupling them to the background solid material as done by our proposed \btsvrc scheme.
For this example, this would not be the case if simplified models for the fibers (\eg modeled as strings without bending stiffness) or for the fiber-solid coupling (\eg \btsvc) were applied.

\begin{figure}
\centering
\includegraphics[resolution=300]{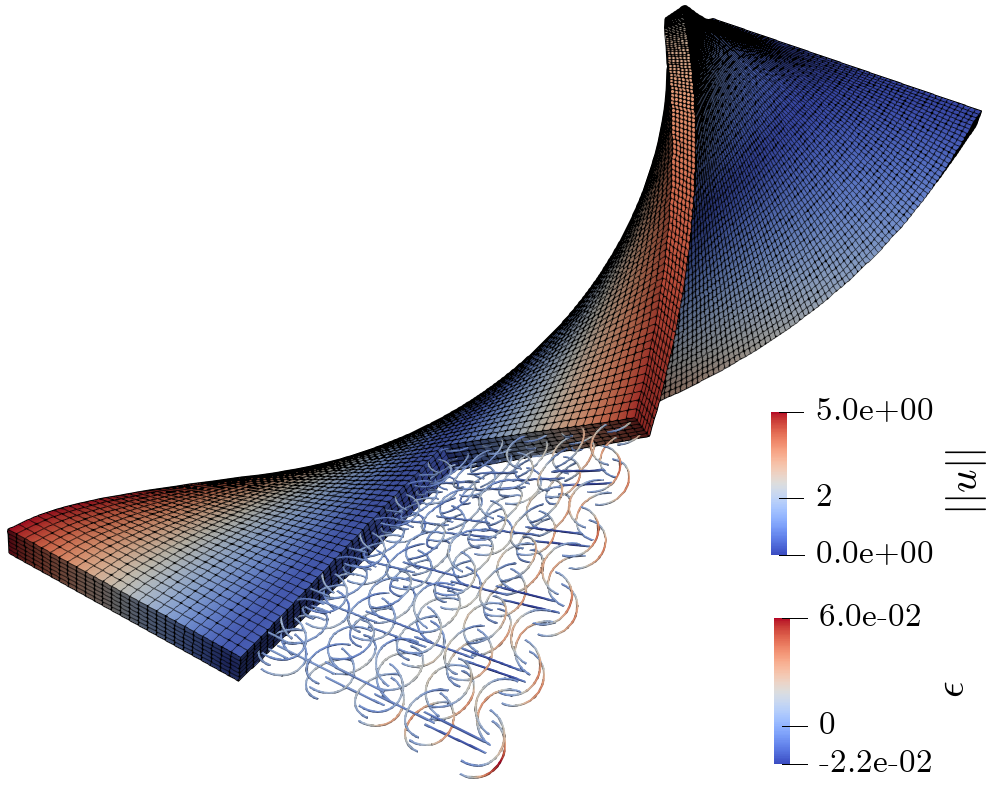}
\caption{
Twisted plate problem with an increased complexity -- deformed configuration after a rotation $\phi=\pi$.
The magnitude of the displacements is plotted in the solid and the axial strains $\epsilon$ are plotted in the beams.
}
\label{fig:examples:plate_twist_results_large}
\end{figure}

This example also showcases the maturity of the implemented \btsvrc method from an algorithmic point of view.
The chosen solid mesh, in combination with the tilted fiber semicircles results in complex 3D intersections between the beam finite elements and the solid finite elements, thus illustrating the robustness of the employed numerical integration algorithm.
As a final example, a more complex model of a fiber-reinforced plate is considered.
Therein, the dimensions of the plate are repeated $5$ times in $\ex$ and $\ey$ direction and $3$ times in $\ez$ direction.
The pattern and size of the fiber-reinforcements is similar to the one illustrated in Figure~\ref{fig:examples:plate_twist_problem}, however, in this case there are 3 layers of fiber-reinforcements over the thickness of the plate.
This results in a total of approximately $53{,}000$ %$52500$
solid finite elements and $1{,}800$ %$1792$
fibers with 4 beam finite elements each, \ie the problem size is scaled by a factor of approximately $75$ compared to the previously considered plate.
The deformed configuration of the plate is visualized in Figure~\ref{fig:examples:plate_twist_results_large} .
This further illustrates the robustness and scalability of the presented \btsvrc method for large-scale problems.

\section{Conclusion}

In this work we have proposed a 1D-3D coupling method to consistently embed 1D \cosserat beams into 3D \boltz continua (solids).
Six constraint equations act on each point along the \cosserat beam centerline, namely three translational constraints and three rotational constraints, thus resulting in a full, mechanically consistent coupling between the 1D beams and the 3D continuum.
Deriving the full 1D-3D coupling on the beam centerline from a 2D-3D coupling on the beam surface via a Taylor series expansion of the solid displacement field would require to fully couple the deformed solid directors with the undeformable beam \cs triad.
It is demonstrated that such an approach, which suppresses all in-plane deformation modes of the solid at the coupling point, might result in severe locking effects in the practically relevant regime of relatively coarse solid mesh sizes.
Therefore, a suitable triad field has to be defined in the 3D \boltz continuum that only represents solid material directions in an average sense without constraining them.
It has been shown that the rotational part of the polar decomposition of the (in-plane projection of the) solid deformation gradient is a natural choice, since it represents the average orientation of material directions of the 3D continuum in a $L_2$-optimal manner.
Additionally, several other solid triad definitions have been presented, which allow for a more efficient numerical evaluation.
Furthermore, existing Lagrange multiplier-based coupling methods for the translational degrees of freedom have been extended for the coupling of rotational degrees of freedom, all within the theory of large rotations.
The coupling equations have been discretized using a mortar-type approach and enforced using a regularized weighted penalty method.
Based on elementary numerical test cases, it has been demonstrated that a consistent spatial convergence behavior can be achieved and potential locking effects can be avoided, if the proposed \btsvrc scheme is combined with a suitable solid triad definition.
Furthermore, numerical experiments have been conducted to show the applicability of the proposed method to real-life engineering applications.

Future work will focus on the extension of the proposed beam-to-solid volume coupling approach to beam-to-solid surface coupling as well as to beam-to-solid surface contact.
Especially in the latter case, the proposed rotational coupling constraints are essential to capture effects such as frictional contact between beam and solid.
Another topic of interest for future research is the combination of 1D-3D and 2D-3D coupling within a unified beam-to-solid coupling approach.
This would allow to use 2D-3D coupling along with a refined solid mesh only in domains where high resolution of solid stress fields is of interest, and using the proposed, highly efficient 1D-3D coupling approach in the remaining problem domain.
Moreover, also a combination of the developed schemes with concepts allowing for a consistent coupling of the beam ends with the solid domain, \cf \cite{Romero2018}, are considered as promising future research direction.

\begin{acknowledgements}
Sketches in this work have been created using the Adobe Illustrator plug-in LaTeX2AI (\url{https://github.com/stoani89/LaTeX2AI}).
\end{acknowledgements}

\appendix
\section{Appendix}
\subsection{Proof of $L_2$-optimality of \solidTriadPolar triad}
\label{sec:appendix:proof_least_squares}

In the following, a proof shall be given for~\eqref{eq:least_square}.
First, an angle $\theta_0 \in [-\pi, \pi]$ is defined that represents the orientation of arbitrary in-plane directors in the reference configuration defined to coincide for solid and beam according to $\tns{g}_{S,0}(\theta_0)=\tns{g}_{B,0}(\theta_0)=\cos{(\theta_0)} \, \tns{g}_{B2,0} +\sin{(\theta_0)} \, \tns{g}_{B3,0}$.
The push-forward to the spatial configuration is given by $\tns{g}_{S}= \FNormal \, \tns{g}_{S,0}(\theta_0)$ for the solid and $\tns{g}_{B}= \tnss{R}_B \, \tns{g}_{B,0}(\theta_0)$ for the beam.
As stated in \Cref{sec:solid_triads_motivation}, it is a desirable property of the (to be defined) solid triad and therefore also of the beam triad that the base vectors $\gtriadbeam{2}$ and $\gtriadbeam{3}$ lie in the $\normal$-plane, \ie the plane spanned by the solid base vectors $\gtriadsolid{2}$ and $\gtriadsolid{3}$.
Thus, in analogy to~\eqref{sec:2D_3D_split} as stated for the solid, it is assumed that the total rotation of the beam \cs $\tnss{R}_B$ is split in a multiplicative manner into two successive rotations
\begin{align}
\label{eq:proof_least_square_0}
\rotMatBeam = \rotMatBeamtwoD \rotMatNormal
\end{align}
where $\rotMatNormal$ describes the 3D rotation from $\triadbeamO$ to $\bar{\triad}$ and $\rotMatBeamtwoD$ the quasi-2D rotation from $\bar{\triad}$ to $\triadbeam$. Thus, after push-forward to the intermediate configuration defined by $\rotMatNormal$ the corresponding directors of solid and beam still coincide:
\begin{align}
\begin{split}
\label{eq:proof_least_square_1}
\bar{\tns{g}}_{S}(\theta_0)=\bar{\tns{g}}_{B}(\theta_0) &=  \rotMatNormal \, {\tns{g}}_{B,0}(\theta_0) \\
&= \cos{(\theta_0)} \, \bar{\tns{g}}_{2} +\sin{(\theta_0)} \, \bar{\tns{g}}_{3}.
\end{split}
\end{align}
In the following, the material and spatial principle axes associated with the polar decomposition~\eqref{eq:polar_decomp_2D} of the in-plane deformation gradient $\FtwoD$ are denoted as $\tns{G}_{P2}$ and $\tns{G}_{P3}$ as well as $\tns{g}_{P2}=\rotMatSolidtwoD \tns{G}_{P2}$ and $\tns{g}_{P3}=\rotMatSolidtwoD \tns{G}_{P3}$. Since the principle axes $\tns{G}_{Pi}$ and the orthonormal base vectors $\bar{\tns{g}}_{i}$ are related by quasi-2D rotations with respect to the normal vector $\normal$, the directors in~\eqref{eq:proof_least_square_1} can alternatively be stated as
\begin{align}
\label{eq:proof_least_square_2}
\bar{\tns{g}}_{S}(\theta_0)=\bar{\tns{g}}_{B}(\theta_0) = \cos{(\tilde{\theta}_0)} \, \tns{G}_{P2} +\sin{(\tilde{\theta}_0)} \, \tns{G}_{P3},
\end{align}
where $\tilde{\theta}_0={\theta}_0- \theta_{\text{diff}}$ is defined via the constant offset value $\theta_{\text{diff}}$ describing the rotation from $\bar{\tns{g}}_{i}$ to  $\tns{G}_{Pi}$.
The final beam \cs triad follows from the second (quasi-2D) rotation $\rotMatBeamtwoD = \rotMat(\theta_B^{\text{2D}} \normal)$ from $\bar{\tns{g}}_{i}$ to $\tns{g}_{Bi}$ described by the scalar rotation angle $\theta_B^{2D}$.
In a similar fashion the (quasi-2D) rotation $\rotMatSolidtwoD = \rotMat(\theta_S^{\text{2D}} \normal)$ from $\tns{G}_{Pi}$ to $\tns{g}_{Pi}$ is described by the scalar rotation angle $ \theta_S^{\text{2D}}$.
Due to the 2D-nature of these rotations, the beam directors $\tns{g}_{B}$ in the spatial configuration can be derived from~\eqref{eq:proof_least_square_2} according to:
\begin{align}
\label{eq:proof_least_square_3}
\begin{split}
{\tns{g}}_{B}(\theta_0) = & \cos{(\tilde{\theta}_0 \!+\! \theta_B^{\text{2D}} \!-\! \theta_S^{\text{2D}})} \, \tns{g}_{P2} \\
&+ \sin{(\tilde{\theta}_0 \!+\! \theta_B^{\text{2D}} \!-\! \theta_S^{\text{2D}})} \, \tns{g}_{P3}.
\end{split}
\end{align}
Let the principle stretch ratios associated with the in-plane deformation gradient $\FtwoD$ be denoted as $\lambda_2$ and $\lambda_3$.
Then, the solid directors $\tns{g}_{S}= \FNormal \, \tns{g}_{S,0}(\theta_0)$ in the spatial configuration can be derived according to:
\begin{align}
\begin{split}
\label{eq:proof_least_square_4}
{\tns{g}}_{S}(\theta_0) \!&= \strechtwoD \rotMatSolidtwoD  [\cos{(\tilde{\theta}_0)} \, \tns{G}_{P2} +\sin{(\tilde{\theta}_0)} \, \tns{G}_{P3}]  \\
&= \lambda_2 \cos{(\tilde{\theta}_0)} \, \tns{g}_{P2} + \lambda_3 \sin{(\tilde{\theta}_0)} \, \tns{g}_{P3}.
\end{split}
\end{align}
Here, the relation $\tns{g}_{Pi}=\rotMatSolidtwoD \tns{G}_{Pi}$ and the diagonal structure $\strechtwoD=\lambda_2 \tns{g}_{P2} \otimes \tns{g}_{P2}  + \lambda_3 \tns{g}_{P3} \otimes \tns{g}_{P3}$ of the spatial stretch tensor has been exploited.
From~\eqref{eq:proof_least_square_3} and~\eqref{eq:proof_least_square_4}, the orientation angles of the spatial beam and solid directors $\tns{g}_{B}(\theta_0)$ and $\tns{g}_{S}(\theta_0)$ relative to the spatial principle axis $\tns{g}_{P2}$ can be identified according to:
\begin{align}
\label{eq:proof_least_square_5}
\theta_B(\theta_0) \!&= \tilde{\theta}_0 \!+\! \theta_B^{\text{2D}} \!-\! \theta_S^{\text{2D}},  \\
\label{eq:proof_least_square_6}
\theta_S(\theta_0) \!&= \arctan \left(\frac{\lambda_3 \sin{(\tilde{\theta}_0)}}{\lambda_2 \cos{(\tilde{\theta}_0)} } \right ).
\end{align}
Now, the difference between the solid director orientations and the beam director orientations, measured in the $L_2$-norm, shall be minimized, \ie:
\begin{align}
\label{eq:deriv_integral}
\int_{-\pi}^{\pi} (\theta_B(\theta_0)-\theta_S(\theta_0))^2 \mathrm d \theta_0 \,\, \rightarrow \,\, \text{min.}
\end{align}
As necessary condition, the first derivative of the integral with respect to $\theta_B^{2D}$ has to vanish, \ie
\begin{align}
\label{eq:deriv_vanish}
\int_{-\pi}^{\pi} (\theta_B(\theta_0)-\theta_S(\theta_0)) \mathrm d \theta_0 \,\, \dot{=} \,\, 0.
\end{align}
By exploiting the property $\theta_S(-\theta_0)=-\theta_S(\theta_0)$ of~\eqref{eq:proof_least_square_6}, it can easily be shown that~\eqref{eq:deriv_vanish} results in the requirement:
\begin{align}
\label{eq:deriv_vanish_requ}
\theta_B^{\text{2D}} = \theta_S^{\text{2D}} \quad \Leftrightarrow \quad \rotMatBeamtwoD = \rotMatSolidtwoD.
\end{align}
This means that the beam directors $\tns{g}_{Bi}$ have to coincide with the principle axes $\tns{g}_{Pi}$ and, thus, the total beam triad has to satisfy $\triadbeam=\rotMatSolidtwoD \rotMatNormal \triadbeamO$, which is identical to the solid triad definition \solidTriadPolar according to~\eqref{eq:triad_def_polar} with the initial condition $\triadbeamO=\triadsolidO$.
By checking the second derivative, it can easily be confirmed that this solid triad choice indeed results in a minimum of the $L_2$-norm in \eqref{eq:deriv_integral}.

\subsection{Full 2D-3D coupling}
\label{sec:appendix:full_2d_3d}
In the example section we compare the 1D-3D (\ie \btsvrc) method to reference solutions obtained with a 2D-3D coupling approach.
For the sake of completeness, we state the kinematic coupling constraints for the employed 2D-3D coupling approach.
The coupling constraints read
\begin{equation}
\label{eq:problem_formulation_2d-3d_coupling}
\rbeam + \rCS - \xsolid = \tnsO \quad \text{on} \quad \domainCouplingFull.
\end{equation}
Therein, $\domainCouplingFull$ is the 2D-3D coupling surface, \ie the part of the beam surface that lies within the solid volume.
Furthermore, $\rCS\in\R{3}$ is the \cs position vector, \ie the vector that points from the \cs centroid to the \cs perimeter.
The \cs position vector can be expressed by the current beam triad basis vectors $\gtriadbeam{2}$ and $\gtriadbeam{3}$, or via the \cs rotation tensor $\triadbeam$ and the Cartesian basis vectors $\ey$ and $\ez$, \ie
\begin{equation}
\rCS = \csa \gtriadbeam{2} + \csb \gtriadbeam{3} = \triadbeam \br{\csa \ey + \csb \ez}.
\end{equation}
Therein, $\csa\in\R{}$ and $\csb\in\R{}$ are the beam \cs coordinates, \ie they parametrize the beam \cs.
In the following, two methods to enforce the 2D-3D coupling conditions \eqref{eq:problem_formulation_2d-3d_coupling} are presented, once with a Lagrange multiplier method and once with a quadratic penalty potential.

\subsubsection{Penalty potential}
The quadratic penalty potential reads
\begin{equation}
\begin{split}
\couplingPotentialFull
=
\frac{\penTwoThree}{2}
\intCouplingFullOpen{
\br[2]{\rbeam + \rCS - \xsolid}\tr
\br[2]{\rbeam + \rCS - \xsolid}
}.
\end{split}
\end{equation}
Here, $\penTwoThree \in \R{}$ is a scalar penalty parameter.
Variation of the penalty potential gives the following contributions to the weak form:
\begin{equation}
\label{eq:problem_formulation:total_gp_potential_variation_2d_3d}
\begin{split}
&\dcouplingPotentialFull
=\\
& \quad
\intCouplingFullOpen{\drbeam\tr \underbrace{\penTwoThree \br[2]{\rbeam + \rCS - \xsolid}}_{\tns{f}_\twotothree}}
\\
& \qquad + \drotmultbeam\tr \underbrace{\penTwoThree \br[2]{
\Sskew[\rCS]
\br{\rbeam - \xsolid}
}}_{\tns{m}_\twotothree}
\\
&\qquad
\intCouplingFullClose{
+\dxsolid\tr \underbrace{\penTwoThree \br[2]{-\rbeam - \rCS + \xsolid}}_{-\tns{f}_\twotothree}
}.
\end{split}
\end{equation}
Therein, the coupling force $\tns{f}_\twotothree$, acting on the beam centerline and solid, can be identified.
Furthermore, $\tns{m}_\twotothree$ is the coupling moment acting on the beam \cs.
This demonstrates the projection of purely positional coupling constraints (on the surface of the beam) onto the beam centerline, and illustrates the arising rotational coupling terms in a 2D-3D coupling approach.

\subsubsection{Lagrange multiplier potential}
The 2D-3D coupling conditions \eqref{eq:problem_formulation_2d-3d_coupling} can also be enforced with a Lagrange multiplier method.
A Lagrange multiplier vector field $\lagrangeRotTtT \in \R3$ is therefore defined on the coupling surface $\domainCouplingFull$.
The total Lagrange multiplier potential for the 2D-3D coupling reads
\begin{equation}
\couplingPotentialFullLagrange = \intCouplingFull{\lagrangeRotTtT\tr\br{ \rbeam + \rCS - \xsolid}}.
\end{equation}
The variation of the total Lagrange multiplier potential gives the following contributions to the weak form:
\begin{equation}
\begin{split}
\dcouplingPotentialFullLagrange &= \intCouplingFullOpen{ \brackets[3]{(}{.}{\dlagrangeRotTtT\tr\br{ \rbeam + \rCS - \xsolid}}} + \drbeam\tr \lagrangeRotTtT
\\
& \qquad \intCouplingFullClose{ \brackets[3]{.}{)}{ + \drotmultbeam\tr \Sskew[\rCS]\lagrangeRotTtT  - \dxsolid\tr \lagrangeRotTtT}}.
\end{split}
\end{equation}
Again, this showcases the projection onto the beam centerline, in this case of the Lagrange multiplier field $\lagrangeRotTtT$, \ie the coupling surface tractions on the beam surface.

\subsection{\gptslong approach for full 2D-3D coupling}
\label{sec:appendix:full_2d_3d_discret}
Evaluating the variation of the total coupling potential \eqref{eq:problem_formulation:total_gp_potential_variation_2d_3d} on the basis of the discretized solid position field and beam \cs rotation field yields the discrete variation of the 2D-3D coupling potential:
\begin{equation}
\label{eq:discretization:2d_3d_variation}
\begin{split}
&\dcouplingPotentialFullh = \penTwoThree \intCouplingFullhOpen{\brackets[3]{(}{.}{}}
\\
&\qquad
\br{\Nbeam\dqbeam + \Sskew[\Lrot \dqbeamrot] \triadbeamh \br{\csa \ey + \csb \ez} - \Nsolid \dqsolid}\tr
\\
&\qquad
\intCouplingFullhClose{
\brackets[3]{.}{)}{
	\br{\Nbeam\qbeam + \triadbeamh \br{\csa \ey + \csb \ez} - \Nsolid \qsolid}
}}.
\end{split}
\end{equation}
Therein, $\domainCouplingFullh$ is the discrete beam surface.
It is important to point out that the beam surface is not directly discretized.
It is an analytical surface defined by the discretized beam centerline, the beam \cs orientations and the beam \cs geometry.
Equation~\eqref{eq:discretization:2d_3d_variation} can be stated in matrix form as
\begin{equation}
\label{eq:discretization:2d_3d_variation_residuum}
\begin{split}
\dcouplingPotentialFullh
&=
\matrix{
{\dqsolid}\tr & {\dqbeam}\tr & {\dqbeamrot}\tr
}
\matrix{
\intCouplingFullh{\fcsolidRotTtT} \\
\intCouplingFullh{\fcbeamPosTtT} \\
\intCouplingFullh{\fcbeamRotTtT} \\
}
\\&=
\matrix{
{\dqsolid}\tr & {\dqbeam}\tr & {\dqbeamrot}\tr
}
\matrix{
\rcsolidRotTtT \\
\rcbeamPosTtT \\
\rcbeamRotTtT \\
},
\end{split}
\end{equation}
with the generalized point-wise 2D-3D coupling forces
\begin{equation}
\begin{split}
\fcsolidRotTtT =&\ \penTwoThree \brackets{(}{.}{ \Nsolid\tr \Nsolid \qsolid - \Nsolid\tr \Nbeam \qbeam} \\
& \qquad \brackets{.}{)}{- \Nsolid\tr \triadbeamh \br{\csa \ey + \csb \ez}},
\\
\fcbeamPosTtT =&\ \penTwoThree \brackets{(}{.}{-\Nbeam\tr \Nsolid \qsolid + \Nbeam\tr \Nbeam \qbeam} \\
& \qquad \brackets{.}{)}{+ \Nbeam\tr \triadbeamh \br{\csa \ey + \csb \ez}},
\\
\fcbeamRotTtT =&\ \penTwoThree \brackets{(}{.}{- \Lrot\tr \Sskew[\triadbeamh \br{\csa \ey + \csb \ez}] \Nsolid \qsolid } \\
& \qquad \brackets{.}{)}{ + \Lrot\tr \Sskew[\triadbeamh \br{\csa \ey + \csb \ez}] \Nbeam\qbeam}.
\end{split}
\end{equation}
Furthermore, $\rcsolidRotTtT$, $\rcbeamPosTtT$ and $\rcbeamRotTtT$ are the local residual vectors.
Again, a linearization of the residual contributions with respect to the discrete \btsFull pair degrees of freedom is required for the \nr algorithm.
The linearization reads:
\begin{equation}
\label{eq:discretization:2d_3d_variation_system}
\begin{split}
&\matrix{\Delta \rcsolidRotTtT \\ \Delta \rcbeamPosTtT \\ \Delta \rcbeamRotTtT}
=
\intCouplingFullhOpen{\brackets[4]{(}{.}{}}
\\
& \qquad\intCouplingFullhClose{\brackets[4]{.}{)}{
\matrix{
\pfrac{\fcsolidRotTtT}{\qsolid} & \pfrac{\fcsolidRotTtT}{\qbeampos} & \pfrac{\fcsolidRotTtT}{\qbeamrot}\Ttrans(\rotvecbeamh) \Irot \\
\pfrac{\fcbeamPosTtT}{\qsolid} & \pfrac{\fcbeamPosTtT}{\qbeampos} & \pfrac{\fcbeamPosTtT}{\qbeamrot}\Ttrans(\rotvecbeamh) \Irot \\
\pfrac{\fcbeamRotTtT}{\qsolid} & \pfrac{\fcbeamRotTtT}{\qbeampos} & \pfrac{\fcbeamRotTtT}{\qbeamrot}\Ttrans(\rotvecbeamh) \Irot \\
}
}}
\matrix{\Dqsolid \\ \Dqbeampos \\ \Dqbeamrot}
.
\end{split}
\end{equation}
The local contributions \eqref{eq:discretization:2d_3d_variation_residuum} and \eqref{eq:discretization:2d_3d_variation_system} to the global residual and the stiffness matrix, respectively, can be assembled in a straightforward manner and will not be stated here for the sake of brevity.
As in the \btsvrc mortar-type coupling, all derivatives explicitly stated in \eqref{eq:discretization:2d_3d_variation_system} are evaluated using forward automatic differentiation (\fad).

\begin{figure}
\centering
\includegraphics[scale=1]{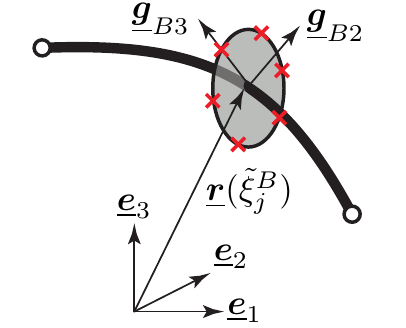}
\caption{
Illustration of the discrete coupling points for 2D-3D coupling along a single \cs \cite{Steinbrecher2020}.
}
\label{fig:examples:convergence_2D-3D}
\end{figure}

In practice, all integrals presented in this section are evaluated using a  \gpts approach as illustrated in Figure~\ref{fig:examples:convergence_2D-3D}, \cf \cite{Steinbrecher2020}.
At each Gauss--Legendre point $\tilde{\xi}_i^B$ along the beam centerline, multiple equally spaced coupling points (illustrated with the symbol '$\times$' in Figure~\ref{fig:examples:convergence_2D-3D}) are defined along the circumference of the corresponding \cs.
Mechanically speaking, each coupling point is tied to the underlying solid via a linear penalty constraint.
\todoDiss{The penalty parameter and the number of Gauss--Legendre points in axial direction and integration points in circumferential direction are chosen according to \cite{Steinbrecher2020}.}

\bibliographystyle{spmpsci}      % mathematics and physical sciences
\bibliography{literature,steinbrecher}

\end{document}